\documentclass[
    aps, pra, reprint,
    twocolumn,
    showkeys,
    xcolor, 
    groupedaddress,
    nofootinbib
    ]{revtex4-2}

\usepackage{graphicx}
\usepackage{silence}
\usepackage[dvipsnames]{xcolor}
\usepackage[colorlinks=true, citecolor=blue, linkcolor=black, urlcolor=purple]{hyperref}
\usepackage{enumitem}
    \setlist[itemize]{noitemsep, topsep=0pt}
    \setlist[enumerate]{noitemsep, topsep=0pt}
\usepackage{verbatim}
\usepackage{booktabs}
\usepackage{array}
\usepackage{float}
\usepackage[normalem]{ulem}
\usepackage{afterpage}
\usepackage{adjustbox}
\usepackage{bibunits}

\usepackage{amsthm}
\usepackage{amssymb}
\usepackage{mathtools,amsmath}
\usepackage{bm}
\usepackage{yhmath}
\usepackage{cancel}
\usepackage{relsize}
\usepackage{algorithm}
\usepackage{algpseudocode}
\usepackage{dsfont}
\usepackage[capitalize]{cleveref}
\usepackage{simplewick}

\usepackage{physics}
\usepackage{tikz}
\usetikzlibrary{decorations.pathreplacing}
\usetikzlibrary{decorations.pathmorphing}
\usetikzlibrary{backgrounds}
\usetikzlibrary{decorations.markings, arrows.meta, math}

\newcommand{\be}{\begin{equation}}
\newcommand{\ee}{\end{equation}}

\algnewcommand\algorithmicinput{\textbf{Input:}}
\algnewcommand\Input{\item[\algorithmicinput]}
\algnewcommand\algorithmicoutput{\textbf{Output:}}
\algnewcommand\Output{\item[\algorithmicoutput]}

\makeatletter
\newsavebox{\@brx}
\newcommand{\llangle}[1][]{\savebox{\@brx}{\(\m@th{#1\langle}\)}%
  \mathopen{\copy\@brx\kern-0.5\wd\@brx\usebox{\@brx}}}
\newcommand{\rrangle}[1][]{\savebox{\@brx}{\(\m@th{#1\rangle}\)}%
  \mathclose{\copy\@brx\kern-0.5\wd\@brx\usebox{\@brx}}}
\makeatother

\newcommand{\PaperTitle}{Fast, accurate, high-resolution simulation of large-scale Fermi-Hubbard models on a digital quantum processor}
\newcommand{\PaperAffiliation}{Q-CTRL, Los Angeles, CA USA and Sydney, NSW Australia}

\newcommand{\AuthorSona}{Khadijeh Sona Najafi}
\newcommand{\AuthorGavin}{Gavin S. Hartnett}
\newcommand{\AuthorAleksei}{Aleksei Khindanov}
\newcommand{\AuthorHaoran}{Haoran Liao}
\newcommand{\AuthorMichaelSchutzman}{Michael Schutzman}
\newcommand{\AuthorMichaelHush}{Michael R. Hush}
\newcommand{\AuthorMichaelBiercuk}{Michael J. Biercuk}
\newcommand{\AuthorYuval}{Yuval Baum}

\newcommand{\PaperAuthorsText}{
    \AuthorGavin,
    \AuthorSona,
    \AuthorAleksei,
    \AuthorHaoran,
    \AuthorMichaelSchutzman,
    \AuthorMichaelHush,
    \AuthorMichaelBiercuk,
    \AuthorYuval
}

\newcommand{\PaperAuthorsRevTeX}{%
    \author{\AuthorGavin}
    \affiliation{\PaperAffiliation}

    \author{\AuthorSona}
    \affiliation{\PaperAffiliation}

    \author{\AuthorAleksei}
    \affiliation{\PaperAffiliation}

    \author{\AuthorHaoran}
    \affiliation{\PaperAffiliation}

    \author{\AuthorMichaelSchutzman}
    \affiliation{\PaperAffiliation}

    \author{\AuthorMichaelHush}
    \affiliation{\PaperAffiliation}

    \author{\AuthorMichaelBiercuk}
    \affiliation{\PaperAffiliation}

    \author{\AuthorYuval}
    \affiliation{\PaperAffiliation}
}

\makeatletter
\newcommand{\ApplyPaperMetadata}[1]{
    \def\PaperMetadataMode{#1}
    \def\MainMode{main}
    \def\SupplementMode{supplement}
    \ifx\PaperMetadataMode\MainMode
        \title{\PaperTitle}
        \PaperAuthorsRevTeX
    \else\ifx\PaperMetadataMode\SupplementMode
        \begin{center}
        {\large \textbf{Supplementary Material for\\ ``\PaperTitle''}}\\
        \vspace{0.5cm}
        \PaperAuthorsText\\
        \vspace{0.25cm}
        \textit{\PaperAffiliation}\\
        \vspace{0.15cm}
        \end{center}
        \vspace{1cm}
    \else
        \PackageError{frontmatter}{Unknown metadata mode `#1'}{Use `main' or `supplement'.}
    \fi\fi
}
\makeatother

\makeatletter 
\renewcommand{\fnum@figure}{\textbf{Fig.~\thefigure}}
\def\@caption@fignum@sep{\textbf{.} }
\makeatother 

\makeatletter
\newcommand{\setbibunitprefix}[1]{%
  \def\@extra@binfo{#1}%
  \def\@extra@b@citeb{#1}%
}
\makeatother

\begin{document}

\begin{bibunit}[apsrev4-2]
\setbibunitprefix{@main}

\ApplyPaperMetadata{main}

\begin{abstract}
    The Fermi-Hubbard model provides a paradigmatic description of strongly correlated electrons relevant to various problems in materials science.  Despite its importance, simulating its non-equilibrium dynamics exactly is extremely challenging beyond modest scales due to exponential expansion of the Hilbert space with increasing system size.  Quantum computers have been postulated as useful tools to address these computational bottlenecks, but have been limited by both machine size and the impact of hardware noise and error.  Here, we report experimental digital quantum simulation of the one-dimensional Fermi-Hubbard model on a superconducting quantum processor at a scale beyond the reach of exact statevector simulation and challenging for state-of-the-art tensor-network methods.  We encode this problem using up to 120 qubits through an efficient mapping that reduces circuit complexity, and we improve accuracy through error suppression to simulate dynamical evolution using up to 90 Trotter steps. From a vacancy defect introduced in the middle of an $L=31$-site (62-qubit) N\'{e}el initial state, we directly observe spin-charge separation to $t=9$ in natural units using up to 90 Trotter steps, and quantitatively extract velocities $v_c$, $v_s$, which match classical simulations across a range of model parameters, and analytics in appropriate regimes.  We then extend experiments to $L=60$ fermions (120 qubits) and evolution times to $t=6$ using 30 Trotter steps. Quantum-processor outputs agree quantitatively with approximate classical simulations performed using a time-dependent variational principle (TDVP) solver and we observe that increasing the TDVP bond dimension through $\chi = 4096$ expands the range of evolution times within which agreement has RMSE $\sim 1\%$ before the approaches diverge.  Owing to the large scale of the simulation and the use of efficient overhead-free error-suppression techniques, for simulated evolution times at the limit of quantum/classical agreement ($t\gtrsim 5$ in natural hopping units), the wall-clock runtime of the quantum processor is approximately three orders of magnitude faster than TDVP simulations with $\chi = 4096$, and implemented with the publicly-available \texttt{ITensor} and \texttt{TeNPy} packages. These results establish contemporary digital quantum processors as a versatile, quantitatively accurate, and competitive platform for the study of fermionic many-body dynamics in regimes where leading classical methods can become prohibitively expensive.
\end{abstract}

\maketitle

\vspace{0.15cm}

Strongly correlated electron systems carry special significance in condensed matter physics, hosting phenomena ranging from Mott-insulator transitions and unconventional superconductivity to strange-metal transport and quantum magnetism~\cite{Imada1998,Lee2006,Keimer2015}. The Fermi-Hubbard model captures the interplay between kinetic energy and onsite Coulomb repulsion with a minimal set of parameters, and provides the canonical theoretical framework for studying these phases of matter~\cite{Hubbard1963,Gutzwiller1963}. In one dimension, the model's spectrum and thermodynamic properties can be calculated exactly via the Bethe ansatz~\cite{LiebWu1968, Essler2005}, though closed-form expressions for dynamical quantities and real-time evolution are generally not obtainable and instead require numerical methods. The low-energy physics away from half-filling is described by Tomonaga--Luttinger liquid theory~\cite{Haldane1981}, while the half-filled system is a Mott insulator with a charge gap for all $U>0$. In two dimensions, the model is widely believed to contain the key ingredients of high-temperature superconductivity in cuprates~\cite{Anderson1987,Zhang1988,Arovas2022}, yet no exact solution exists and its phase diagram at finite doping remains incompletely understood.

In either setting, numerical simulations are critical to understanding the broad range of real-time dynamics in the Fermi-Hubbard model---how correlations build up after a quantum quench, how energy redistributes between spin and charge sectors, and how local charge defects and the associated spin disturbances propagate at finite energy density. Simulation via exact diagonalization is limited to small system sizes due to the unfavorable growth of the Hilbert space with particle count~\cite{Dagotto1994}: $L\lesssim 14$ sites for the 1D Fermi-Hubbard model~\cite{innerberger2020electron}. Recent advances in exact simulation methods such as the \texttt{ffsim} package can improve the memory prefactor and reach larger systems at low filling, but still face the same exponential scaling limitation~\cite{sung2026ffsim}. As a result, a range of approximate numerical methods have been developed to overcome this challenge, including quantum Monte Carlo~\cite{Troyer2005}, dynamical mean-field theory~\cite{Georges1996}, approximate Heisenberg simulation of Pauli operators~\cite{beguvsic2024fast, beguvsic2025real, beguvsic2025simulating, rudolph2025pauli, fontana2025classical, angrisani2025simulating, lin2026utility}, and tensor-network methods~\cite{Haegeman2016,Paeckel2019}.  In the latter class, time-dependent variational principle (TDVP) solvers are particularly powerful in one dimension~\cite{Haegeman2016,Paeckel2019}; the \texttt{ITensor} and \texttt{TeNPy} packages in particular~\cite{itensor, hauschild2018TeNPy, hauschild2024TeNPy} have become workhorses in the computational condensed-matter research community, with over 1{,}500 published papers using these packages over the last decade.
Despite their popularity, the computational cost of accurate simulation via TDVP grows exponentially with time and system size after a quench.  This arises because the required bond dimension grows exponentially with the entanglement entropy, which in turn grows linearly in time before saturating at a volume-law value $\sim L$.

Quantum computers provide a well-established framework for the efficient simulation of local many-body Hamiltonians. Analog quantum simulators based on ultracold atoms have realized Fermi-Hubbard and spin models at large scales, including single-site resolution of spin and charge degrees of freedom~\cite{Britton2012,Bakr2009,Sherson2010,las2015fermionic,reiner2016emulating,Mazurenko2017,Vijayan2020,Arute2020SpinCharge,Koepsell2021,Stanisic2022,celeri2023digital,gonzalez2023fermionic,srinivasan2024trapped, khodaeva2024quantum,michel2024hubbard,wolf2026}, but generally face limits on tunability imposed by the physical mechanisms used to realize interaction Hamiltonians.  In the alternative digital approach, the time-evolution operator is decomposed into a sequence of discrete circuit elements using Trotter techniques, enabling programmable dynamical control over model parameters~\cite{Cade2020,Stanisic2022}.  It is known that the real-time evolution generated by a local Hamiltonian acting on $n$ degrees of freedom for time $t$ can be approximated to accuracy $\epsilon$ using a number of quantum gates that scales polynomially in $n$, $t$, and $1/\epsilon$~\cite{Lloyd1996,Abrams1997,Childs2021}, making digital quantum simulation attractive from a computational complexity standpoint. Early experimental efforts demonstrated digital quantum simulation of the Fermi-Hubbard model on a handful of fermionic modes~\cite{barends2015digital}, and this approach has recently advanced to impressive scales, with digital quantum simulations executed on superconducting processors approaching 100 qubits~\cite{Alam2025_2D, Evered2025, Chowdhury2026, lee2026} as well as trapped-ion platforms reaching 56 qubits~\cite{Alam2025_ion, Granet2025}.

Here, we report digital quantum simulation of the 1D Fermi-Hubbard model on a superconducting processor for particle numbers and evolution times that are far beyond the reach of exact-diagonalization methods, and challenging for state-of-the-art simulation methods.  Our approach introduces an efficient compilation scheme that encodes the problem using a fermion-to-qubit mapping placing all hopping and interaction gates on adjacent pairs of qubits, plus use of a fermionic SWAP ($\mathrm{fSWAP}$) network. These techniques are combined with a comprehensive set of runtime error-reduction strategies which do not incur execution overhead and push the limits of our simulations up to: (widest) $L=60$, $120$ qubits, $t=6$ in natural units of the inverse hopping amplitude $t_h^{-1}$, and 30 Trotter steps; (deepest) $L=31$, $62$ qubits, $t=9$, and 90 Trotter steps.  Our simulations begin by confirming known phenomena using $L=31$, demonstrating spin-charge separation by tracking the propagation of charge and spin wavefronts from a single-vacancy N\'{e}el initial state.  Quantitative extraction of charge and spin velocities $v_c$, $v_s$ over a range of interaction parameters yields results in good agreement with TDVP-based classical simulations, and with the Bethe ansatz and free-fermion models in appropriate limits of interaction strength. We then enter an otherwise unexplored regime, studying relaxation dynamics of initial Fock states for $L=60$ using 120 qubits. Over all lattice sites we observe agreement between quantum and classical TDVP simulations within root-mean-square error (RMSE) $\lesssim 1\%$, up to $t\approx 5.2$ using bond dimension $\chi=4096$.  Beyond this evolution time, the agreement between quantum and classical simulations diverges, with RMSE reaching $\sim4\%$ at $t\approx 6$; in this range of evolution times the accuracy of the digital quantum simulation is indeterminate. We include a wall-clock runtime comparison between quantum and classical simulation methods, showing orders of magnitude faster execution using the digital quantum simulator compared against widely-used and publicly-available implementations of TDVP, with a speedup of over $3000\times$ compared to \texttt{ITensor} ~\cite{itensor} and $\sim500\times$ compared to \texttt{TeNPy} ~\cite{TeNPy2024} while achieving similar accuracy.  We also provide a detailed analysis of alternative classical-simulation methods based on Heisenberg methods and compare performance against TDVP.

\begin{figure*}[hpt!]
     \centering
    \includegraphics[width=\linewidth]{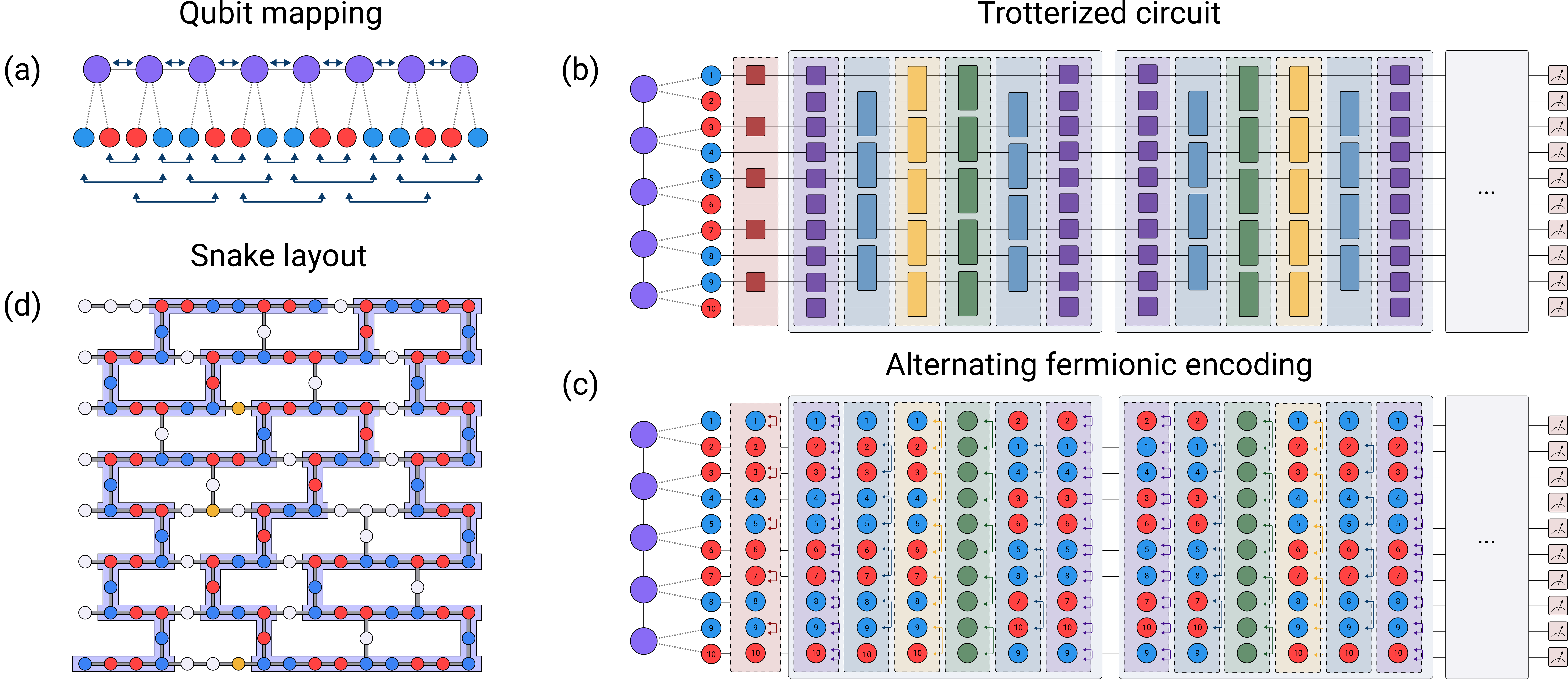}
    \caption{
    Application-aware compilation for fermionic simulation.
    (a) The fermion-to-spin mapping represents a length-$L$ chain of sites with $L=8$ here (top row, purple circles) as a system of $2L$ qubits, one for each site/spin combination (bottom row, red and blue circles). We use a pair-interleaved ordering $\downarrow \uparrow \uparrow \downarrow \downarrow \uparrow \uparrow \downarrow \cdots$. The nearest-neighbor hopping interactions (indicated by the $\leftrightarrow$ arrows in the top row) map to ``short-hop'' and ``long-hop'' terms (bottom arrows).
    (b) The Trotterized time-evolution circuit with layers color-coded for simplicity. Red indicates state preparation.  Purple indicates the single-qubit $R_Z$ for onsite and chemical potential.  Blue indicates short hopping terms, onsite interactions are yellow, and $\mathrm{fSWAP}$ for spin exchange are green.  Subsequent Trotter layers are mirrored to improve the Trotter error scaling of the overall circuit.
    (c) As a companion to (b), this panel tracks the degree of freedom encoded by a given qubit. The $\mathrm{fSWAP}$ layer exchanges adjacent spin-up and spin-down qubits.  Alternating red and blue shading indicates qubit pairs used to encode a single fermion.
    (d) Embedding the qubitized spin chain into a heavy-hex coupling graph on an IBM quantum device. Qubits are colored according to which spin they encode---spin-up (red) or spin-down (blue). Qubits in use have blue shading around the chain.   The snaking layout is chosen to avoid qubits with low gate fidelity (indicated by orange color coding) relative to the median.
    }
    \label{fig:fermion_mapping_and_circuit}
\end{figure*}

The Fermi-Hubbard model, defined over a general graph $G = (V, E)$, is:
\begin{equation}
\begin{aligned}
  H ={} & -\sum_{(i,j)\in E} t_{ij} \sum_{\sigma}
        \Bigl(c^{\dagger}_{i\sigma}c_{j\sigma}
        + c^{\dagger}_{j\sigma}c_{i\sigma}\Bigr) \\
      & \quad + \sum_{i\in V} U_i \, n_{i\uparrow} \, n_{i\downarrow} - \sum_{i\in V}\sum_\sigma \mu_{i\sigma} \, n_{i\sigma} \,.
  \label{eq:H_general}
\end{aligned}
\end{equation}
Here, $c_{i\sigma}^{\dagger}$ is the creation operator for lattice site $i$ and spin $\sigma \in \{ \uparrow, \downarrow\}$, and $n_{i\sigma} := c^{\dagger}_{i\sigma}c^{\phantom{\dagger}}_{i\sigma}$ is the number operator for site $i$ and spin $\sigma$. The first term represents a hopping kinetic interaction, with strength controlled by a (possibly inhomogeneous) coupling $t_{ij}$. The second term represents the onsite interaction, controlled by the coupling $U_i$. For $U_i>0$, doubly occupied orbitals are energetically penalized, and the interaction is repulsive; for $U_i<0$ the situation is reversed, and the interaction is attractive. Lastly, $\mu_{i\sigma}$ is the chemical potential, which encodes onsite energies due to external potentials or disorder.

Equation~\eqref{eq:H_general} represents a broad family of electronic systems. In this work, we focus on the case of a homogeneous one-dimensional chain of fermions to facilitate the direct benchmarking of quantum computer performance against classical alternatives.  The model is formulated as
\begin{equation}
    \begin{aligned}
    H ={} & -t_h \sum_{i=0}^{L-2} \sum_\sigma \left( c^{\dagger}_{i\sigma} c^{\phantom{\dagger}}_{i+1,\sigma} +  c^{\dagger}_{i+1,\sigma} c_{i\sigma} \right) \\
    & \quad {}+ U \sum_{i=0}^{L-1} n_{i\uparrow} n_{i\downarrow} - \mu \sum_{i,\sigma} n_{i\sigma} \,,
    \end{aligned}
    \label{eq:H_1D}
\end{equation}
where $t_h$ is the uniform hopping strength. We will implicitly measure time in units of $1/t_h$ and thus often set $t_h = 1$ in what follows. We emphasize that none of the following claims rely on the existence of translation symmetry or the integrability of the model in this parameter regime.

Our framework for digital quantum simulation is as follows. First, an initial state $\ket{\psi_0}$ is prepared; in this work, we will always take $\ket{\psi_0}$ to be a Fock state characterized by the set of occupation numbers for each of the $2L$ orbitals:
\begin{equation}
    \ket{n_{0,\uparrow} n_{0, \downarrow} \cdots n_{L-1, \downarrow}} = \prod_{i=0}^{L-1} \prod_{\sigma \in \{\uparrow, \downarrow\}} \left( c_{i \sigma}^{\dagger} \right)^{n_{i, \sigma}} \ket{\mathrm{vac}} \,,  
\end{equation}
where $\ket{\mathrm{vac}}$ is the zero-particle vacuum state, and $n_{i, \sigma} \in \{0, 1\}$ are the occupation numbers (Fig.~\ref{fig:fermion_mapping_and_circuit}).

In our studies we explore relaxation dynamics following a quantum quench from a half-filled N\'{e}el state $\ket{\psi_0} = \ket{\downarrow \uparrow \downarrow \uparrow \cdots}$ (including instances with an inserted local vacancy), where ${\ket{\downarrow} \equiv \ket{0,1}}$ and ${\ket{\uparrow} \equiv \ket{1,0}}$ are shorthands for single-particle occupations. Quench dynamics are observed by approximately time-evolving the initial state under the action of the model Hamiltonian by applying an $n_{\mathrm{step}}$ Trotter circuit. Each Trotter step approximates the time evolution for a step size of $\Delta t$; the entire circuit therefore evolves the system up to a total time of $t = n_{\mathrm{step}} \, \Delta t$ (up to maximum value $T=60\times \Delta t$) resulting in the state
\begin{equation}
    \ket{ \psi_t } = \prod_{k=1}^{n_{\mathrm{step}}} U_k \ket{\psi_0} \,,
\end{equation}
where $U_k$ is the unitary corresponding to the $k$-th Trotter layer. Once $\psi_t$ has been prepared, the state is measured in a suitable basis depending on the desired expectation values we wish to estimate. Here, we are primarily interested in observables in the occupation (Fock) basis (equivalently, $Z$-basis), which allows access to both particle and spin densities and correlations.

Our experimental implementation on quantum hardware relies on \emph{application-aware compilation}.  By framing the simulation problem directly in terms of the relevant physical quantities---the Hamiltonian, initial state, simulation time, and target observables---we co-optimize the circuit design and compilation procedure. Specifically, accounting for device topology as part of the Trotterization scheme yields significant performance gains, ultimately contributing to the viability of the large-scale simulations executed in this work.

Application-aware compilation begins with the task of mapping fermionic variables to spin variables, which is accomplished through the Jordan--Wigner transformation~\cite{JordanWigner1928}:
\begin{equation}
    c_{J} = \frac{1}{2} \left( \prod_{k=0}^{J-1} Z_k \right) ( X_J + i Y_J) \,.
\end{equation}
To fully specify the transformation, the two fermion indices, the site index $i$ and the spin index $\sigma$, must be mapped to the single index $J=J(i,\sigma)$, resulting in a qubitized Hamiltonian. We seek a mapping that avoids introducing both high-weight and non-local Paulis in $H$ to ensure short-depth Trotter circuits that may be efficiently compiled for IBM Heron devices with heavy-hex topology~\cite{chamberland2020topological} (Fig.~\ref{fig:fermion_mapping_and_circuit}(a)). The \textit{pair-interleaved} ordering ${\{ c_{0 \downarrow}, c_{0 \uparrow}, c_{1 \uparrow}, c_{1 \downarrow}, c_{2 \downarrow}, c_{2 \uparrow}, c_{3 \uparrow}, c_{3 \downarrow}, \cdots \}}$ satisfies this criterion when used in conjunction with an fSWAP network~\cite{Jiang2018, Kivlichan2018}. The qubitized Hamiltonian is then
\begin{align}
    H &= -\frac{t_h}{2} \sum_{j=0}^{L-2} \Bigl( X_{2j+1} X_{2j+2} + Y_{2j+1} Y_{2j+2} \nonumber \\
    &\qquad +  
    X_{2j} Z_{2j+1} Z_{2j+2} X_{2j+3} + Y_{2j} Z_{2j+1} Z_{2j+2} Y_{2j+3} \Bigr) \nonumber \\
    &\quad + \frac{U}{4} \sum_{i=0}^{L-1} \left( \mathds{1} - Z_{2i}\right) \left( \mathds{1} - Z_{2i+1} \right) \,.
\end{align}

Next, we address the implementation of Hamiltonian terms as circuit elements.  For our 1D system, the pair-interleaved ($\downarrow \uparrow \uparrow \downarrow \downarrow \uparrow \uparrow \downarrow \cdots$) ordering introduces an asymmetry in the kinetic term leading to weight-2 ``short-hopping'' terms of the form $XX + YY$, and weight-4 ``long-hopping'' terms of the form $XZZX + YZZY$. The short-hopping terms may be readily implemented in a Trotter scheme via two-qubit $R_{XX+YY}$ rotation gates.  The long-hopping terms require additional care; we employ a layer of $\mathrm{fSWAP}$s to exchange the spin-up and spin-down fermions, $c_{i \uparrow} \leftrightarrow c_{i \downarrow}$, which transforms short-hops into long-hops, and vice versa. The onsite interaction and chemical potential terms together yield two-qubit $R_{ZZ}$ rotations between the spin-up and spin-down qubits at each site $(2i,2i+1)$ and single-qubit $R_Z$ rotations on every qubit, with the latter combining the $Z$ contributions from both terms.

The fermion ordering utilized in our experiments leads to an efficient circuit construction, with the complete circuit as implemented illustrated in Fig.~\ref{fig:fermion_mapping_and_circuit}(b-c).  The first step consists of a sequence of $X$ gates on select qubits to prepare a desired Fock state. A single Trotter layer is then applied, beginning with a one-qubit layer that implements the $R_Z$ rotations needed for the onsite and chemical potential terms (purple). Next, the short-hopping terms are implemented by a layer of $R_{XX + YY}$ gates (blue), and a layer of $R_{ZZ}$ gates implements the onsite interactions (yellow). A subsequent layer of $\mathrm{fSWAP}$ gates (green) performs the spin-exchange operation which permutes the fermions such that the ordering alternates every Trotter layer from $\downarrow \uparrow \uparrow \downarrow \cdots$ to $\uparrow \downarrow \downarrow \uparrow \cdots$, and so on. This converts the as-yet unimplemented long-hopping interactions into short-hopping interactions, which are then executed with a second layer of $R_{XX + YY}$ gates (blue), followed by a symmetrized $R_Z$ layer (purple). Subsequent Trotter layers are mirrored to improve the Trotter error scaling of the overall circuit.

In this construction, a single Trotter step is a first-order approximation with error $\mathcal{O}(\Delta t^2)$; however, the mirrored contraction of two adjacent steps enjoys an enhanced symmetry that yields second-order accuracy with $\mathcal{O}(\Delta t^3)$ error. Thus, the total Trotter error after $n$ steps is $\mathcal{O}(n \Delta t^3)$ for $n$ even and $\mathcal{O}(\Delta t^2) + \mathcal{O}(n \Delta t^3)$ for $n$ odd, where the quadratic term arises from the single un-paired step.  Due to the local nature of the qubitized Hamiltonian, each step requires only constant depth in the system size $L$.  For each choice of Hamiltonian parameters, we empirically select the Trotter step $\Delta t$ that balances total evolution time against algorithmic error (see Supplementary Material, Sec.~\ref{app:qc_contruction_compilation_error_suppresion} for detailed circuit compilation and Sec.~\ref{sec:trotter_step_selection} for Trotter error analysis.)

With the Trotter circuits defined, we utilize the \mbox{Q-CTRL} compilation pipeline in \texttt{Fire Opal}~\cite{Mundada2023, kakkar2025no, fire_opal} to transpile and schedule the circuits for execution on the IBM quantum processor. First, the circuits are compiled using the native gate set on IBM Heron devices.  This includes fractional gates $R_{ZZ}(\theta)$, $R_Z(\theta)$, in addition to the fixed-angle gates $\mathrm{C}Z$, $X$, and $\sqrt{X}$, but does not include the $\mathrm{fSIM}(\theta,\phi)$ gate specifically tailored for fermionic simulations and available on other platforms~\cite{Kivlichan2018}. Next, we embed the ordered fermions into the device connectivity as a snaking path that is natively compatible with the heavy-hex topology of IBM Heron processors (Fig.~\ref{fig:fermion_mapping_and_circuit}(d)). A layout selection process~\cite{Hartnett2024learningtorank, Wang2026} avoids poorly performing qubits while maintaining the 1D chain connectivity.  The subsequent workflow used here integrates deterministic error suppression~\cite{fire_opal, Mundada2023, Coote2025} and randomized compiling to reduce runtime errors, and a lightweight post-processing step to compensate for both readout errors and incoherent decay (see Supplementary Material, Sec.~\ref{app:postprocessing}).  With dynamical decoupling included, our largest circuits use 120 qubits and consist of $9{,}057$ two-qubit gates organized into 152 layers. All experiments are executed on \texttt{ibm\_boston}, a 156-qubit IBM Heron device, with 20{,}000 shots per circuit.

\begin{figure*}[t!]
  \centering
  \includegraphics[width=0.9\linewidth]{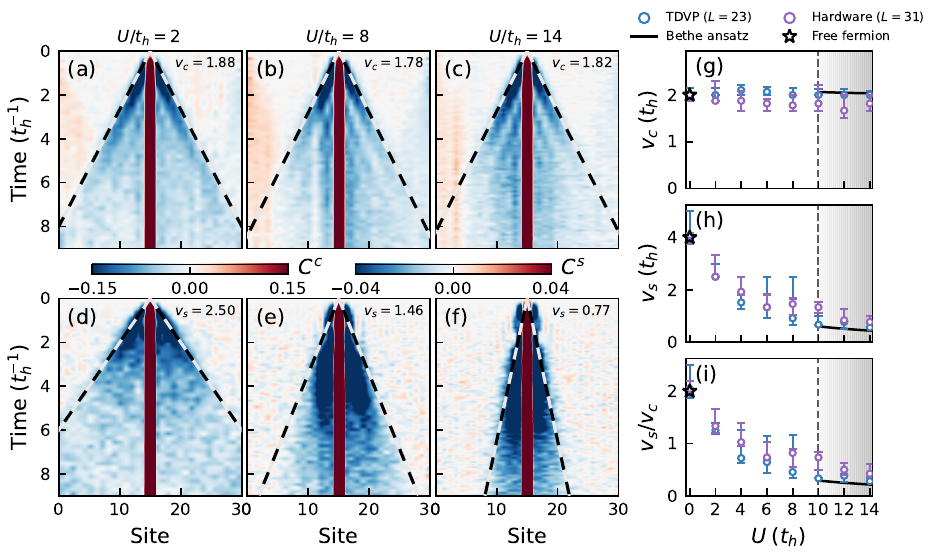}
  \caption{
    Spin-charge separation exhibited in the evolution of a central vacancy in a N\'{e}el initial state for $L=31$ over a range of repulsive couplings in each column ($U > 0$).  Simulations in panels (a,d) use 46 Trotter steps ($\Delta t = 0.2\, t_h^{-1}$), of which the first 45 are shown (total simulation time $t = 9\, t_h^{-1}$); (b,e) use 60 steps ($\Delta t = 0.15\, t_h^{-1}$, $t = 9\, t_h^{-1}$); and (c,f) use 90 steps ($\Delta t = 0.1\, t_h^{-1}$, $t = 9\, t_h^{-1}$).
    (a--c) Heatmaps of the charge tracer correlator ${C^{\rm c}_i(t) := \langle n_{i,\uparrow}(t) + n_{i,\downarrow}(t)\rangle  - \langle n_{i,\uparrow}(0) + n_{i,\downarrow}(0)\rangle}$. Time flows top-to-bottom, and the horizontal axis indicates site location on the one-dimensional chain. The wavefront boundary, as identified by the protocol detailed in Supplementary Material, Sec.~\ref{sec:wavefront}, is shown as a dashed line, and the velocity of the wavefront is reported in each panel.
    (d--f) Heatmaps of the spin tracer observable, ${C^{\rm s}_i(t) := 4(\langle S_i^z(t) S_{i_*}^z(t)\rangle - \langle S_i^z(t) \rangle \langle S_{i_*}^z(t)\rangle)}$, where $i_*$ is the site of the initial vacancy. Axes oriented as above. The same wavefront extraction has been applied as above. 
    (g--i) Extracted wavefront velocities as a function of $U/t_h$. All TDVP simulations run on a lattice of size $L=23$ to a total time $t = 9\, t_h^{-1}$, using a time step size of $0.1\, t_h^{-1}$ and max bond dimension of $\chi=1024$. For the digital quantum simulations we use the following conditions: for $U/t_h \in \{0, 2, 4\}$, Trotter step size $\Delta t = 0.2\, t_h^{-1}$ ($t = 9.2\, t_h^{-1}$); for $U/t_h \in \{6, 8, 10\}$, $\Delta t = 0.15\, t_h^{-1}$ ($t = 9\, t_h^{-1}$); for $U/t_h \in \{12, 14\}$, $\Delta t = 0.1\, t_h^{-1}$ ($t = 9\, t_h^{-1}$). Error bars reflect the sensitivity of the extracted velocity to the wavefront-detection and velocity-fitting parameters rather than statistical uncertainty. Also shown are the analytic free-fermion results (exact at $U = 0$) and the Bethe-ansatz predictions, valid in the strong-coupling regime $U/t_h \gg 1$, which we take to be $U/t_h \ge 10$ here (indicated via shading). See Supplementary Material Sec.~\ref{sec:wavefront} for details.
    }
    \label{fig:charge_spin_separation_hardware_and_tdvp}
\end{figure*}

The elementary excitations of the Fermi-Hubbard model in 1D are spinons (charge $0$, spin $1/2$) and holons (charge $\pm e$, spin $0$), which propagate independently with distinct velocities~\cite{Tomonaga1950, Luttinger1960, Haldane1981, LiebWu1968, Essler2005}. This leads to spin-charge separation as the dynamical signature of fractionalization in 1D interacting systems. Our first study probes this phenomenon, following earlier experimental work by Ref.~\cite{Arute2020SpinCharge}. The initial state is taken to be the N\'{e}el state with a localized vacancy defect at the central site, corresponding to the absence of an electron. Time evolution under the Trotterized Hamiltonian generates charge and spin disturbances that spread through the chain at different velocities (Fig.~\ref{fig:charge_spin_separation_hardware_and_tdvp}). To isolate this effect, we utilize correlators which are separately sensitive to charge or spin. For charge, we use ${C_i^c(t) := \langle n_{i,\uparrow}(t) + n_{i,\downarrow}(t)\rangle  - \langle n_{i,\uparrow}(0) + n_{i,\downarrow}(0)\rangle}$, the change in the per-site electron density, and for spin we use ${C_i^s(t) :=4\langle S^z_i(t) S^z_{i_*}(t)\rangle_c}$, where ${S^z_i = (n_{i,\uparrow} - n_{i, \downarrow})/2}$ is the $z$-spin at site $i$, $i_*$ is the site of the central vacancy, and the $c$ subscript indicates a connected correlator. Because the initial N\'{e}el state lacks connected correlations, this observable exclusively captures the dynamic magnetic disturbance associated with the spin sector.  Because these two-point correlators involve only operators that are diagonal in the Fock basis, they can be extracted from the same quantum simulation of fermionic dynamics, requiring no additional hardware execution.

Visual examination of the hardware results presented in the upper (charge) and lower (spin) rows of Fig.~\ref{fig:charge_spin_separation_hardware_and_tdvp}(a-f) clearly indicates that the observed disturbance propagation occurs at differing rates from the site of the initial defect. The sudden quench of the highly ordered N\'{e}el background induces a transient equilibration period before ballistic propagation emerges~\cite{lee2026} and remains visible out to the longest evolution times indicating the persistence of coherent dynamics over the deepest circuits in use here. This phenomenology is confirmed by auxiliary simulations (see Supplementary Material, Sec.~\ref{sec:additional_results}).

We compare hardware results against analytic calculations and numerical simulations by extracting wavefront velocities within the ballistic regime using a simple detection algorithm detailed in the Supplementary Material (Sec.~\ref{sec:wavefront}). Fig.~\ref{fig:charge_spin_separation_hardware_and_tdvp}(g--i) shows the extracted velocities and their ratio compared against TDVP calculations and exhibits quantitative agreement over all values of $U$, within experimental error. First, we consider the weakly interacting regime; at $U = 0$ the model reduces to a chain of free fermions, for which the tracer correlators and their wavefront velocities can be derived analytically. Here we find good agreement between the analytic and hardware-derived experimental values. In the strong-coupling regime $U/t_h \gg 1$, the N\'{e}el state becomes an approximate ground state -- at large $U$ the Hamiltonian reduces to the potential term $U\sum_i n_{i\uparrow}n_{i\downarrow}$ for which the doublon-free N\'{e}el configuration is a zero energy eigenstate. In this regime, the correlator wavefront velocities should approximately agree with the Bethe-ansatz quasiparticle-velocity predictions, with  agreement becoming exact in the strict $U \rightarrow \infty$ limit (see Supplemental Material Sec.~\ref{sec:bethe} for details). In this regime again we see good agreement between hardware, numerics, and analytics, within experimental error.  Figure~\ref{fig:charge_spin_separation_hardware_and_tdvp} displays the analytic predictions for $U \ge 10$ to emphasize that agreement should only hold for large $U$.  Overall, the small, systematic deviations between quantum hardware and either TDVP classical simulation or analytic calculation results are largely attributable to hardware noise.

\begin{figure*}[t!]
  \centering
  \includegraphics[width=0.9\linewidth]{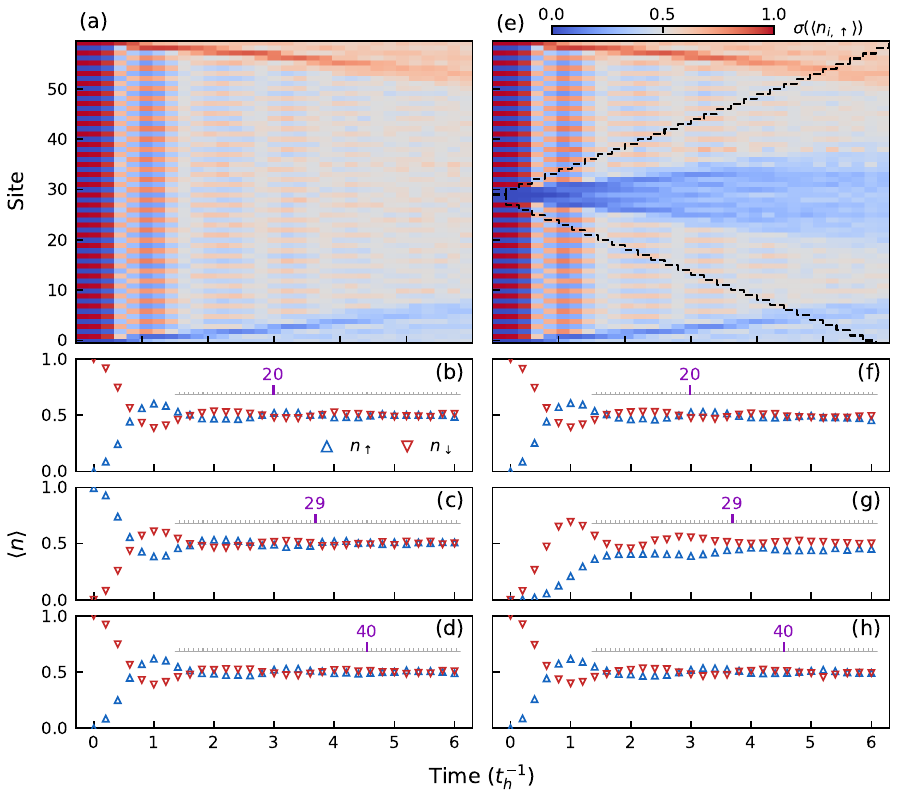}
  \caption{
  Digital quantum simulation of the Fermi-Hubbard model for two initial states of an $L=60$ chain with interaction strength $U/t_h = -2$: a N\'{e}el state (\emph{left}) and a N\'{e}el state with a central vacancy defect (\emph{right}). 
  (a) Heatmap showing the evolution of the per-site spin-up occupation $\langle n_{i,\uparrow} \rangle$. To enhance the visibility of small late-time occupations, the colormap applies the transformation $\sigma(\langle n_{i,\uparrow} \rangle)$, where $\sigma(x) = [1 + \tanh(k(x-1/2))/\tanh(k/2)]/2$. 
  (b)--(d) Time evolution of the per-site spin occupations, $\langle n_{i,\uparrow} \rangle$ and $\langle n_{i,\downarrow} \rangle$, for selected sites. The site shown in each panel is indicated by the purple marker in the one-dimensional chain schematic shown in the inset. 
  (e) Heatmap analogous to (a) for an initial state in which the central site ($i=29$) is a vacancy. The black line indicates the causal light cone. 
  (f)--(h) Time evolution plots analogous to (b)--(d) for the same set of sites. Error bars are smaller than the markers and have been omitted for clarity.}
  \label{fig:quantum_simulator_output}
\end{figure*}
We next extend digital quantum simulations to scales beyond prior demonstrations using systems with $L=60$ (120 qubits). We execute simulations that begin in a Fock state and evolve under Trotterized time evolution for a total simulation time up to $t = 6\,t^{-1}_h$, corresponding to simulations employing up to 30 Trotter steps. Fig.~\ref{fig:quantum_simulator_output}(a) and (e) show the average per-site spin-up occupancy arising from digital quantum simulation as a function of evolution time for initial states excluding and including a localized vacancy.  The digital quantum simulations exhibit complex dynamical evolution of the spin-up occupation profile $\langle n_{i,\uparrow}\rangle$ that remains visible by eye out to the maximum evolution time.  This includes prominent oscillation of the staggered spin state, edge effects, and (where appropriate) spreading of localized vacancies.  Such effects are highlighted in 1D time-evolution slices taken at individual sites (Fig.~\ref{fig:quantum_simulator_output}(b-d) and (f-h)) showing the oscillatory behavior for both spin species, $\langle n_{i,\uparrow} \rangle$ and $\langle n_{i,\downarrow} \rangle$.

At this system scale, the only relevant method to determine the accuracy of the digital quantum simulation is to compare against approximation methods. We rely on TDVP as the primary high-accuracy classical baseline, utilizing maximum bond dimensions up to $\chi=4096$ to validate the digital quantum simulation results. For further details on the classical-simulation benchmarking approach, see Supplementary Material Sec.~\ref{sec:classical_benchmark_methods} where we also present detailed comparisons against Heisenberg-picture simulation methods including Pauli path propagation (PPP), Majorana propagation (MP), and an MPO-based technique.

\begin{figure}[tp]
    \centering
    \includegraphics[width=0.98\linewidth]{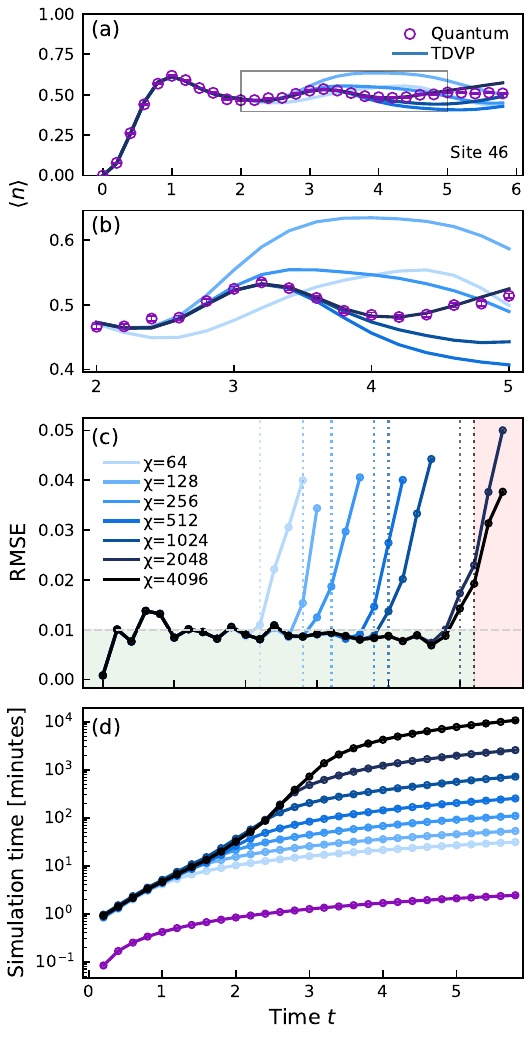}
    \caption{
    Quantum simulator outputs for $U/t_h=-2$ and the N\'{e}el initial state, benchmarked against TDVP for a range of bond dimensions, $\chi \in \{64, 128, 256, 512, 1024, 2048, 4096\}$.
    (a) Time evolution of the occupation expectation value of a representative spin-orbital, $\langle n_{46,\uparrow}\rangle(t)$ for site $i=46$. Quantum hardware results are shown as purple circles; TDVP results are shown as solid lines, with color indicating bond dimension $\chi$.  The light rectangular box indicates the region of zoomed-in dynamics treated in panel (b).
    (b) Zoom of (a), showing that larger-$\chi$ TDVP simulations track the quantum data to later times before diverging.
    (c) RMSE calculated between the quantum and TDVP occupation values across all 120 spin orbitals, shown for all tested values of $\chi$ at the discrete times for which quantum measurements exist (markers). The dashed horizontal line and green shading indicate the estimated noise floor, $\mathrm{RMSE}=0.01$, and vertical dotted lines mark the last step where each curve represents $<2\%$ error.  Red shading at right for $t\gtrsim 5.2 t^{-1}_{h}$ indicates the zone in which correctness of quantum and classical simulations is indeterminate.
    (d) Wall-clock runtime required to simulate evolution to time $t$ using quantum hardware (purple lines and markers) and TDVP (blue lines and markers, shaded by $\chi$).
    }
    \label{fig:TDVP_summary_plot}
\end{figure}

The general performance of TDVP in capturing the dynamics exhibited by the digital quantum simulator is illustrated in Fig.~\ref{fig:TDVP_summary_plot}(a--b).  First, we compare the digital quantum simulation outputs for an example single site and spin against TDVP with different bond dimensions up to $\chi=4096$. The quantum data and TDVP simulations exhibit good agreement to increasing evolution times as $\chi$ grows, before eventually diverging.  This is consistent with the expectation that TDVP incurs singular value decomposition (SVD) truncation errors that restrict its ability to capture highly entangled states at long simulation times and finite $\chi$, and that the method systematically approaches exactness as $\chi\to\infty$.

We quantitatively evaluate this phenomenology over the entire 60-site lattice by calculating the root-mean-square error (RMSE) between the different simulation approaches averaged over all sites and spins: $\mathrm{RMSE}(\langle \bm{n} \rangle_1, \langle \bm{n} \rangle_2) = \sqrt{\frac{1}{2L} \sum_{i,\sigma}\left(\langle n_{i\sigma}\rangle_1 - \langle n_{i\sigma} \rangle_2 \right)^2}$. As shown in Fig.~\ref{fig:TDVP_summary_plot}(c), the RMSE remains within $\sim1\%$ to increasing evolution times up to $t\approx 5.2\,(1/t_h)$ as $\chi$ increases up to $4096$.  The level of agreement at shorter times is consistent with the expected computational error arising from shot noise and Trotterization in the digital quantum simulation ($\sim1\%$), and even at the most extreme evolution time $t= 6\,(1/t_h)$, agreement remains within $4\%$.  For evolution times beyond $t\approx 5.2\,(1/t_h)$, where agreement between quantum and classical simulations diverges at the largest values of $\chi$, it is not possible to know which simulation methodology most accurately reflects the true system dynamics.  We note the visual continuity of digital quantum simulation results shown over all lattice sites (Fig.~\ref{fig:quantum_simulator_output}(a) and (e)) as evidence that it is not likely that accumulated errors arising from decoherence or gate error have undermined the potential correctness of the digital quantum simulation at the longest evolution times treated.

An additional point of comparison between the simulation methodologies relates to the wall-clock runtime required for execution. The execution time for the digital quantum and TDVP simulations implemented here using \texttt{ITensor} as a function of evolution time is shown in Fig.~\ref{fig:TDVP_summary_plot}(d). In these experiments, the digital quantum simulation over all time steps requires approximately 2 minutes and 46 seconds of wall-clock QPU execution for all circuits and repetition over 20{,}000 shots. The runtime is independent of the number of sites simulated, $L$, and only grows linearly with the number of simulated Trotter steps. By contrast, the runtime of TDVP scales as $\mathcal{O}(L\chi^3)$ per step, indicating explicit dependence on both evolution time and system size.

Reflecting the fact that beyond a particular evolution time we cannot assert ``correctness'' for either simulation methodology, we explicitly compare runtimes at the point of quantum/classical divergence, where $\mathrm{RMSE}>2\%$. This occurs near evolution time $t=5.2\,(1/t_h)$ for TDVP using $\chi=4096$. For these specific parameters, the quantum wall-clock runtime is around two minutes and TDVP execution exceeds 100 hours. The longest TDVP execution for $t=6\,(1/t_h)$ and $\chi=4096$ exceeds 160 hours.  An alternative implementation of TDVP using \texttt{TeNPy} (not shown in figure) requires approximately 23 hours to evolve the entire state and provide access to all observables using the same compute resources.

\begin{table*}[t]
\caption{A comparison of system sizes, evolution times, and Trotter steps across recent demonstrations of simulating the Fermi-Hubbard model on digital quantum simulators. The last two rows correspond to this work.
$^{\dagger\dagger}$Executed over 8 different days with a total device runtime (for all 8 twirling instances) of 36 hours, using $\sim 4\times$ fewer samples, at 1,280 shots, than this work.  
$^{**}$The mirrored construction of our $30\,(90)$-step Trotter circuits makes them equivalent to a circuit with $15\,(45)$ second-order Trotter steps. See Sec.~\ref{subsec:trotterization} for details.}
\begin{ruledtabular}
\begin{tabular}{lccccc}
\textbf{System} &
\textbf{Lattice sites} &
\textbf{Evolution time ($t_h^{-1}$)} &
\textbf{Trotter steps} &
\textbf{Trotter order} &
\textbf{Platform / Ref.} \\
\colrule
2D $6{\times}6$ square &36       & $2$   & 4  &   1st \& 2nd  & Quantinuum \cite{Granet2025} \\
2D $7{\times}4$ torus &28    & $2$   & 4 & 2nd  & Quantinuum$^{\dagger\dagger}$\cite{Alam2025_ion} \\
2D $6{\times}6$ square& 36      & $1.2$ & 3 & 2nd  & Google \cite{Alam2025_2D} \\
1D chain &52                 & $5$   & 10 & 1st \& 2nd & IBM \cite{Chowdhury2026} \\
1D chain (this work, deepest) & 31                 & $9$   & 90$^{**}$ & 1st (mirrored$^{**}$) & IBM \\
1D chain (this work, widest) & 60                 & $6$   & 30$^{**}$ & 1st (mirrored$^{**}$) & IBM \\
\end{tabular}

\end{ruledtabular}
\label{tab:comparisons}
\end{table*}

Our \texttt{ITensor} implementation of TDVP is optimized according to discussions with the \texttt{ITensor}~\cite{itensor} developers.  We employ the \texttt{ITensor} Julia implementation on a 32vCPU instance (AWS c7i.8xlarge, 64GB RAM), and use all computational parallelization supported by \texttt{ITensorMPS.jl}. We identify computational bottlenecks in the algorithm arising from essential singular value decomposition and Krylov exponentiation steps, which for $\chi=4096$ inhibit full use of the available computational cores. We also empirically validate negligible improvements doubling the classical compute engine to 64 cores.  Accordingly, large-scale parallelization is likely not possible or at best highly constrained for \texttt{ITensor} or other TDVP implementations, consistent with previous studies that show rapidly declining benefit from parallelization at the scales of classical CPU cores used here \cite{Secular2020}. \texttt{TeNPy} calculation performance is qualitatively similar, with a slight speedup overall relative to \texttt{ITensor} when leveraging the same computational resources. At the time of publication of this work, there is no readily available support for GPU acceleration that incorporates the particle and spin conservation symmetries required to reproduce the runtimes achieved here~\cite{ITensorGPUDocs2025}.  Further details of the computational configuration and commentary on the limits of parallelization for TDVP are presented in Supplementary Material~Sec.~\ref{sec:tdvp}.

TDVP as implemented using either package is a Schr\"odinger-picture simulation algorithm -- the wavefunction is evolved under the action of the Hamiltonian, and the expectation value of any observable may be obtained from the output.  This has the advantage that any multipoint density correlator may be directly extracted from the single numerical simulation, similar to the behavior of the quantum circuit simulation. Heisenberg-picture simulations are an alternative class of methods wherein a target observable is back-propagated through the time-evolution unitary to estimate the expectation value.  Because the observable calculations are independent, this permits parallelization over target observables. In Supplementary Material Sec.~\ref{sec:heisenberg}, we present extended  benchmarking to include three such Heisenberg methods: a tensor-network method that represents the target operator as a matrix product operator (MPO); Majorana Propagation (MP), which expands the operator in a basis of Majorana fermions and evolves them; and Pauli Path Propagation (PPP), which expands the operator in the Pauli basis and evolves them.

We compare the Heisenberg methods against a TDVP simulation at bond dimension $\chi = 4096$, over $t \in [0, 5.2]\,t_h^{-1}$, the range in which agreement between TDVP and the quantum computer is validated to $<1\%$ RMSE. Overall, the non-string-based MPO simulation method is the most accurate, with its error exceeding $1\%$ only near the end of the relevant evolution-time interval; the MP and PPP errors are larger, agreement with the quantum-hardware data diverges earlier in the evolution, and variability between performance for different individual operators is higher.

Making meaningful runtime comparisons across method classes require care because the methods produce fundamentally different outputs. Each Heisenberg simulation yields a single expectation value, whereas one TDVP run produces an MPS from which any operator expectation value can be computed, and a quantum execution yields a bitstring distribution from which any of the $2^{N_Q}$ $Z$-basis expectation values can be estimated. Within each class, moreover, not all available expectation values are equally accurate or reached in comparable runtimes. A final consideration is that the Heisenberg methods are trivially parallelizable while TDVP is not. With these caveats in mind, we now compare the runtimes directly for the classical simulation methods explored here.

For sequential execution of the best-performing MPO procedure using fixed computational resources, computing the full set of observable expectation values would require $\sim 51$, $170$, and $1700$ \textit{days} at bond dimensions $100$, $400$, and $1600$, respectively (assuming a per-observable budget of 4 logical CPUs and 4 Julia threads).  These values can of course be reduced by exploiting the trivially parallel nature of Heisenberg simulations; here the total compute time scales inversely with the available compute nodes. Full parallelization across all $120$ one-point and $7140$ two-point density correlators for a system of $L = 60$ fermionic sites would require $7260$ independent jobs run concurrently---necessitating on the order of tens of thousands of compute cores, depending on the per-job allocation. In our benchmarking, no Heisenberg simulation method executed the calculation across all three example operators tested faster than the corresponding complete quantum execution time of $2\,\text{min}\ 46\,\text{s}$, Supplementary Fig.~\ref{fig:heisenberg_runtime_vs_error_rmse}(a).  In the maximally parallelized case the minimal compute time is set by the maximal time required to evolve a single observable; the $\chi = 200$ MPO calculation would thus require a minimum of approximately 1000s over the aforementioned tens of thousands of compute cores. For the benchmarks conducted in this work, for which we extract one- and two-point expectation values, we conclude that TDVP remains the most resource-efficient publicly-available classical baseline.

We cannot exclude the possibility that the classical computational runtime could be improved via future modifications to the underlying algorithm, development of new support for enhanced computing resources, or complete replacement with a novel computational method, including the development of narrow-purpose tooling capable of only simulating the specific conditions and parameters treated in this study. To wit, following the initial appearance of this work as a preprint, Rausch \emph{et al.}~\cite{rausch2026pushingclassicalfrontier1d} developed an improved TDVP implementation that exploits a larger symmetry group and adds previously unavailable GPU support. Running on four NVIDIA H200 GPUs, these enabled bond dimensions as large as $\chi \approx 62{,}000$ -- roughly fifteen times the $\chi = 4096$ we employed -- yielding converged results across the entire quantum-hardware simulation window.  Importantly, this combination of innovations allowed them to certify our hardware results throughout $t \in [0,6]\,t_h^{-1}$, including the high-entanglement regime $t \in [5.2,6]\,t_h^{-1}$ that our own TDVP benchmark could not resolve within practically relevant runtimes.

The combination of a novel algorithmic implementation and GPU acceleration reduced the wall-clock cost of the classical TDVP benchmark to approximately $100$ minutes at $\chi \approx 4880$ -- a bond dimension of comparable expressive power to the $\chi = 4096$ we employed -- lowering the effective classical-versus-quantum runtime ratio from $\sim 500-3000\times$ to $\sim 36\times$ at that bond dimension.  We view this as a constructive sharpening of the quantum-classical comparison: the present quantum results are validated as accurate, the runtime challenge of the classical simulation is independently confirmed over an entire dataset, and the authors implicitly verify that novel tools were required in order to build further competitiveness for classical numerical simulations.

The results we present here demonstrate that digital quantum simulation of the 1D Fermi-Hubbard model can now be performed at scales and in regimes that challenge or exceed the reach of the best available classical methods with accuracy and speed meeting end-user expectations.  Achieving strong quantitative agreement between the outputs from both simulation methods for $L=31$ builds confidence in the accuracy of the digital quantum simulation results before pushing to more challenging scales and models. To the best of our knowledge, the measurements we have performed---characterized by the maximum dynamical evolution time ($t=9$), number of lattice sites calculated ($L=60$), number of qubits ($120$),  number of Trotter steps employed ($90$), circuit depth ($452$ layers), number of two-qubit gates ($>13,800$), and quantitative agreement with classical benchmarks (RMSE $\lesssim 1\%$)---represent the largest and most accurate digital quantum simulations of the 1D Fermi-Hubbard model to date.

Achieving these results has required significant innovation in the execution of Trotterized time evolution on digital quantum simulators relative to previous experiments (see Table~\ref{tab:comparisons}). For the largest circuits we execute, our specialized compilation reduces the overall circuit depth (measured in two-qubit-gate layers) by $>80\%$ and the number of two-qubit gates by $>60\%$ relative to a baseline \texttt{Qiskit Nature} Jordan--Wigner transformation and the \texttt{Qiskit} transpiler (level-3 optimization)~\cite{qiskitnature}. Relative to special-purpose methods for Trotterized circuits e.g., the \textit{interleaved} ordering used in Ref.~\cite{Chowdhury2026}, and for comparable circumstances (equivalent to $L=52$ sites and 10 second-order Trotter steps), our approach uses $40.5\%$ fewer two-qubit gates and has $60\%$ lower circuit depth. Further, our demonstrations include some of the largest circuits executed on quantum computers overall, with over 13{,}800 two-qubit gates ($L=31$, $t=9$, 90 Trotter steps), enabled by a customized error-reduction pipeline that incurs no execution overhead. Ensuring there is no exponential growth in the number of circuit executions commonly required for error mitigation~\cite{van_den_Berg_2023} has proven essential to preserving potential time-to-solution benefits with digital quantum simulators. Details on circuit characteristics and error suppression strategies are presented in Supplementary Material Sec.~\ref{app:qc_contruction_compilation_error_suppresion}.

The digital quantum simulation approach we employ enables wide flexibility in calculated observables; measurement at arbitrary $U/t_h$, and including spatially dependent hopping, interactions, and chemical potential are achievable by simply adjusting the appropriate angles of the $R_Z$, $R_{ZZ}$, and $R_{XX+YY}$ gates in the quantum circuit.  Accordingly, this platform allows the full interaction dependence of any physical observable to be explored in a single experiment. This tunability---combined with the ability to use arbitrary initial product states, measure any local observable, and scale the chain length in a straightforward and penalty-free manner---is a structural advantage of the use of digital quantum simulators over analytic methods as well as classical and analog-quantum simulation alternatives.

Natural next steps to consider include extensions to address impurity doping and two-dimensional lattices. Digital quantum simulation of two-dimensional Fermi-Hubbard models~\cite{Kivlichan2018} with efficient two-dimensional fSWAP networks can greatly enhance the investigation of regimes such as finite doping, where conventional classical methods are often severely limited by the fermionic sign problem. Ongoing work must also address residual challenges in digital quantum simulators for them to realize their full potential as new computational tools in condensed matter physics. Measurement of higher-order correlation functions, including the single-particle spectral function and the dynamical structure factor, can provide stringent tests of Luttinger liquid theory, but also requires lower noise floors than current hardware provides.  A modest expansion of these results may also enable access to the regime believed to harbor the mechanism of high-temperature superconductivity in the cuprates~\cite{Lee2006,Keimer2015,Arovas2022}---a problem that has resisted exact classical treatment for nearly four decades.

Bolstered by the independent validation of the accuracy of our quantum-computer simulations by Rausch \emph{et al.}, we are encouraged that these results support the view that useful digital quantum simulation of certain physically relevant condensed matter problems with pre-fault-tolerant machines is now possible.

\vspace{5mm}
\emph{Author Contributions}: 
K.S.N. proposed the initial Fermi-Hubbard concept, and co-managed the project with G.S.H. and Y.B.
H.L. designed the pair-interleaved ordering and fSWAP network, which H.L. and A.K. implemented alongside A.K.'s work on circuit compilation.
A.K. and M.S. executed hardware experiments, managed data collection, and analyzed data.
M.S. and A.K ran the TDVP benchmarks.
For Heisenberg simulations, K.S.N. implemented MP, and G.S.H. implemented PPP and MPO.
G.S.H. performed free-fermion and Bethe ansatz calculations and worked with M.R.H. to compare theory with experiment.
M.S. developed the wavefront detection approach, supported by G.S.H. and K.S.N.
H.L. designed and implemented decay recovery (conceptualized by Y.B.) and A.K. integrated twirling, readout mitigation, and broader error suppression techniques.
All authors collaboratively contributed to data analysis, writing, and reviewing the final manuscript.

\vspace{5mm}
\emph{Acknowledgments}: 
We thank Garnet Chan for insightful conversations regarding Pauli path propagation and tensor network simulation methods, and Miles Stoudemire for valuable discussions concerning TDVP and the \texttt{ITensor} package. We also thank Alexey Gorshkov, Zhi-Yuan Wei, and Mathi Raja for illuminating discussions on the physics of the Fermi-Hubbard model and for ongoing collaborations on related topics.  We acknowledge the use of IBM Quantum Credits via the IBM Quantum Startups Program for this work. The views expressed are those of the authors and do not reflect the official policy or position of IBM or the IBM Quantum Platform team.  We acknowledge and thank Michael Brett from AWS for the provision of access to a computational cluster used in the conduct of classical benchmarking simulations.  Finally, we are grateful to all our colleagues at Q-CTRL whose technical work has supported the results presented in this paper.



%

\end{bibunit}


\clearpage 
\onecolumngrid 

\setcounter{page}{1}
\setcounter{equation}{0}
\setcounter{figure}{0}
\setcounter{table}{0}

\renewcommand{\theequation}{S.\arabic{equation}}
\renewcommand{\thefigure}{S\arabic{figure}}
\renewcommand{\thetable}{S\arabic{table}}
\renewcommand*{\thepage}{S\arabic{page}}

\renewcommand{\theHequation}{S\arabic{equation}}
\renewcommand{\theHfigure}{S\arabic{figure}}
\renewcommand{\theHtable}{S\arabic{table}}

\ApplyPaperMetadata{supplement}

\begin{center}
{\large \textbf{SUPPLEMENTARY CONTENTS}}
\end{center}

\makeatletter
\@starttoc{stoc}

\let\oldaddcontentsline\addcontentsline
\renewcommand{\addcontentsline}[3]{%
    \def\@dest{#1}%
    \def\@tocstr{toc}%
    \ifx\@dest\@tocstr
        \oldaddcontentsline{stoc}{#2}{#3}%
    \else
        \oldaddcontentsline{#1}{#2}{#3}%
    \fi
}
\makeatother
\vspace{1cm}

\setcounter{section}{0}
\setcounter{subsection}{0}
\setcounter{subsubsection}{0}

\renewcommand{\thesection}{\Roman{section}}
\renewcommand{\thesubsection}{\Alph{subsection}}

\renewcommand{\theHsection}{supp.\Roman{section}}
\renewcommand{\theHsubsection}{supp.\Roman{section}.\Alph{subsection}}

\begin{bibunit}[apsrev4-2]
\setbibunitprefix{@supp}

\section{Application-aware compilation and error suppression} \label{app:qc_contruction_compilation_error_suppresion}
In this section, we describe the end-to-end procedure used to perform the digital quantum simulation of the one-dimensional Fermi-Hubbard model. This workflow includes mapping the fermionic degrees of freedom to qubits, constructing Trotterized time-evolution circuits, and compiling those circuits for execution on IBM heavy-hex devices. We refer to this broader workflow as \emph{application-aware compilation}, distinguishing it from standard compilation pipelines, which typically treat the quantum circuit as a ``black box" input, employing general-purpose optimization techniques to produce a hardware-native output. In contrast, our approach leverages the specific structure of the Fermi-Hubbard model to optimize the circuit construction and compilation.

To make this supplement self-contained and establish notation, we begin by restating the model Hamiltonian:
\begin{equation}
    \label{eq:Hamiltonian_appendix}
    H_{\mathrm{FH}} = -t_h \sum_{i=0}^{L-2} \sum_\sigma \bigl(c^{\dagger}_{i\sigma} c^{\phantom{\dagger}}_{i+1,\sigma} + \mathrm{h.c.}\bigr) + U \sum_{i=0}^{L-1} n_{i\uparrow} n_{i\downarrow} - \mu \sum_{i,\sigma} n_{i\sigma}\equiv H_{\mathrm{K}} + H_{U} + H_{\mu} \,.
\end{equation}
Here, $L$ is the length of the chain, with sites indexed as $i=0, 1, \ldots, L-1$. Each site supports spin-up and spin-down orbitals. The $c^{\dagger}_{i\sigma}$ ($c^{\phantom{\dagger}}_{i\sigma}$) are creation (annihilation) operators at site $i$ and for spin $\sigma \in \{\uparrow, \downarrow\}$, and $n_{i\sigma} = c^{\dagger}_{i\sigma}c^{\phantom{\dagger}}_{i\sigma}$ is the corresponding site/spin number operator. The first term, $H_{\mathrm{K}}$, represents the kinetic energy due to nearest-neighbor hopping, the second term, $H_{U}$, is an onsite interaction (with the Coulomb interaction strength $U$) that only contributes if both orbitals at a given site are occupied, and the final term, $H_{\mu}$, is the chemical potential term (with the chemical potential $\mu$). 
In the following, we adopt units where the hopping amplitude $t_h = 1$, effectively measuring energy in units of $t_h$ and time in units of $t_h^{-1}$.

\subsection{Jordan--Wigner transformation}
\label{subsec:j-w}

To simulate fermionic dynamics on a qubit-based quantum processor, the fermionic creation and annihilation operators must be mapped to qubit Pauli operators.
To that end, we employ the standard Jordan--Wigner (JW) transformation, which is particularly efficient for one-dimensional models as it maintains the locality of nearest-neighbor terms~\cite{JordanWigner1928}. 
For an ordered set of fermionic modes indexed by $J \in \{0, 1, \dots, 2L-1\}$, the JW transformation is defined as
\begin{equation}
c_{J} = \left( \prod_{k=0}^{J-1} Z_k \right) \sigma^{-}_{J}\, , \qquad c_{J}^{\dagger} = \left( \prod_{k=0}^{J-1} Z_k \right) \sigma^+_J\, ,
\label{eq:j-w}
\end{equation}
where $\sigma^{\pm}_J = \frac{1}{2}(X_J \mp iY_J)$.
Implementing this mapping requires an explicit ordering $J(i, \sigma)$, which assigns a unique qubit index $J$ to each pair of physical site $i \in \{0, \dots, L-1\}$ and spin $\sigma \in \{\uparrow, \downarrow\}$ indices.
In this work, we introduce a \textit{pair-interleaved} ordering, specifically designed to minimize routing overhead on heavy-hex device topologies.
We define this ordering as the sequence
\begin{equation}
    \{ c_{0\downarrow}, c_{0\uparrow}, c_{1\uparrow}, c_{1\downarrow}, c_{2\downarrow}, c_{2\uparrow}, c_{3\uparrow}, c_{3\downarrow}, \ldots \} \to \{ c_{0}, c_{1},\ldots c_{2L-1} \}\,,
    \label{eq:pair_interleaved_ordering}
\end{equation}
and it is related to an equivalent ordering by a global spin-flip ($\uparrow \, \leftrightarrow \, \downarrow$).
For simplicity, we focus on the convention defined above, as the results are invariant under this choice.
A detailed comparison between this ordering and standard JW orderings---focusing on their respective routing costs and circuit depths---is presented in Sec.~\ref{sec:other_orderings_comparison}.

Using the pair-interleaved ordering, we perform the JW transformation \eqref{eq:j-w} of the Fermi-Hubbard Hamiltonian Eq.~\eqref{eq:Hamiltonian_appendix}, upon which the kinetic part of the Hamiltonian $H_{\mathrm{K}}$ splits into ``short-hopping'' ($H_{\mathrm{S}}$) and ``long-hopping'' ($H_{\mathrm{L}}$) components.
As such, we write $H_{\mathrm{K}} = H_{\mathrm{S}} + H_{\mathrm{L}}$, where
\begin{subequations}
\begin{align}
    &H_{\mathrm{S}} = -\frac{1}{2} \sum_{J=0}^{L-2} \left( 
    X_{2J+1} X_{2J+2} 
    + Y_{2J+1} Y_{2J+2} \right) \, ,\label{eq:H_S}\\
    &H_{\mathrm{L}} = - \frac{1}{2} \sum_{J=0}^{L-2} \left( X_{2J} Z_{2J+1} Z_{2J+2} X_{2J+3} 
    + Y_{2J} Z_{2J+1} Z_{2J+2} Y_{2J+3} \right) \,.\label{eq:H_L}
\end{align}
\label{eq:H_S_H_L}
\end{subequations}
In Eq.~\eqref{eq:H_L}, the $Z_{2J+1}Z_{2J+2}$ operators are JW ``$Z$-strings", which account for the fermionic anti-commutation relations between non-adjacent modes in our ordering.

Following the JW transformation, the interaction and chemical potential terms of the Hamiltonian \eqref{eq:Hamiltonian_appendix} can be grouped into one- and two-qubit components: $H_U+H_{\mu} = H_{1\mathrm{Q}}+H_{U,2\mathrm{Q}}$.
After discarding irrelevant additive constants, these components can be written as
\begin{equation}
    H_{1\mathrm{Q}} = \frac{1}{2}\left(\mu -\frac{U}{2} \right) \sum_{J=0}^{2L-1} Z_J \,, 
    \qquad
    H_{U,2\mathrm{Q}} = \frac{U}{4} \sum_{J=0}^{L-1} Z_{2J} Z_{2J+1} \,,
    \label{eq:mu_and_U_terms}
\end{equation}
and the full Fermi-Hubbard Hamiltonian \eqref{eq:Hamiltonian_appendix} is thus transformed into
\begin{equation}
H_{\mathrm{FH}}=H_{\mathrm{L}}+H_{\mathrm{S}}+H_{1\mathrm{Q}}+H_{U,2\mathrm{Q}} \,.
\label{eq:fh_ham_transformed}
\end{equation}

\subsection{Trotterization}
\label{subsec:trotterization}

To simulate the dynamics governed by the Hamiltonian in Eq.~\eqref{eq:fh_ham_transformed}, we employ a standard Trotterization technique, discretizing the full evolution operator $e^{-iH_{\mathrm{FH}}t}$ into $n_{\mathrm{step}}$ Trotter steps of duration $\Delta t$, such that $t=n_{\mathrm{step}}\Delta t$.
For each individual Trotter step, its evolution operator is decomposed using a first-order Trotterization scheme
\begin{align}
    e^{-iH_{\mathrm{FH}}\Delta t} &= e^{-i\frac{\Delta t}{2}H_{1\mathrm{Q}}}e^{-i\Delta t H_{\mathrm{L}}}e^{-i\Delta t H_{U,2\mathrm{Q}}}e^{-i\Delta t H_{\mathrm{S}}}e^{-i\frac{\Delta t}{2}H_{1\mathrm{Q}}}\nonumber\\
    &+\frac{\Delta t^2}{2}\left([H_{\mathrm{S}}-H_{\mathrm{L}}, H_{U,2\mathrm{Q}}]-[H_{\mathrm{L}},H_{\mathrm{S}}]\right) + \mathcal{O}(\Delta t^3)\, ,
    \label{eq:first_order_trotter}
\end{align}
leading to an $\mathcal{O}(\Delta t^2)$ scaling of the Trotterization error.
Since each exponentiated component in Eq.~\eqref{eq:first_order_trotter} consists of mutually commuting terms, it can be further decomposed into product sequences of individual qubit rotations.

Specifically, the $H_{1\mathrm{Q}}$ and $H_{U,2\mathrm{Q}}$ terms are implemented via $R_Z$ and $R_{ZZ}$ rotations with angles $(\mu - U/2)\Delta t/2$ and $U\Delta t/2$, respectively. The kinetic terms, $H_{\mathrm{S}}$ and $H_{\mathrm{L}}$, are mapped to pairs of $R_{XX}$/$R_{YY}$ and $R_{XZZX}$/$R_{YZZY}$ rotations with angles $\Delta t$. Notably, the $H_{1\mathrm{Q}}$ contribution is symmetrized in the style of a second-order Trotter expansion. This targeted symmetrization reduces the error contribution from the $H_{1\mathrm{Q}}$ terms without increasing the physical gate depth, as $R_Z$ rotations are implemented as ``virtual" phase shifts in superconducting hardware.

The four-qubit rotations $R_{XZZX}$ and $R_{YZZY}$ associated with $H_{\mathrm{L}}$ cannot be straightforwardly implemented on hardware and must be further decomposed into deeper sequences of two-qubit gates.
To perform this decomposition efficiently, we employ fermionic swap networks~\cite{Kivlichan2018}.
In this framework, the fSWAP operator
\begin{equation}
    \mathrm{fSWAP}=\mathrm{SWAP}\cdot \mathrm{C}Z=
    \begin{bmatrix}
    1 & 0 & 0 & 0 \\
    0 & 0 & 1 & 0 \\
    0 & 1 & 0 & 0 \\
    0 & 0 & 0 & -1 \\
    \end{bmatrix}
    =\mathrm{fSWAP}^{\dagger} = \mathrm{fSWAP}^{-1}
\end{equation}
is utilized to exchange fermionic modes.
Specifically, given our pair-interleaved ordering Eq.~\eqref{eq:pair_interleaved_ordering},
a layer of fSWAP gates,
\begin{equation}
    \mathcal{F} = \bigotimes_{J=0}^{L-1}\mathrm{fSWAP}_{2J, 2J+1} = \mathcal{F}^{\dagger} = \mathcal{F}^{-1} \,,
\end{equation}
effectively swaps the spin-$\uparrow$ and spin-$\downarrow$ modes on every site ($\mathcal{F} \, c_{i,\uparrow(\downarrow)} \, \mathcal{F}^{\dagger} = c_{i,\downarrow(\uparrow)}$).
Crucially, this transformation maps the short- and long-hopping terms onto one another:
\begin{equation}
    \mathcal{F} \, H_{S(L)} \, \mathcal{F}^{\dagger} = H_{L(S)} \,.
    \label{eq:short_long_swap}
\end{equation}
This identity can be verified by observing how $\mathrm{fSWAP}$ conjugates the constituent Pauli strings in Eq.~\eqref{eq:H_S_H_L},
\begin{equation}
    \mathrm{fSWAP}(XZ)\mathrm{fSWAP}^{\dagger} = IX \,, \quad \mathrm{fSWAP}(YZ)\mathrm{fSWAP}^{\dagger} = IY \,, \quad \mathrm{fSWAP}(IZ)\mathrm{fSWAP}^{\dagger} = ZI \,.
    \label{eq:paulis_under_fSWAP}
\end{equation}

By inserting Eq.~\eqref{eq:short_long_swap} into the Trotter step Eq.~\eqref{eq:first_order_trotter}, we can replace the hardware-intensive ``long-hops'' with ``short-hops'', yielding
\begin{align}
    e^{-iH_{\mathrm{FH}}\Delta t} &= \mathcal{F}e^{-i\frac{\Delta t}{2}H_{1\mathrm{Q}}}e^{-i\Delta t H_{\mathrm{S}}}\mathcal{F}e^{-i\Delta t H_{U,2\mathrm{Q}}}e^{-i\Delta t H_{\mathrm{S}}}e^{-i\frac{\Delta t}{2}H_{1\mathrm{Q}}}\nonumber\\
    &+\frac{\Delta t^2}{2}\left([H_{\mathrm{S}}-H_{\mathrm{L}}, H_{U,2\mathrm{Q}}]-[H_{\mathrm{L}},H_{\mathrm{S}}]\right) + \mathcal{O}(\Delta t^3)\, ,
    \label{eq:first_order_trotter_fSWAPs}
\end{align}
where we have used the fact that $\mathcal{F}$ commutes with $H_{1\mathrm{Q}}$ (as can seen from the last identity of Eq.~\eqref{eq:paulis_under_fSWAP}).
Consequently, through the strategic use of two $\mathrm{fSWAP}$ layers, we reduce the complex four-qubit rotations to a sequence of simpler two-qubit $R_{XX}$ and $R_{YY}$ rotations, significantly decreasing the total circuit depth.

An equivalent, \textit{mirrored} Trotter decomposition is obtained by interchanging $H_{\mathrm{S}}$ and $H_{\mathrm{L}}$ in Eq.~\eqref{eq:first_order_trotter},
\begin{align}
    e^{-iH_{\mathrm{FH}}\Delta t} &= e^{-i\frac{\Delta t}{2}H_{1\mathrm{Q}}}e^{-i\Delta t H_{\mathrm{S}}}e^{-i\Delta t H_{U,2\mathrm{Q}}}\mathcal{F}e^{-i\Delta t H_{\mathrm{S}}}e^{-i\frac{\Delta t}{2}H_{1\mathrm{Q}}}\mathcal{F}\nonumber\\
    &+\frac{\Delta t^2}{2}\left([H_{\mathrm{L}}-H_{\mathrm{S}}, H_{U,2\mathrm{Q}}]-[H_{\mathrm{S}},H_{\mathrm{L}}]\right) + \mathcal{O}(\Delta t^3)\, ,
    \label{eq:first_order_trotter_fSWAPs_long_short}
\end{align}
where $H_{\mathrm{L}}$ has been mapped to $H_{\mathrm{S}}$ via the $\mathcal{F}$ layers. To minimize circuit depth, we alternate between Eq.~\eqref{eq:first_order_trotter_fSWAPs} for odd steps and Eq.~\eqref{eq:first_order_trotter_fSWAPs_long_short} for even steps. Consequently, an odd-even pair of Trotter steps decomposes as
\begin{align}
    e^{-2iH_{\mathrm{FH}}\Delta t} &= e^{-i\frac{\Delta t}{2}H_{1\mathrm{Q}}}e^{-i\Delta t H_{\mathrm{S}}}\mathcal{F}e^{-i\Delta t H_{U,2\mathrm{Q}}}e^{-i\Delta t H_{\mathrm{S}}}e^{-i\frac{\Delta t}{2}H_{1\mathrm{Q}}}\nonumber\\
    &\times e^{-i\frac{\Delta t}{2}H_{1\mathrm{Q}}}e^{-i\Delta t H_{\mathrm{S}}}\mathcal{F}e^{-i\Delta t H_{U,2\mathrm{Q}}}e^{-i\Delta t H_{\mathrm{S}}}e^{-i\frac{\Delta t}{2}H_{1\mathrm{Q}}} + \mathcal{O}(\Delta t^3)\nonumber\\
    &\equiv U_{\mathrm{step}}^2+ \mathcal{O}(\Delta t^3)\, ,
\end{align}
where we have utilized the fact that $\mathcal{F}$ commutes with $H_{U,2\mathrm{Q}}$ and defined an operator approximating a single Trotter step,
\begin{equation}
    U_{\mathrm{step}} = e^{-i\frac{\Delta t}{2}H_{1\mathrm{Q}}}e^{-i\Delta t H_{\mathrm{S}}}\mathcal{F}e^{-i\Delta t H_{U,2\mathrm{Q}}}e^{-i\Delta t H_{\mathrm{S}}}e^{-i\frac{\Delta t}{2}H_{1\mathrm{Q}}}\, .
    \label{eq:trotter_step}
\end{equation}
Notably, the leading-order commutators from Eq.~\eqref{eq:first_order_trotter_fSWAPs} and Eq.~\eqref{eq:first_order_trotter_fSWAPs_long_short} cancel out when steps are paired. This results in an $\mathcal{O}(\Delta t^3)$ local Trotter error scaling, consistent with a second-order Trotterization scheme achieved through mirroring.  Therefore, for an even number of steps, the full evolution operator is
\begin{equation}
    e^{-iH_{\mathrm{FH}}t} = U_{\mathrm{step}}^{n_{\mathrm{step}}} + \mathcal{O}(n_{\mathrm{step}}\Delta t^3)\,, \quad n_{\mathrm{step}} = 2k\, .
    \label{eq:full_evol_even}
\end{equation}
For an odd number of steps, the final step (which is odd) is appended to the even-step evolution of Eq.~\eqref{eq:full_evol_even} using Eq.~\eqref{eq:first_order_trotter_fSWAPs}. The concluding layer of fSWAPs in this case can be implemented virtually by a classical relabeling of the fermionic modes (measurement outcomes) at the end of the evolution; we denote this relabeling as $\mathcal{P}$. The full evolution operator for an odd number of steps thus yields
\begin{equation}
    e^{-iH_{\mathrm{FH}}t} = \mathcal{P}U_{\mathrm{step}}^{n_{\mathrm{step}}} + \mathcal{O}(\Delta t^2) + \mathcal{O}[(n_{\mathrm{step}}-1)\Delta t^3]\,, \quad n_{\mathrm{step}} = 2k+1\, ,
    \label{eq:full_evol_odd}
\end{equation}
where the quadratic Trotter error $\mathcal{O}(\Delta t^2)$ stems from the error in Eq.~\eqref{eq:first_order_trotter_fSWAPs}. While the quadratic error dominates at short times [$(n_{\mathrm{step}}-1)\Delta t \approx t \ll 1$], the cubic error becomes dominant as the number of steps increases for long-time simulations ($t\gg 1$). Thus, for the timescales of interest, we effectively achieve second-order accuracy.

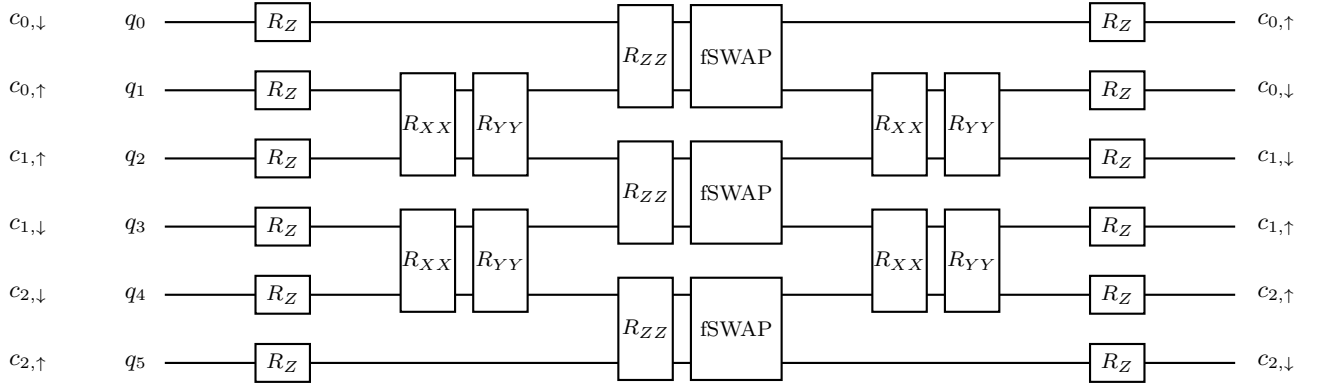
\begin{figure}[t]
\centering
\begin{tikzpicture}[x=1.2cm,y=0.9cm]

\tikzset{
  wire/.style={thick},
  gate/.style={draw, thick, fill=white, minimum width=0.72cm, minimum height=0.5cm, inner sep=1pt, font=\footnotesize},
  qlabel/.style={anchor=east, font=\small},
  rlabel/.style={anchor=west, font=\small}
}

%

\def\xleft{0}
\def\xright{11.8}

\def\xRZa{1.3}
\def\xRXXa{2.9}
\def\xRYYa{3.7}
\def\xRZZ{5.3}
\def\xFSWAP{6.3}
\def\xRXXb{8.1}
\def\xRYYb{8.9}
\def\xRZb{10.5}

\def\ya{0}
\def\yb{-1}
\def\yc{-2}
\def\yd{-3}
\def\ye{-4}
\def\yf{-5}

\def\yab{-0.5}
\def\ybc{-1.5}
\def\ycd{-2.5}
\def\yde{-3.5}
\def\yef{-4.5}

\draw[wire] (\xleft,\ya) -- (\xright,\ya);
\draw[wire] (\xleft,\yb) -- (\xright,\yb);
\draw[wire] (\xleft,\yc) -- (\xright,\yc);
\draw[wire] (\xleft,\yd) -- (\xright,\yd);
\draw[wire] (\xleft,\ye) -- (\xright,\ye);
\draw[wire] (\xleft,\yf) -- (\xright,\yf);

\node[qlabel] at (-0.1,\ya) {$q_0$};
\node[qlabel] at (-0.1,\yb) {$q_1$};
\node[qlabel] at (-0.1,\yc) {$q_2$};
\node[qlabel] at (-0.1,\yd) {$q_3$};
\node[qlabel] at (-0.1,\ye) {$q_4$};
\node[qlabel] at (-0.1,\yf) {$q_5$};

\node[qlabel] at (-1.2,\ya) {$c_{0,\downarrow}$};
\node[qlabel] at (-1.2,\yb) {$c_{0,\uparrow}$};
\node[qlabel] at (-1.2,\yc) {$c_{1,\uparrow}$};
\node[qlabel] at (-1.2,\yd) {$c_{1,\downarrow}$};
\node[qlabel] at (-1.2,\ye) {$c_{2,\downarrow}$};
\node[qlabel] at (-1.2,\yf) {$c_{2,\uparrow}$};

\node[rlabel] at (\xright+0.15,\ya) {$c_{0,\uparrow}$};
\node[rlabel] at (\xright+0.15,\yb) {$c_{0,\downarrow}$};
\node[rlabel] at (\xright+0.15,\yc) {$c_{1,\downarrow}$};
\node[rlabel] at (\xright+0.15,\yd) {$c_{1,\uparrow}$};
\node[rlabel] at (\xright+0.15,\ye) {$c_{2,\uparrow}$};
\node[rlabel] at (\xright+0.15,\yf) {$c_{2,\downarrow}$};

\foreach \row in {\ya,\yb,\yc,\yd,\ye,\yf} {
  \node[gate] at (\xRZa,\row) {$R_Z$};
}

\draw[thick, fill=white] (\xRXXa-0.3,\yb+0.25) rectangle (\xRXXa+0.3,\yc-0.25);
\node at (\xRXXa,\ybc) {\footnotesize $R_{XX}$};
\draw[thick, fill=white] (\xRYYa-0.3,\yb+0.25) rectangle (\xRYYa+0.3,\yc-0.25);
\node at (\xRYYa,\ybc) {\footnotesize $R_{YY}$};
\draw[thick, fill=white] (\xRXXa-0.3,\yd+0.25) rectangle (\xRXXa+0.3,\ye-0.25);
\node at (\xRXXa,\yde) {\footnotesize $R_{XX}$};
\draw[thick, fill=white] (\xRYYa-0.3,\yd+0.25) rectangle (\xRYYa+0.3,\ye-0.25);
\node at (\xRYYa,\yde) {\footnotesize $R_{YY}$};

\draw[thick, fill=white] (\xRZZ-0.3,\ya+0.25) rectangle (\xRZZ+0.3,\yb-0.25);
\node at (\xRZZ,\yab) {\footnotesize $R_{ZZ}$};
\draw[thick, fill=white] (\xRZZ-0.3,\yc+0.25) rectangle (\xRZZ+0.3,\yd-0.25);
\node at (\xRZZ,\ycd) {\footnotesize $R_{ZZ}$};
\draw[thick, fill=white] (\xRZZ-0.3,\ye+0.25) rectangle (\xRZZ+0.3,\yf-0.25);
\node at (\xRZZ,\yef) {\footnotesize $R_{ZZ}$};
\draw[thick, fill=white] (\xFSWAP-0.5,\ya+0.25) rectangle (\xFSWAP+0.5,\yb-0.25);
\node at (\xFSWAP,\yab) {\footnotesize fSWAP};
\draw[thick, fill=white] (\xFSWAP-0.5,\yc+0.25) rectangle (\xFSWAP+0.5,\yd-0.25);
\node at (\xFSWAP,\ycd) {\footnotesize fSWAP};
\draw[thick, fill=white] (\xFSWAP-0.5,\ye+0.25) rectangle (\xFSWAP+0.5,\yf-0.25);
\node at (\xFSWAP,\yef) {\footnotesize fSWAP};

\draw[thick, fill=white] (\xRXXb-0.3,\yb+0.25) rectangle (\xRXXb+0.3,\yc-0.25);
\node at (\xRXXb,\ybc) {\footnotesize $R_{XX}$};
\draw[thick, fill=white] (\xRYYb-0.3,\yb+0.25) rectangle (\xRYYb+0.3,\yc-0.25);
\node at (\xRYYb,\ybc) {\footnotesize $R_{YY}$};
\draw[thick, fill=white] (\xRXXb-0.3,\yd+0.25) rectangle (\xRXXb+0.3,\ye-0.25);
\node at (\xRXXb,\yde) {\footnotesize $R_{XX}$};
\draw[thick, fill=white] (\xRYYb-0.3,\yd+0.25) rectangle (\xRYYb+0.3,\ye-0.25);
\node at (\xRYYb,\yde) {\footnotesize $R_{YY}$};

\foreach \row in {\ya,\yb,\yc,\yd,\ye,\yf} {
  \node[gate] at (\xRZb,\row) {$R_Z$};
}

\end{tikzpicture}
\caption{
Schematic quantum circuit implementing a single Trotter step $U_{\mathrm{step}}$ of Eq.~\eqref{eq:trotter_step}, together with the operation $\mathcal{P}$ relabeling fermionic modes at the end, shown for $L=3$ sites.
The two layers of $R_Z$ rotations realize the one-qubit term $H_{1\mathrm{Q}}$ of Eq.~\eqref{eq:mu_and_U_terms}, the two layers of $R_{XX}$ and $R_{YY}$ rotations realize the ``short hops" in the Hamiltonian $H_\mathrm{S}$ of Eq.~\eqref{eq:H_S}, and the layer of $R_{ZZ}$ rotations implements the two-qubit onsite interaction term $H_{U,2\mathrm{Q}}$ of Eq.~\eqref{eq:mu_and_U_terms}.
The ``long hops'' $H_\mathrm{L}$ of Eq.~\eqref{eq:H_L} are converted into the ``short hops'' using the layer of fSWAPs, $\mathcal{F}$, and the final virtual permutation of the fermionic modes $\mathcal{P}$, denoted by the mismatch between the input and the output qubit wire labels.
}
\label{fig:fh1d_trotter}
\end{figure}
The quantum circuit implementing the Trotter step $U_{\mathrm{step}}$ and the virtual permutation $\mathcal{P}$ at the end is schematically depicted in Fig.~\ref{fig:fh1d_trotter}. Crucially, the pair-interleaved ordering allows for the compilation of circuits onto a one-dimensional topology with minimal fSWAP overhead. By extension, this efficiency translates directly to heavy-hex architectures, which natively embed the required one-dimensional connectivity.

We emphasize that the error analysis presented in this subsection is performed for the evolution operators and thus provides a worst-case error bound for physical observables. In practice, the error in an observable depends on the initial state and the form of that observable and may be significantly lower. The appropriate value for the Trotter step should thus be determined and validated empirically; such an analysis is presented in Sec.~\ref{sec:trotter_step_selection}.

\subsection{Circuit compilation: Gate decompositions and layout selection}
\label{sec:compilation_layout_selection}

To execute the Trotterized evolution on quantum hardware, the abstract circuit, such as the one depicted in Fig.~\ref{fig:fh1d_trotter}, must be transpiled into the device's native gate set and mapped onto its topology. We utilize the \texttt{ibm\char`_boston} processor, which features a heavy-hex topology. The pair-interleaved JW ordering, introduced in Sec.~\ref{subsec:j-w}, allows for an optimal embedding of the fermionic chain into a 1D qubit line topology that is natively supported by the heavy-hex lattice. Examples of this embedding for $120$ qubits are illustrated in Fig.~\ref{fig:layouts}, where different colors distinguish the spin-$\uparrow$ (marked red) and spin-$\downarrow$ (marked blue) species.  Crucially, this mapping enables the implementation of the required connectivity without incurring any routing overhead beyond the fSWAPs already integrated into the Trotter steps (Sec.~\ref{subsec:trotterization}).

The native instruction set on \texttt{ibm\char`_boston} includes fractional $R_X(\theta)$ and $R_{ZZ}(\theta)$ gates, alongside fixed-angle $X$, $\sqrt{X}$, and $\mathrm{C}Z$ gates.
During compilation, we utilize these fractional gates to numerically decompose circuit blocks into the most efficient native representations, thereby minimizing the total gate count and circuit depth.
All circuit identities below are defined up to an irrelevant global phase.

\subsubsection{Kinetic term decompositions}

Each $R_{XX}(\theta)R_{YY}(\theta)$ block in the Trotter circuit (see Fig.~\ref{fig:fh1d_trotter}) is decomposed as follows,
\begin{figure}[H]
\centering
\begin{tikzpicture}[x=1cm,y=0.9cm]

\tikzset{wire/.style={thick}}

%

\draw[wire] (0,0)  -- (13.5,0);
\draw[wire] (0,-1) -- (13.5,-1);

\draw[thick,fill=white] (0.5,-0.25)  rectangle (1.7,0.25);  \node[font=\footnotesize] at (1.1,0)  {$R_Z(\tfrac{\pi}{2})$};
\draw[thick,fill=white] (0.5,-1.25)  rectangle (1.7,-0.75); \node[font=\footnotesize] at (1.1,-1) {$R_Z(\tfrac{\pi}{2})$};
\draw[thick,fill=white] (2.0,-0.25)  rectangle (3.2,0.25);  \node[font=\footnotesize] at (2.6,0)  {$\sqrt{X}$};
\draw[thick,fill=white] (2.0,-1.25)  rectangle (3.2,-0.75); \node[font=\footnotesize] at (2.6,-1) {$\sqrt{X}$};
\draw[thick,fill=white] (3.5,-0.25)  rectangle (4.7,0.25);  \node[font=\footnotesize] at (4.1,0)  {$R_Z(\tfrac{\pi}{2})$};
\draw[thick,fill=white] (3.5,-1.25)  rectangle (4.7,-0.75); \node[font=\footnotesize] at (4.1,-1) {$R_Z(\tfrac{\pi}{2})$};

\draw[thick,fill=white] (5.2,-1.25)  rectangle (6.4,0.25);  \node[font=\footnotesize] at (5.8,-0.5) {$R_{ZZ}(\theta)$};

\draw[thick,fill=white] (6.9,-0.25)  rectangle (8.1,0.25);  \node[font=\footnotesize] at (7.5,0)  {$R_X(-\tfrac{\pi}{2})$};
\draw[thick,fill=white] (6.9,-1.25)  rectangle (8.1,-0.75); \node[font=\footnotesize] at (7.5,-1) {$R_X(-\tfrac{\pi}{2})$};
\draw[thick,fill=white] (8.4,-0.25)  rectangle (9.6,0.25);  \node[font=\footnotesize] at (9.0,0)  {$R_Z(-\tfrac{\pi}{2})$};
\draw[thick,fill=white] (8.4,-1.25)  rectangle (9.6,-0.75); \node[font=\footnotesize] at (9.0,-1) {$R_Z(-\tfrac{\pi}{2})$};

\draw[thick,fill=white] (10.1,-1.25) rectangle (11.3,0.25);  \node[font=\footnotesize] at (10.7,-0.5) {$R_{ZZ}(\theta)$};

\draw[thick,fill=white] (11.8,-0.25) rectangle (13.0,0.25);  \node[font=\footnotesize] at (12.4,0)  {$R_X(-\tfrac{\pi}{2})$};
\draw[thick,fill=white] (11.8,-1.25) rectangle (13.0,-0.75); \node[font=\footnotesize] at (12.4,-1) {$R_X(-\tfrac{\pi}{2})$};

\end{tikzpicture}.
\label{fig:r_xx_r_yy_decomposition}
\end{figure}
\noindent 
Furthermore, the adjacent blocks of $R_{XX}$, $R_{YY}$ and $R_Z$ gates at the boundary between the $k$-th and $(k+1)$-th Trotter steps,
\begin{figure}[H]
\centering
\begin{tikzpicture}[x=1cm,y=0.9cm]

\tikzset{wire/.style={thick}}

%

\draw[wire] (0,0)  -- (10.1,0);
\draw[wire] (0,-1) -- (10.1,-1);

\draw[thick,fill=white] (0.5,-1.25)  rectangle (1.7,0.25);  \node[font=\footnotesize] at (1.1,-0.5) {$R_{XX}(\theta)$};
\draw[thick,fill=white] (2.0,-1.25)  rectangle (3.2,0.25);  \node[font=\footnotesize] at (2.6,-0.5) {$R_{YY}(\theta)$};

\draw[thick,fill=white] (3.7,-0.25)  rectangle (4.9,0.25);  \node[font=\footnotesize] at (4.3,0)  {$R_Z(\theta_Z)$};
\draw[thick,fill=white] (3.7,-1.25)  rectangle (4.9,-0.75); \node[font=\footnotesize] at (4.3,-1) {$R_Z(\theta_Z)$};
\draw[thick,fill=white] (5.2,-0.25)  rectangle (6.4,0.25);  \node[font=\footnotesize] at (5.8,0)  {$R_Z(\theta_Z)$};
\draw[thick,fill=white] (5.2,-1.25)  rectangle (6.4,-0.75); \node[font=\footnotesize] at (5.8,-1) {$R_Z(\theta_Z)$};

\draw[dashed,thick] (5.05,0.5) -- (5.05,-1.5);

\draw[thick,fill=white] (6.9,-1.25)  rectangle (8.1,0.25);  \node[font=\footnotesize] at (7.5,-0.5) {$R_{XX}(\theta)$};
\draw[thick,fill=white] (8.4,-1.25)  rectangle (9.6,0.25);  \node[font=\footnotesize] at (9.0,-0.5) {$R_{YY}(\theta)$};

\end{tikzpicture},
\label{fig:r_xx_r_yy_boundary}
\end{figure}
\noindent are combined and resynthesized into a consolidated block,
\begin{figure}[H]
\centering
\begin{tikzpicture}[x=1cm,y=0.9cm]

\tikzset{wire/.style={thick}}

%

\draw[wire] (0,0)  -- (15.0,0);
\draw[wire] (0,-1) -- (15.0,-1);

\draw[thick,fill=white] (0.5,-0.25)   rectangle (1.7,0.25);   \node[font=\footnotesize] at (1.1,0)  {$R_Z(\theta_1)$};
\draw[thick,fill=white] (0.5,-1.25)   rectangle (1.7,-0.75);  \node[font=\footnotesize] at (1.1,-1) {$R_Z(\theta_1)$};
\draw[thick,fill=white] (2.0,-0.25)   rectangle (3.2,0.25);   \node[font=\footnotesize] at (2.6,0)  {$R_X(\tfrac{\pi}{2})$};
\draw[thick,fill=white] (2.0,-1.25)   rectangle (3.2,-0.75);  \node[font=\footnotesize] at (2.6,-1) {$R_X(\tfrac{\pi}{2})$};
\draw[thick,fill=white] (3.5,-0.25)   rectangle (4.7,0.25);   \node[font=\footnotesize] at (4.1,0)  {$R_Z(-\tfrac{\pi}{2})$};
\draw[thick,fill=white] (3.5,-1.25)   rectangle (4.7,-0.75);  \node[font=\footnotesize] at (4.1,-1) {$R_Z(-\tfrac{\pi}{2})$};

\draw[thick,fill=white] (5.2,-1.25)   rectangle (6.4,0.25);   \node[font=\footnotesize] at (5.8,-0.5) {$R_{ZZ}(2\theta)$};

\draw[thick,fill=white] (6.9,-0.25)   rectangle (8.1,0.25);   \node[font=\footnotesize] at (7.5,0)  {$R_X(\tfrac{\pi}{2})$};
\draw[thick,fill=white] (6.9,-1.25)   rectangle (8.1,-0.75);  \node[font=\footnotesize] at (7.5,-1) {$R_X(\tfrac{\pi}{2})$};

\draw[thick,fill=white] (8.6,-1.25)   rectangle (9.8,0.25);   \node[font=\footnotesize] at (9.2,-0.5) {$R_{ZZ}(2\theta)$};

\draw[thick,fill=white] (10.3,-0.25)  rectangle (11.5,0.25);  \node[font=\footnotesize] at (10.9,0)  {$R_Z(-\tfrac{\pi}{2})$};
\draw[thick,fill=white] (10.3,-1.25)  rectangle (11.5,-0.75); \node[font=\footnotesize] at (10.9,-1) {$R_Z(-\tfrac{\pi}{2})$};
\draw[thick,fill=white] (11.8,-0.25)  rectangle (13.0,0.25);  \node[font=\footnotesize] at (12.4,0)  {$R_X(\tfrac{\pi}{2})$};
\draw[thick,fill=white] (11.8,-1.25)  rectangle (13.0,-0.75); \node[font=\footnotesize] at (12.4,-1) {$R_X(\tfrac{\pi}{2})$};
\draw[thick,fill=white] (13.3,-0.25)  rectangle (14.5,0.25);  \node[font=\footnotesize] at (13.9,0)  {$R_Z(\theta_2)$};
\draw[thick,fill=white] (13.3,-1.25)  rectangle (14.5,-0.75); \node[font=\footnotesize] at (13.9,-1) {$R_Z(\theta_2)$};

\end{tikzpicture},
\label{fig:r_xx_r_yy_boundary_decomposition}
\end{figure}
\noindent where the angles $\theta_1$ and $\theta_2$ are determined numerically from the input angles $\theta$ and $\theta_Z$.

\subsubsection{Interaction term and \textnormal{fSWAP} decompositions}

Each combined $R_{ZZ}(\theta)\cdot\text{fSWAP}$ block is compiled as
\begin{figure}[H]
\centering
\resizebox{\linewidth}{!}{\begin{tikzpicture}[x=1cm,y=0.9cm]

\tikzset{wire/.style={thick}}

%

\draw[wire] (0,0)  -- (20.1,0);
\draw[wire] (0,-1) -- (20.1,-1);

\draw[thick,fill=white] (0.5,-0.25)   rectangle (1.7,0.25);   \node[font=\footnotesize] at (1.1,0)  {$R_Z(\theta_1)$};
\draw[thick,fill=white] (0.5,-1.25)   rectangle (1.7,-0.75);  \node[font=\footnotesize] at (1.1,-1) {$R_Z(\theta_2)$};
\draw[thick,fill=white] (2.0,-0.25)   rectangle (3.2,0.25);   \node[font=\footnotesize] at (2.6,0)  {$R_X(\tfrac{\pi}{2})$};
\draw[thick,fill=white] (2.0,-1.25)   rectangle (3.2,-0.75);  \node[font=\footnotesize] at (2.6,-1) {$R_X(\tfrac{\pi}{2})$};
\draw[thick,fill=white] (3.5,-0.25)   rectangle (4.7,0.25);   \node[font=\footnotesize] at (4.1,0)  {$R_Z(\tfrac{\pi}{2})$};
\draw[thick,fill=white] (3.5,-1.25)   rectangle (4.7,-0.75);  \node[font=\footnotesize] at (4.1,-1) {$R_Z(-\tfrac{\pi}{2})$};

\draw[thick,fill=white] (5.2,-1.25)   rectangle (6.4,0.25);   \node[font=\tiny] at (5.8,-0.5) {$R_{ZZ}(-\tfrac{\pi}{2})$};

\draw[thick,fill=white] (6.9,-0.25)   rectangle (8.1,0.25);   \node[font=\footnotesize] at (7.5,0)  {$R_X(-\tfrac{\pi}{2})$};
\draw[thick,fill=white] (6.9,-1.25)   rectangle (8.1,-0.75);  \node[font=\footnotesize] at (7.5,-1) {$R_X(-\tfrac{\pi}{2})$};
\draw[thick,fill=white] (8.4,-0.25)   rectangle (9.6,0.25);   \node[font=\footnotesize] at (9.0,0)  {$R_Z(-\pi)$};
\draw[thick,fill=white] (8.4,-1.25)   rectangle (9.6,-0.75);  \node[font=\footnotesize] at (9.0,-1) {$R_Z(-\pi)$};

\draw[thick,fill=white] (10.1,-1.25)  rectangle (11.3,0.25);  \node[font=\tiny] at (10.7,-0.5) {$R_{ZZ}(-\tfrac{\pi}{2})$};

\draw[thick,fill=white] (11.8,-0.25)  rectangle (13.0,0.25);  \node[font=\footnotesize] at (12.4,0)  {$R_Z(\tfrac{\pi}{2})$};
\draw[thick,fill=white] (11.8,-1.25)  rectangle (13.0,-0.75); \node[font=\footnotesize] at (12.4,-1) {$R_Z(-\tfrac{\pi}{2})$};
\draw[thick,fill=white] (13.3,-0.25)  rectangle (14.5,0.25);  \node[font=\footnotesize] at (13.9,0)  {$R_X(\tfrac{\pi}{2})$};
\draw[thick,fill=white] (13.3,-1.25)  rectangle (14.5,-0.75); \node[font=\footnotesize] at (13.9,-1) {$R_X(-\tfrac{\pi}{2})$};

\draw[thick,fill=white] (15.0,-1.25)  rectangle (16.2,0.25);  \node[font=\footnotesize] at (15.6,-0.5) {$R_{ZZ}(\theta)$};

\draw[thick,fill=white] (16.7,-0.25)  rectangle (17.9,0.25);  \node[font=\footnotesize] at (17.3,0)  {$R_Z(\theta_3)$};
\draw[thick,fill=white] (16.7,-1.25)  rectangle (17.9,-0.75); \node[font=\footnotesize] at (17.3,-1) {$R_Z(\theta_1)$};

\draw[thick,fill=white] (18.4,-1.25)  rectangle (19.6,-0.75); \node[font=\footnotesize] at (19.0,-1) {$X$};

\end{tikzpicture}}
\label{fig:rzz_fSWAP}
\end{figure}
\noindent where angles $\theta_{1,2,3}$ are determined numerically based on the input angle $\theta$.

\subsubsection{Total depth and gate complexity}

Following these decomposition rules, the total two-qubit circuit depth ($D_{2\mathrm{Q}}$) and gate count ($N_{2\mathrm{Q}}$) can be straightforwardly evaluated. For $n_{\mathrm{step}}$ Trotter steps, the two-qubit depth is
\begin{equation}
    D_{2\mathrm{Q}} = 5n_{\mathrm{step}} + 2.
\end{equation} 
Crucially, $D_{2\mathrm{Q}}$ is independent of the system size $L$, allowing us to scale the simulation without increasing its execution time.
The total number of two-qubit gates scales linearly with $L$,
\begin{equation}
    N_{2\mathrm{Q}} = (5L-2)n_{\mathrm{step}} + 2(L-1) - 1,
\end{equation}
where the final $-1$ accounts for the cancellation of a constituent $\mathrm{C}X$ gate within the fSWAP sequence. This compiler optimization occurs because the control qubit of that gate is initialized in a known computational basis state ($|0\rangle$ or $|1\rangle$). Table~\ref{tab:gate_complexity} shows $D_{2\mathrm{Q}}$, $N_{2\mathrm{Q}}$, and the aggregated QPU execution times for various experimental configurations \{$L$, $n_{\mathrm{step}}$\}. We distinguish between the time required for the primary circuits alone ($T_{\mathrm{QPU}}$) and the total time including noise characterization overhead ($T_{\mathrm{QPU}}^{\mathrm{REM+DR}}$). For our largest-scale simulations by system size ($L=60$, $n_{\mathrm{step}}=30$), the circuits utilize $N_{2\mathrm{Q}}=9{,}057$ two-qubit gates at a depth of $D_{2\mathrm{Q}}=152$. Meanwhile, our deepest simulations ($L=31$, $n_{\mathrm{step}}=90$) reach a two-qubit depth of $D_{2\mathrm{Q}}=452$ with $N_{2\mathrm{Q}}=13{,}829$ two-qubit gates.

\begin{table*}[t] 
\centering
\caption{\label{tab:gate_complexity} 
Two-qubit circuit depths ($D_{2\mathrm{Q}}$), total gate counts ($N_{2\mathrm{Q}}$), and aggregated QPU execution times ($T_{\mathrm{QPU}}$) for various experimental configurations. Em-dashes (---) indicate configurations not explicitly reported in this study. The reported $T_{\mathrm{QPU}}$ represents the cumulative wall-clock execution time for the main circuits, while $T_{\mathrm{QPU}}^{\mathrm{REM+DR}}$ includes the additional noise characterization overhead for post-processing (readout error mitigation and decay recovery, see Sec.~\ref{app:postprocessing}). Both $T_{\mathrm{QPU}}$ and $T_{\mathrm{QPU}}^{\mathrm{REM+DR}}$ report actual QPU usage only, and do include time needed for compilation, data retrieval from IBM Cloud, and post-processing.
}
\begin{ruledtabular}
\begin{tabular}{l @{\hspace{0.7em}} cccc @{\hspace{0.7em}} cccc @{\hspace{0.7em}} cccc}
 & \multicolumn{4}{c}{$n_{\mathrm{step}} = 30$} & \multicolumn{4}{c}{$n_{\mathrm{step}} = 60$} & \multicolumn{4}{c}{$n_{\mathrm{step}} = 90$} \\
\cmidrule(lr){2-5} \cmidrule(lr){6-9} \cmidrule(lr){10-13}
System size $L$ & $D_{2\mathrm{Q}}$ & $N_{2\mathrm{Q}}$ & $T_{\mathrm{QPU}}$ & $T_{\mathrm{QPU}}^{\mathrm{REM+DR}}$ & $D_{2\mathrm{Q}}$ & $N_{2\mathrm{Q}}$ & $T_{\mathrm{QPU}}$ & $T_{\mathrm{QPU}}^{\mathrm{REM+DR}}$ & $D_{2\mathrm{Q}}$ & $N_{2\mathrm{Q}}$ & $T_{\mathrm{QPU}}$ & $T_{\mathrm{QPU}}^{\mathrm{REM+DR}}$ \\
\midrule
10 sites (20 qubits)  & 152 & 1{,}457 & 2m46s & 4m25s & 302 & 2{,}897 & 5m36s & 8m42s & 452 & 4,337 & 8m46s & 13m26s \\
31 sites (62 qubits)  & 152 & 4{,}649 & 2m46s & 4m25s & 302 & 9{,}239 & 5m36s & 8m42s & 452 & 13{,}829 & 8m46s & 13m26s \\
60 sites (120 qubits) & 152 & 9{,}057 & 2m46s & 4m25s &---&---&---&---&---&---&---&---\\
\end{tabular}
\end{ruledtabular}
\end{table*}

\subsubsection{Layout selection}
\label{sec:layout}

\begin{figure}
    \centering
    \includegraphics[width=\linewidth]{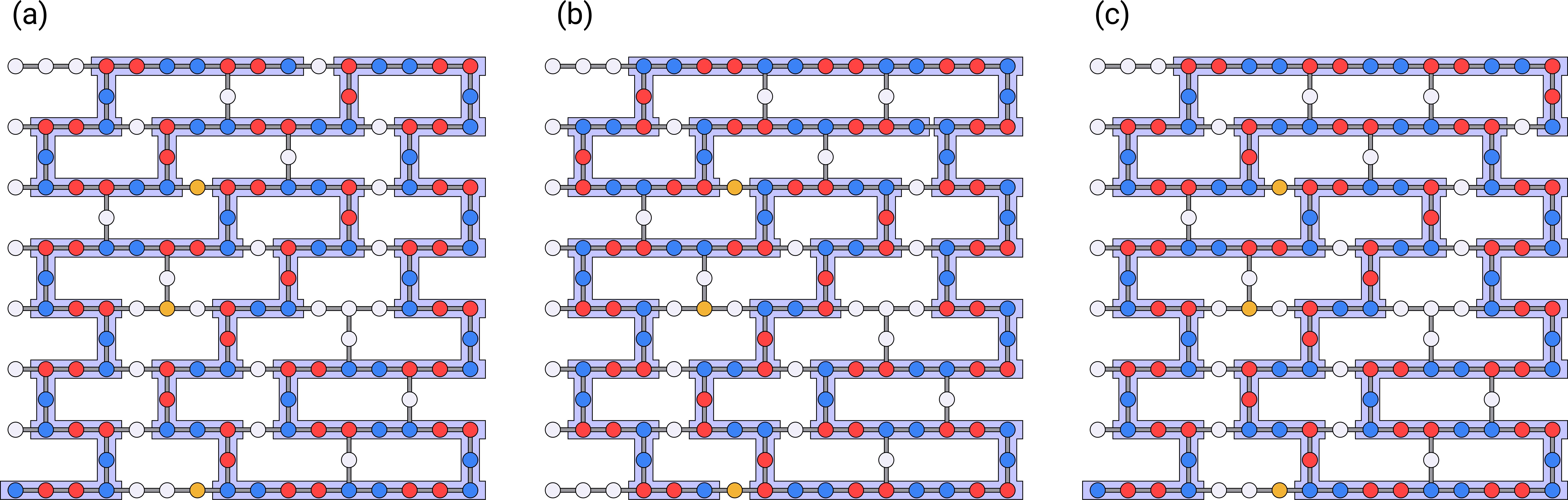}
    \caption{
    Device layout and qubit mapping on \texttt{ibm\char`_boston}. Representative layouts for $L=60$ sites ($120$ qubits) utilized in our experiments. Blue and red circles denote qubits encoding spin-up and spin-down fermionic modes, respectively. These alternate every Trotter step due to the spin-exchange $\mathrm{fSWAP}$ layer. White circles indicate unused qubits, while orange circles highlight qubits excluded from the layout due to gate and readout fidelities significantly below the device median. The chosen layouts avoid these qubits.}
    \label{fig:layouts}
\end{figure}

Once the circuit is transpiled into hardware-native gates and its routing is established, it must be mapped onto a specific physical qubit register. This task is known as layout selection, and it can be framed as a subgraph isomorphism problem on the device coupling graph. Typically, this is addressed using the VF2++ algorithm \cite{Juttner2018, qiskit}, which identifies viable isomorphic subgraphs based on the circuit's connectivity. These candidates are then ranked using a scoring function that incorporates the device's backend calibration data---retrieved from the most recent calibration of \texttt{ibm\char`_boston}---including gate error rates, coherence times ($T_1, T_2$), and readout fidelities.

For simulations up to $2L<60$ qubits, we find that VF2++ efficiently identifies high-fidelity layouts. Specifically, the algorithm consistently avoids qubits with fidelities significantly below the device median. We observed that the \texttt{ibm\char`_boston} processor contains three such persistently high-noise qubits (marked in orange in Fig.~\ref{fig:layouts}) whose locations remain fixed.

However, for circuits with $2L \ge 60$ qubits (such as the 62- and 120-qubit experiments we perform), the increased complexity of the search space often causes VF2++ to converge on suboptimal solutions that include high-noise qubits.
Due to the nature of error propagation in Trotterized circuits, the presence of even a single high-noise qubit can significantly degrade the simulation's overall fidelity. To avoid this, we employ a variation of the recently introduced $\Delta$-Motif algorithm \cite{Wang2026}, which offers a highly parallelizable approach to the subgraph isomorphism problem. For $120$-qubit circuits, this approach allows us to exhaustively identify and score all $108,988$ possible 1D-chain layouts on the \texttt{ibm\char`_boston} heavy-hex lattice.

At the time of our experiments, the \texttt{ibm\char`_boston} device contained three notably high-noise qubits. An exhaustive search revealed that only 21 of the 108,988 possible layouts successfully excluded these faulty components. From this filtered subset of high-fidelity candidates, we selected the layout with the highest aggregate fidelity score for runtime execution
The exact chosen layout depends on the device's backend calibration data at the time of the circuit execution. Examples of the configurations we used in our experiments are shown in Fig.~\ref{fig:layouts}.

\subsection{Error suppression and Pauli twirling}
\label{sec:error_suppression}

State-of-the-art quantum hardware remains limited by decoherence and residual error sources, even when employing sophisticated application-aware compilation to minimize circuit depth and bypass high-noise qubits. These noise processes accumulate during execution, reducing fidelity and effectively limiting the circuit depth over which a meaningful signal can be resolved.

To counteract these effects, we utilize circuit-level error suppression and noise tailoring via Q-CTRL's \texttt{Fire Opal} pipeline~\cite{Mundada2023, Coote2025, Seif_2024, Hartnett2024learningtorank}. Unlike the post-processing mitigation techniques discussed in Sec.~\ref{app:postprocessing}, these methods---including hardware-optimized dynamical decoupling~\cite{Coote2025}---require no additional sampling overhead or the execution of auxiliary noise characterization circuits.  Due to the large gate counts employed in circuits executed here and limitations on the IBM Platform's support for direct analog-level gate-waveform optimization~\cite{carvalho2021, baum2021, Mundada2023}, we further layer Pauli twirling\cite{Wallman2016, Winick2022} in the compilation stage, in addition to our standard pipeline.  

Pauli twirling is a noise tailoring strategy that applies random single-qubit Pauli gates before and after each two-qubit gate. These transformations are chosen such that each resulting gate (and thus the entire circuit) remain unitarily equivalent to the original.
The randomly sampled Pauli gates are then absorbed into the existing single-qubit rotations of the circuit. This ensures that each circuit randomization---or ``twirl"---maintains a gate count and execution time virtually identical to those of the original circuit.
This process effectively converts coherent gate errors---which can constructively interfere and lead to significant systematic biases---into stochastic Pauli errors, and works well in circumstances where the effective orientation of noise processes in the Pauli basis is not well characterized.
Beyond making the noise more amenable to subsequent mitigation, this transformation suppresses the accumulation of residual coherent errors not otherwise suppressed, typically reducing the overall error magnitude by ensuring that imperfections add incoherently rather than constructively. For the parameterized $R_{ZZ}(\theta)$ gates utilized in this work, we employ pseudo-twirling (or partial twirling), restricting to the subset of the Pauli group that commutes with the $ZZ$ interaction.

In our implementation, a target circuit requiring $N_{\mathrm{shot}}$ is compiled into $N_{\mathrm{twirl}}$ randomized variations, each executed with $N_{\mathrm{shot}}/N_{\mathrm{twirl}}$ shots. After execution, the results from these variations are aggregated into a single distribution. Consequently, Pauli twirling introduces no additional shot overhead. For all experiments, we used $N_{\mathrm{twirl}}=10$, which provides an optimal balance between coherent error reduction and compilation time. Using $N_{\mathrm{shot}} = 20{,}000$ total shots, we executed $10$ unique circuit randomizations with $2{,}000$ shots each.

\begin{figure}
    \centering
    \includegraphics[width=0.75\linewidth]{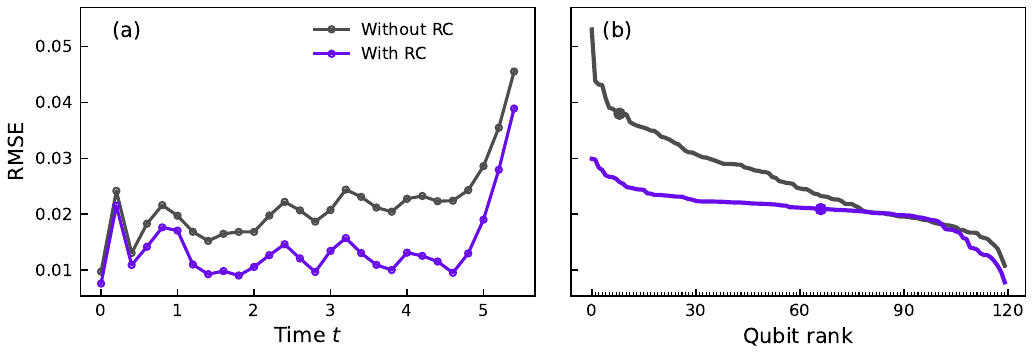}
    \caption{Impact of Pauli twirling (AKA Randomized compiling, RC). Data is shown for the $U=-2$, $L=60$ system prepared in a Néel initial state. The RMSE of the fermionic occupations $\langle n_{i\sigma} \rangle$ is calculated relative to classical TDVP benchmarks (maximum bond dimension $\chi=2048$) and presented as a function of evolution time (a) and qubit rank (b). All experiments were performed with dynamical decoupling enabled. No post-processing or error mitigation techniques were applied.
    (a) The RMSE is computed across all qubits for each time step.
    (b) The RMSE is computed across all time steps for each qubit, with both datasets (with and without RC) separately sorted by magnitude. To emphasize that the sorting is independent for each dataset, a single representative qubit (index 30) is marked with a distinct marker on both lines. Error bars are smaller than the markers and have been omitted for clarity.}
    \label{fig:twirling_effect}
\end{figure}

Figure~\ref{fig:twirling_effect} quantifies the impact of Pauli twirling on the measured fermionic occupations $\langle n_{i\sigma} \rangle$. In Fig.~\ref{fig:twirling_effect}(a), we evaluate the RMSE relative to classical TDVP benchmarks as a function of evolution time. We observe that Pauli twirling consistently suppresses the RMSE across the entire time domain by an average of $\sim1\%$. The performance gain from Pauli twirling generally increases with the number of Trotter steps until $t \approx 5$, at which point the RMSE begins to drastically rise due to either the effects of hardware decoherence, the breakdown of the TDVP approximation, or both (as detailed in the main text). This behavior is consistent with suppression of the compounding of coherent errors with increasing circuit depth.  

Figure~\ref{fig:twirling_effect}(b) presents the RMSE calculated for each individual qubit across the full evolution period. To illustrate the global improvement in fidelity, both datasets with and without Pauli twirling are independently sorted by magnitude. This visualization demonstrates that the benefits of noise tailoring are distributed across the entire register, rather than being confined to a subset of qubits.

Quantitatively, the application of Pauli twirling yields a $21.7$\% reduction in the total RMSE across all time steps; within the $t \in [0, 5.2]$ interval---the regime where the simulation remains most reliable---the reduction in RMSE is $34.5$\%.

\subsection{Comparative analysis of fermion-to-qubit orderings and routing efficiency}
\label{sec:other_orderings_comparison}

In this subsection, we compare our proposed \textit{pair-interleaved} fermion-to-qubit ordering in the JW transformation [Eq.~\eqref{eq:pair_interleaved_ordering}] with commonly-used alternative orderings for digital quantum simulations of the Fermi-Hubbard model. In one dimension, the pair-interleaved ordering yields the lowest routing overhead on heavy-hex device topologies.

A common choice in the literature is the \textit{block} ordering~\cite{Arute2020SpinCharge, Stanisic2022, Alam2025_2D}, which sequentially groups all spin-up modes followed by all spin-down modes:
\begin{equation}
    \{ c_{0\uparrow}, c_{1\uparrow},\ldots,c_{(L-1)\uparrow}, c_{0\downarrow}, c_{1\downarrow},\ldots, c_{(L-1)\downarrow} \} \to \{ c_{0}, c_{1},\ldots, c_{L-1}, c_{L}, c_{L+1},\ldots, c_{2L-1} \} \,.
\end{equation}
Applying the JW transformation to the Fermi-Hubbard Hamiltonian under this ordering yields the following expression for the kinetic term
\begin{equation}
    H_{\mathrm{K}} = -\frac{1}{2} \sum_{\substack{J=0 \\ J \neq L-1}}^{2L-2} \left( 
    X_{J} X_{J+1} 
    + Y_{J} Y_{J+1} \right) \,.
\end{equation}
Unlike the pair-interleaved approach [see Eq.~\eqref{eq:H_S_H_L}], this kinetic term is free of ``$Z$-strings". Consequently, the Trotterized evolution of $H_{\mathrm{K}}$ can be executed on a (linear) JW chain using strictly local two-qubit gates, circumventing the need for fSWAP gates.

However, this simplification for the kinetic term comes at a cost for the interaction term. While the one-qubit component $H_{1\mathrm{Q}}$ remains identical to that in Eq.~\eqref{eq:mu_and_U_terms}, the two-qubit interaction term becomes
\begin{equation}
    H_{U,2\mathrm{Q}} = \frac{U}{4} \sum_{J=0}^{L-1} Z_{J} Z_{J+L} \,.
\end{equation}
This term requires executing $R_{ZZ}$ rotations between qubits separated by a distance of $L$ along the JW chain. On hardware with a square-grid topology, a JW chain can be folded such that the spin-up and spin-down sectors run parallel, allowing indices $J$ and $J+L$ to remain physically adjacent~\cite{Arute2020SpinCharge, Alam2025_2D}. On a heavy-hex architecture, however, embedding such a perfectly folded structure is topologically prohibited. As a result, implementing these long-range inter-sector interactions necessitates extensive fSWAP routing, rendering the block ordering highly inefficient for heavy-hex devices. 

Additionally, our pair-interleaved ordering naturally accommodates spin-flip terms of the form \(c_{i,\uparrow}^\dagger c_{j,\downarrow}+c_{i,\downarrow}^\dagger c_{j,\uparrow}\), should such interactions be present in the fermionic Hamiltonian, for example due to magnetic impurities. By contrast, in a block ordering where the spin-up and spin-down sectors are represented as two \textit{disconnected} JW chains~\cite{Alam2025_2D}, such spin-flip interaction terms cannot be incorporated directly.

Another common choice in the literature is the \textit{interleaved} ordering~\cite{Chowdhury2026}, defined by the sequence
\begin{equation}
    \{ c_{0\uparrow}, c_{0\downarrow}, c_{1\uparrow}, c_{1\downarrow}, \ldots,c_{(L-1)\uparrow},c_{(L-1)\downarrow} \} \to \{ c_{0}, c_{1}, c_{2}, c_{3},\ldots, c_{2L-2}, c_{2L-1} \} \,.
\end{equation}
Under this mapping, the one-qubit component $H_{1\mathrm{Q}}$ remains identical to Eq.~\eqref{eq:mu_and_U_terms}. However, unlike the block ordering, this interleaved sequence enables the direct, nearest-neighbor implementation of the two-qubit interaction term on a heavy-hex topology
\begin{equation}
    H_{U,2\mathrm{Q}} = \frac{U}{4} \sum_{J=0}^{L-1} Z_{2J} Z_{2J+1} \,.
\end{equation}
Conversely, the kinetic hopping term transforms into weight-three Pauli operators, as the ``$Z$-strings" reduce to a single intermediate Pauli $Z$ operator
\begin{equation}
    H_{\mathrm{K}} = -\frac{1}{2} \sum_{J=0}^{2L-3} \left( 
    X_{J} Z_{J+1} X_{J+2} 
    + Y_{J} Z_{J+1} Y_{J+2} \right) \,.
\end{equation}
Implementing these next-nearest-neighbor hoppings on a heavy-hex lattice requires routing through the intermediate qubits via fSWAP gates. Specifically, the interleaved ordering necessitates two layers of fSWAP gates per Trotter step. In contrast, our proposed pair-interleaved ordering reduces this routing overhead to just a single fSWAP layer per Trotter step, rendering it significantly more hardware-efficient.

To quantify this advantage, we compare our pair-interleaved ordering against the interleaved ordering utilized in Ref.~\cite{Chowdhury2026} for 1D Fermi-Hubbard dynamical simulations on IBM Heron devices. 
Reference~\cite{Chowdhury2026} reports a two-qubit depth of $D_{2\mathrm{Q}} \approx 26$ per second-order Trotter step. In comparison, to achieve the equivalent algorithmic evolution using our symmetrized Trotterization (two mirrored first-order steps), our compilation requires a two-qubit depth of only $D_{2\mathrm{Q}} = 10$.

This efficiency scales highly favorably for deeper circuits. For the largest simulation reported in Ref.~\cite{Chowdhury2026} ($L=52$ sites, $10$ second-order Trotter steps), the interleaved ordering in that study yields $D_{2\mathrm{Q}} = 263$ and $N_{2\mathrm{Q}} = 8{,}844$, as reported in Tab. III of that study. For the same system size and equivalent Trotter sequence, our pair-interleaved ordering requires only $D_{2\mathrm{Q}} = 102$ and $N_{2\mathrm{Q}} = 5{,}261$. This represents a substantial $61.2$\% reduction in two-qubit circuit depth and a $40.5$\% reduction in total two-qubit gate count.

\section{Classical baselines \label{sec:classical_benchmark_methods}}
In this section, we detail the classical simulation algorithms used to benchmark our quantum hardware results. We consider both Schr\"odinger-picture and Heisenberg-picture simulation schemes. In the former, the quantum state $\ket{\psi(t)} = U(t)\ket{\psi(0)}$ is evolved while observables remain fixed and where $U(t)$ is the time evolution operator, while in the latter observables $O(t) = U^\dagger(t)\, O(0)\, U(t)$ are evolved while the state is held fixed. The two pictures are formally equivalent but lead to qualitatively different classical algorithms: Schr\"odinger methods are limited by the entanglement of the evolved state, whereas Heisenberg methods are limited by the operator-space complexity (e.g., the operator entanglement or operator weight, depending on the method). A key distinction is that Schr\"odinger-picture methods produce an approximation to the full wavefunction, from which the expectation value of any observable can be computed, whereas Heisenberg-picture methods target a single observable by construction, evolving that operator to obtain one expectation value per run. This asymmetry must be accounted for when comparing the two classes: a single Schr\"odinger-picture run amortizes its cost across all observables, while a Heisenberg-picture run must be repeated for each.

For Schr\"odinger-picture simulation, we employ the time-dependent variational principle (TDVP)~\cite{haegeman2011time, haegeman2016unifying}, a leading tensor network approach for real-time dynamics. For Heisenberg-picture simulation, we consider three methods. Pauli Path Propagation (PPP) and Majorana Propagation (MP) both expand the operator of interest $O$ in a particular basis and evolve the basis operators, with the exponential growth of terms mitigated by truncating small-coefficient or high-weight contributions; PPP uses the Pauli basis and MP uses the Majorana basis. We also consider a tensor network method wherein the Heisenberg-picture simulation of a Matrix Product Operator (MPO) is performed. For each method, we present results at the large system sizes accessed in our experiments (up to $L = 60$) and across the different initial states considered. In addition, for validation and benchmarking purposes, we consider several methods that do not scale to large system sizes---namely exact diagonalization (ED) and direct simulation of Trotter circuits, which are discussed separately in Sec.~\ref{sec:validation}.

\subsection{Time-dependent variational principle \label{sec:tdvp}} 

The TDVP algorithm~\cite{haegeman2011time, haegeman2016unifying} is a widely-used algorithm for classically simulating the time evolution of quantum systems. In this section, we first provide some background information for the algorithm, discuss in more detail the specific implementations used for our classical benchmarking, and conclude with a discussion about improved implementations.

\subsubsection{Background}
TDVP evolves a Matrix Product State (MPS) by projecting the Schrödinger equation onto the tangent space of the variational manifold and splitting the tangent-space projector into local terms that can each be integrated exactly. Two variants exist, distinguished by whether the projector is built from single-site or two-site tangent vectors; we use the two-site version, which adapts the bond dimension on the fly at the cost of an SVD truncation. These local flows are composed in a DMRG-like sweep: at each bond, the two-site tensor is forward-evolved under an effective Hamiltonian, SVD-factorized with small singular values truncated, and the resulting single-site tensor is back-evolved before the sweep proceeds; a reverse sweep then yields a symmetric second-order integrator. The method applies to any Hamiltonian with an efficient MPO representation, incurs no projection error for nearest-neighbor Hamiltonians, and mirrors a DMRG sweep with the local eigensolver replaced by a local matrix exponential.

In more detail, two-site TDVP for a one-dimensional MPS evolves the state by sweeping through the chain, applying a Krylov exponential of the effective two-site Hamiltonian at each bond and truncating the resulting SVD. Five parameters control the cost. The chain length $L$ and the on-site physical dimension $d$ are fixed by the problem -- in our case, $d=4$ for Fermi-Hubbard, as there are four states per site. The MPO bond dimension $w$, i.e.\ the size of the auxiliary index in the matrix-product representation of $H$, is fixed by the Hamiltonian -- in our case, $w=6$. The MPS bond dimension $\chi$ sets how much entanglement the state can represent. This dimension varies throughout the simulation, $\chi = \chi(t)$, and grows as needed until a maximum value is reached, $\texttt{maxdim}$, which is the value we directly control and scan over in our simulations.  Truncation in the SVD step discards as many of the smallest singular values $s_i$ as possible subject to $\sum_{i \in \mathrm{discarded}} s_i^2 \le \varepsilon$, where the $s_i$ are normalized such that $\sum_i s_i^2 = 1$. Finally, the Krylov subspace dimension $k$ is the size of the Lanczos subspace built at each local matrix-exponential solve; it is adaptive, governed by a maximum cap with default value 30. The dominant per-bond cost is the matvec (matrix-vector multiplication) inside the Krylov solver, which scales as $\mathcal{O}(k d^2 w \chi^3)$, plus an $\mathcal{O}(d^3 \chi^3)$ truncation SVD; environment updates are subleading at $\mathcal{O}(d w \chi^3)$. Of these parameters, $L, d, w, k$ are effectively constants of a given run. A full sweep visits $L-1$ bonds, so a symmetric step (one forward + one backward sweep, second-order in $\Delta t_{\mathrm{TDVP}}$) costs $\mathcal{O}(L \chi^3)$ in time and $\mathcal{O}(L \chi^2)$ in memory, with prefactors set by $k, d, w$. The binding constraint is $\chi$. Since under a quench entanglement grows linearly ($S(t) \sim v_E t$), forcing $\chi(t) \sim e^{v_E t}$ at fixed truncation error and pushing the total runtime to $\mathcal{O}(L T e^{3 v_E T} / \Delta t_{\mathrm{TDVP}})$. 

\subsubsection{Benchmarking results}

Next, we detail the implementation used in our benchmarking simulations. We used the \texttt{ITensorMPS.jl} Julia implementation within the \texttt{ITensor} package~\cite{itensor}. TDVP involves a time step $\Delta t_{\mathrm{TDVP}}$ which is \textit{a priori} independent of the Trotter step $\Delta t_{\mathrm{Trotter}}$. To enable pointwise comparison between the TDVP and quantum-hardware simulations, we require the two to be commensurate: $\Delta t_{\mathrm{Trotter}} = \ell \, \Delta t_{\mathrm{TDVP}}$ for some positive integer $\ell$, so that every $\ell$-th TDVP time point coincides with a Trotter time point. Holding $\Delta t_{\mathrm{Trotter}}$ fixed, increasing $\ell$ amounts to refining the TDVP time step. In our simulations, we take $\ell=1$, so the two sets of time points coincide. For the SVD truncation, we used a cutoff value of $\varepsilon = 10^{-8}$, which was applied at each step. With one exception discussed below, all simulations were executed on a single 32vCPU instance (AWS \texttt{c7i.8xlarge}, 64\,GB RAM). We save the full MPS after each step and extract expectation values as needed to compare against the output of the quantum computer. The results of our TDVP benchmarking simulations are reported in the main text, in particular Fig.~\ref{fig:TDVP_summary_plot}. 

\begin{figure}[b]
    \centering
    \includegraphics[width=0.75\linewidth]{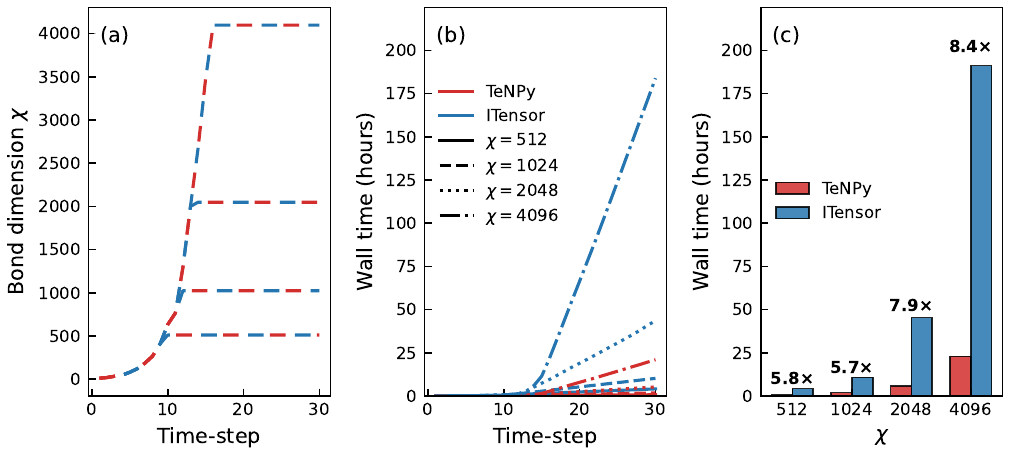}
    \caption{
    \texttt{TeNPy} vs \texttt{ITensor} implementation of the TDVP algorithm for the simulation of the 1D Fermi-Hubbard model. Data corresponds to the same $L=60$, $U/t_h=-2$, N\'{e}el initial state simulation depicted in Fig.~\ref{fig:TDVP_summary_plot} in the main text. 
    (a) The bond-dimension as a function of time-step. 
    (b) The cumulative wall time for both implementations across multiple maximum bond dimensions as a function of time-step.
    (c) The total wall time for both methods and different maximum bond dimensions.
    }    
    \label{fig:itensor_vs_tenpy}
\end{figure}

After the first version of this work appeared, we became aware of \texttt{TeNPy}, another publicly-available implementation of TDVP \cite{hauschild2024tenpy}. Because the runtime advantage of quantum hardware is one of the main claims of this work, we reimplemented the same Fermi-Hubbard simulations in \texttt{TeNPy} and benchmarked them against \texttt{ITensor}. We first verified that the two codes agree across observables to within the expected tolerance. We then compared their runtimes, after ensuring that both algorithms are truncating small singular values according to the same criterion. Both implementations accept a maximum bond dimension $\texttt{maxdim}$ as a convergence parameter. As shown in Fig.~\ref{fig:itensor_vs_tenpy}(a), the bond dimension grows throughout the simulation at the same rate for both methods. Despite this matched growth in bond dimension, the \texttt{TeNPy} implementation is faster across the board, as shown in Fig.~\ref{fig:itensor_vs_tenpy}(b,c). The runtime is dominated by steps occurring after the bond dimension saturates the maximal value, as shown in Panel (b). We find that \texttt{TeNPy} is overall faster than \texttt{ITensor} by a factor of 6--8$\times$. We speculate that the reason for this discrepancy likely has to do with the details of how matrix algebra operations are implemented in light of the particle conservation symmetry we impose for both implementations. Whatever the explanation, the comparative speedup afforded by \texttt{TeNPy} is dwarfed by the improvements made in Ref.~\cite{rausch2026pushingclassicalfrontier1d} (which appeared after the first version of this work), rendering that work the state-of-the-art classical baseline to compare our quantum hardware results against. All simulations ran on a 64-core/128-thread workstation (AMD Ryzen Threadripper PRO 3995WX, 1 TB RAM), each inside a Docker container configured with a 4-CPU time quota and 4 Julia threads, for reproducibility and dependency isolation.

\begin{figure}[b]
    \centering
    \includegraphics[width=0.75\linewidth]{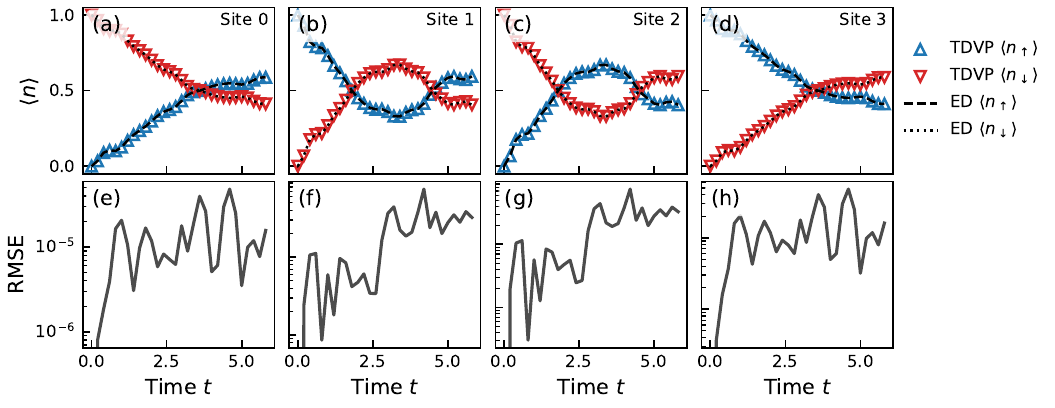}
    \caption{
    Validation of TDVP for small system size ($L=4$).
    (a-d) The occupation number expectation values $\langle n_{i,\uparrow} \rangle$, $\langle n_{i,\downarrow}\rangle$, as computed by exact dynamics (ED) (triangular markers) and TDVP with $\chi = 256$ (dashed and dotted line) for each site and spin, as a function of time. 
    (e-h) The per-site RMSE between the TDVP and ED simulated values.
    }    
    \label{fig:28_ed_tdvp_L4_comparison}
\end{figure}

Separately, to confirm the correctness of our TDVP implementations, we compared against exact diagonalization, as  shown in Fig.~\ref{fig:28_ed_tdvp_L4_comparison} for a small system of $L=4$ sites. The top panels Fig.~\ref{fig:28_ed_tdvp_L4_comparison}(a-d) depict the number operator expectation values $\langle n_{i, \sigma} \rangle$ for each site $i$ and spin $\sigma$. The bottom panels Fig.~\ref{fig:28_ed_tdvp_L4_comparison}(e-h) show the RMSE, which remains small throughout the entire evolution. The failure of the two methods to exactly agree is primarily attributable to Trotter error.

\subsubsection{Simulation improvements}
\begin{figure}[t]
    \centering
    \includegraphics[width=0.75\linewidth]{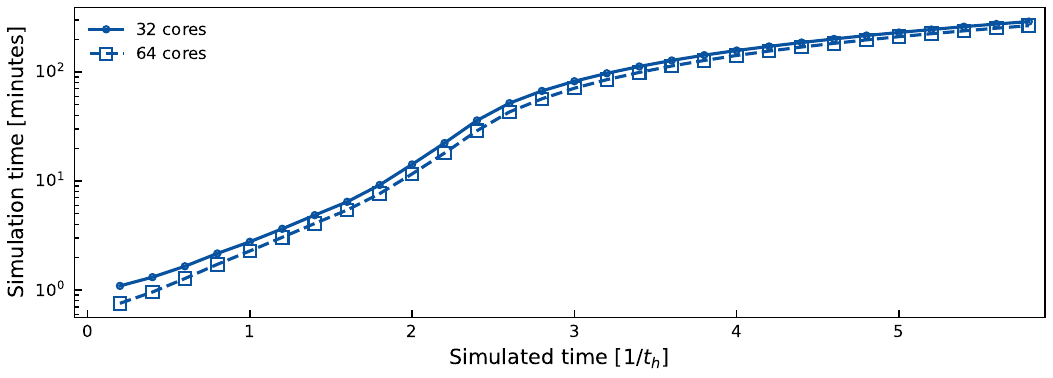}
    \caption{
    Impact of CPU parallelization on overall TDVP runtime for $L=30$ and $\chi=1024$.  There are small differences in how MPS is written to RAM vs Disk in these plots, but this has negligible impact on overall runtime or scaling.
    }    
    \label{fig:corecomparison}
\end{figure}

Lastly, we discuss how the simulation could be made more efficient. Parallelization across multiple CPU cores provides the most straightforward potential route to acceleration, but in practice yields only modest gains~\cite{Secular2020} due to the structure of the TDVP algorithm and the details of the publicly-available implementations we had access to. We validate this directly by expanding the computational cluster used throughout this study to employ 64 cores, demonstrating negligible impact on overall execution time in Fig.~\ref{fig:corecomparison}, with diminishing enhancement at the longest simulation times of interest here.  As described above, TDVP execution requires a loop within which one must perform two matrix exponentiations, SVD, and a large tensor contraction; these steps may be individually optimized but must then be serialized, and then serialized again over multiple sweeps. We empirically confirm these limitations in Table~\ref{SMTab:SVDscaling}, which directly reveals the diminishing impact of paralellization on the SVD step up to 32 cores, indicating little opportunity for further speedups at larger cluster sizes.  Because the SVD implementation in TDVP uses a linear algebra library (BLAS within LAPACK) that scales as $\sim n_{\text{min}}/32$, with $n_{\text{min}}$ the minimum matrix dimension, for the largest matrices treated we only employ 26 of 32 available cores in the SVD step. 

GPU-accelerated tensor network libraries such as cuTensorNet \cite{GaoS2023} report substantial speedups for \emph{dense} tensor contractions at large bond dimension, but these gains do not carry over to the symmetry-adapted TDVP simulations used here. Enforcing the $U(1) \times U(1)$ symmetry of separate particle-number conservation for spin-up and spin-down electrons decomposes each tensor into many small symmetry sectors, so the computation is dominated by numerous small matrix multiplications and decompositions---a workload that utilizes the GPU poorly and yields limited practical speedup. This limitation reflects the current state of publicly-available tensor network software, in which efficient GPU support for block-sparse symmetric tensors and their decompositions remains under development \cite{ITensorGPUDocs2025}. After the first version of this work appeared, Ref.~\cite{rausch2026pushingclassicalfrontier1d} surmounted these limitations with an efficient GPU implementation exploiting the full $U(1) \times SU(2)$ symmetry of our simulations; see the main text for discussion.

\begin{table}[hb]
\centering
\caption{Impact of CPU parallelization on the SVD step of TDVP for $\chi=4096$, corresponding to manipulation of a $1483\times 837$ element matrix}
\label{tab:performance_data}
\begin{tabular}{rccc}
\hline
\textbf{Threads} & \textbf{Time (s)} & \textbf{Speedup} & \textbf{Efficiency} \\ \hline
1  & 0.8877 & 1.00x & 100.0\% \\
2  & 0.6129 & 1.45x & 72.4\%  \\
4  & 0.4481 & 1.98x & 49.5\%  \\
8  & 0.3439 & 2.58x & 32.3\%  \\
12 & 0.3468 & 2.56x & 21.3\%  \\
16 & 0.3349 & 2.65x & 16.6\%  \\
20 & 0.3274 & 2.71x & 13.6\%  \\
24 & 0.3565 & 2.49x & 10.4\%  \\
28 & 0.3448 & 2.57x & 9.2\%   \\
32 & 0.3336 & 2.66x & 8.3\%   \\ \hline
\label{SMTab:SVDscaling}
\end{tabular}
\end{table}

\subsection{Heisenberg-picture simulation algorithms\label{sec:heisenberg}}
We next discuss Heisenberg-picture simulation algorithms. In general, the computational complexity of Heisenberg simulation methods is determined by the complexity of the evolved operator $O(t)$. This in turn is determined by the operator representation employed. For matrix product operator (MPO) evolution, the relevant quantity is the operator-space entanglement~\cite{prosen2007operator, dubail2017entanglement, zhou2020entanglement}; for Pauli path propagation (PPP) and Majorana propagation (MP), $O$ is expanded in a basis and the complexity is given in terms of the spatial support and weight of the basis elements this decomposition. We will consider each method in turn, and then provide benchmarking results for all three methods.

\subsubsection{Pauli path propagation \label{sec:ppp}}

PPP evolves observables in the Heisenberg picture by expanding them in the Pauli basis, $O = \sum_{P \in \mathcal{P}} c_P P$, where $\mathcal{P}$ is the set of all $n_Q$-qubit Pauli operators, and $c_P$ is the coefficient of Pauli $P$. Once expanded, the basis operators appearing in the support of the chosen operator (i.e., those terms with $|c_P|>0$) are propagated through the Trotterized circuit. For a Pauli rotation $U(\theta) = e^{-i\theta\sigma/2}$ generated by a Pauli $\sigma$, each basis element transforms as
\begin{equation}
    U^\dagger(\theta)\, P\, U(\theta) =
    \begin{cases}
        P, & [P, \sigma] = 0, \\[2pt]
        \cos(\theta)\, P + i\sin(\theta)\, \sigma P, & \{P, \sigma\} = 0.
    \end{cases}
    \label{eq:P_rot}
\end{equation}
Each anticommuting rotation thus induces a binary branching, and the number of Pauli strings will in general grow exponentially with circuit depth. An important exception are Clifford rotations ($\theta \in \tfrac{\pi}{2}\mathbb{Z}$), which simply permute the basis. For generic non-Clifford evolution, truncation is required: at each Trotter step, we discard Pauli strings whose weight exceeds a threshold $\mathrm{mw}$ or whose coefficient satisfies $|c_P| < \epsilon$. Thus the two key parameters controlling the strength of the simulation are $\mathrm{mw}$ and $\epsilon$. Increasing $\mathrm{mw}$ enlarges the set of basis elements used to represent $O$, while decreasing $\epsilon$ retains basis elements with smaller coefficients; in both cases the memory and time cost of the algorithm grows accordingly.

In our implementation, we additionally constrain the $XY$-weight $\mathrm{mw}_{xy}$ (the number of single-qubit factors that are $X$ or $Y$) to $\mathrm{mw}_{xy} \leq \mathrm{mw}/2$, motivated by the fact that the final measurement is in the $Z$-basis: any Pauli string retaining nonzero $XY$-weight at the end of the evolution has zero expectation. (As a concrete example, the string $IZZYZXI$ has $\mathrm{mw}=5$ and $\mathrm{mw}_{xy} = 2$.) While off-diagonal terms can in principle rotate back to diagonal under further evolution, we make the assumption that this becomes increasingly improbable as the $XY$-weight grows---an assumption that prunes statistically irrelevant trajectories but is not systematically controlled.

We use the \texttt{PauliPropagation.jl} library~(v0.4.1)~\cite{rudolph2025pauli}, with the Fermi--Hubbard Hamiltonian discretized as first-order Trotter steps of $\Delta t = 0.2$ over $30$ layers ($t \in [0, 6]$). Fermionic operators are mapped to qubits via the Jordan--Wigner (JW) transformation with interleaved ordering. The family of related methods to which PPP belongs---including sparse Pauli dynamics (SPD), Pauli path simulation, and the variants studied in Refs.~\cite{beguvsic2024fast, beguvsic2025real, beguvsic2025simulating, rudolph2025pauli, fontana2025classical, angrisani2025simulating, lin2026utility}---has been applied to systems ranging from the kicked Ising model at 127 qubits to two- and three-dimensional transverse-field Ising dynamics at over 1000 qubits.

\subsubsection{Majorana propagation \label{sec:MP}}

MP is the natural fermionic counterpart of PPP -- rather than expanding $O$ in the Pauli basis, a basis of Majorana fermions is employed instead~\cite{miller2025majorana, danna2025majorana}. This is a somewhat natural basis to use when considering fermionic models, as it avoids the non-local Jordan--Wigner strings that the Pauli representation incurs for fermionic operators in general. In the case of the 1D Fermi--Hubbard model studied here, the JW non-locality is mild across different fermion orderings. For instance, in the block fermion ordering where all up and down spin-orbitals are grouped together, i.e., $\{c_{0,\uparrow}, \ldots, c_{L-1,\uparrow}, c_{0,\downarrow}, \ldots, c_{L-1,\downarrow}\}$, the qubitized Hamiltonian contains only quadratic Pauli terms---see for instance~\cite{Arute2020SpinCharge}. For the local tiling ordering used in our quantum experiments, weight-2 Z-strings are introduced for half of the hopping terms (the so-called long-hopping terms), although we avoid the need to implement these directly through our fermionic SWAP-based circuit construction (see Sec.~\ref{sec:compilation_layout_selection}). Nonetheless, MP is particularly well-suited to the 1D Fermi-Hubbard model because the operator complexity does not increase throughout time evolution in the free model ($U=0$), and as a result MP is exact without truncation in this limit.

For each fermionic mode $(i, \sigma)$ with $i \in \{1, \ldots, L\}$ and $\sigma \in \{\uparrow, \downarrow\}$, we define two Majorana operators
\begin{equation}
    \gamma_{i,\sigma} = c_{i,\sigma} + c^\dagger_{i,\sigma}, \qquad
    \gamma'_{i,\sigma} = i\bigl(c^\dagger_{i,\sigma} - c_{i,\sigma}\bigr),
    \label{eq:majorana_def}
\end{equation}
satisfying the relations
\begin{equation}
    \gamma_{i,\sigma}^\dagger = \gamma_{i,\sigma} \,, \qquad
    \gamma_{i,\sigma}^2 = \mathbb{I} \,, \qquad 
    \{\gamma_{i,\sigma}, \gamma_{j,\tau}\} = \{\gamma'_{i,\sigma}, \gamma'_{j,\tau}\} = 2\delta_{ij} \delta_{\sigma \tau} \mathbb{I} \,, \qquad
    \{\gamma_{i,\sigma}, \gamma'_{j,\tau}\} = 0 \,.
\end{equation}
The $2L$ fermionic modes thus yield $4L$ Majorana operators, naturally split into ``unprimed'' and ``primed'' families. The Majorana basis is the set of all operators $\{\mu_S\}_S$, where each $S = \{(i_1,\sigma_1,p_1),\ldots,(i_w,\sigma_w,p_w) \}$ is a set of $w$ tuples, each specifying a mode $(i,\sigma)$ together with a flag $p$ indicating an unprimed ($p=0$) or primed ($p=1$) Majorana. Fixing a canonical ordering of the entries removes the redundancy from reordering. The basis element associated with $S$ is
\begin{equation}
    \mu_S := i^{w(w-1)/2}\, \gamma_{i_1,\sigma_1}^{(p_1)}\,\gamma_{i_2,\sigma_2}^{(p_2)}\cdots\gamma_{i_w,\sigma_w}^{(p_w)},
    \label{eq:mu_def}
\end{equation}
where $\gamma_{i,\sigma}^{(0)} \equiv \gamma_{i,\sigma}$ and $\gamma_{i,\sigma}^{(1)} \equiv \gamma'_{i,\sigma}$, and $w = |S|$ is the weight of the string. The phase $i^{w(w-1)/2}$ renders each $\mu_S$ Hermitian.

The Fermi--Hubbard Hamiltonian admits a compact representation in this basis. With $n_{i,\sigma} = \tfrac{1}{2}(\mathbb{I} + i\gamma_{i,\sigma}\gamma'_{i,\sigma})$, the on-site interaction becomes a sum of weight-two and weight-four strings on the four Majoranas of site $i$:
\begin{equation}
    \label{eq:hubbard_interaction_majorana}
    U n_{i,\uparrow} n_{i,\downarrow} = \tfrac{U}{4}\bigl(\mathbb{I} + i\gamma_{i,\uparrow}\gamma'_{i,\uparrow} + i\gamma_{i,\downarrow}\gamma'_{i,\downarrow} - \gamma_{i,\uparrow}\gamma'_{i,\uparrow}\gamma_{i,\downarrow}\gamma'_{i,\downarrow}\bigr) \,,
\end{equation}
while the hopping kinetic term becomes a sum of two weight-two strings on the four Majoranas of the same-spin nearest-neighbor pair:
\begin{equation}
    \label{eq:hubbard_hopping_majorana}
    -t_h\bigl(c^\dagger_{i,\sigma} c_{i+1,\sigma} + \mathrm{h.c.}\bigr) = -\tfrac{t_h}{2}\bigl(i\gamma_{i,\sigma}\gamma'_{i+1,\sigma} - i\gamma'_{i,\sigma}\gamma_{i+1,\sigma}\bigr) \,.
\end{equation}
The full Hamiltonian is therefore a sum of geometrically local Majorana strings of weight at most four, independent of $L$.

Trotter evolution is implemented as a sequence of Majorana rotations $V_S(\theta) = e^{-i\theta\mu_S/2}$, with $\theta \propto t_h\Delta t$ for hopping bilinears and $\theta \propto U\Delta t$ for interactions. Each rotation acts on an arbitrary basis element $\mu_{S'}$ via
\begin{equation}
    V_S^\dagger(\theta)\, \mu_{S'}\, V_S(\theta) =
    \begin{cases}
        \mu_{S'}, & [\mu_{S'}, \mu_S] = 0, \\[2pt]
        \cos(\theta)\, \mu_{S'} + i\sin(\theta)\, \mu_S \mu_{S'}, & \{\mu_{S'}, \mu_S\} = 0,
    \end{cases}
    \label{eq:M_rot}
\end{equation}
where $\mu_S \mu_{S'}$ is, up to an overall phase $\pm 1, \pm i$, the basis operator $\mu_{S \triangle S'}$ indexed by the symmetric difference $S \triangle S'$ of $S$ and $S'$. Equation~\eqref{eq:M_rot} is the direct fermionic analogue of Eq.~\eqref{eq:P_rot}.

The exactness of this method for $U=0$ is a consequence of the fact that hopping bilinears generate weight-preserving rotations. As a result, MP is \emph{exact} when the maximum weight of all retained operators, mw, is greater than or equal to the weight of the initial observable. For $U \neq 0$, the on-site interaction shifts weight by up to $\pm 4$, so generic Hubbard dynamics require mw$ > 2$ and will feature a growth of weights until the limit is reached. In our simulations, we truncate at each Trotter step by retaining only basis operators with $|S| \leq \mathrm{mw}$ and $|c_S| \geq \epsilon$. In our benchmarking simulations, we made use of the \texttt{MajoranaPropagation.jl} Julia package \cite{danna2025majorana}.

\subsubsection{Matrix product operator evolution \label{sec:MPO}}

MPO evolution is the operator-space counterpart of the state-based tensor-network method of Sec.~\ref{sec:tdvp}: rather than compressing the wavefunction with an MPS ansatz that caps the state entanglement, it represents the evolved Heisenberg observable as an MPO, compressing it by exploiting limited operator-space entanglement~\cite{prosen2007operator, dubail2017entanglement}. The evolved operator is represented as
\begin{equation}
    O = \sum_{\{s_k\},\{s'_k\}} \mathrm{Tr}\!\left[ M^{[1]}_{s_1 s'_1} \cdots M^{[L]}_{s_L s'_L} \right]\,
    \ket{s_1 \cdots s_L}\bra{s'_1 \cdots s'_L},
    \label{eq:MPO_def}
\end{equation}
with bond dimension $\chi = \max_k \chi_k$ controlling expressive power, and storage/operation cost scaling as $\mathcal{O}(L d^2 \chi^2)$ and $\mathcal{O}(L d^2 \chi^3)$ respectively.

We propagate $O$ in the Heisenberg picture using a second-order Suzuki--Trotter (Strang) splitting built from the Schr\"odinger-picture gates $g_j = e^{-i\tfrac{\Delta t}{2} h_j}$, where $h_j$ is the two-site Fermi--Hubbard term on bond $(j, j{+}1)$ -- that is, the time-evolution operator consists of a forward sweep followed by a reverse sweep:
\begin{equation}
    U(\Delta t) \approx \left( \prod_{j=1}^{L-1} g_j \right) \left( \prod_{j=L-1}^{1} g_j \right) \,.
\end{equation}
The Heisenberg update conjugates $O$ by this sequence, $O \to g_j^\dagger\, O\, g_j$ applied in the same forward-then-reverse order, with each two-site gate increasing the bond dimension across its cut by a factor of at most $d^2 = 16$. After each gate application, the MPO is locally recompressed by SVD. The smallest singular values $s_i$ at each bond are discarded up to a cumulative weight $\sum_{i \in \mathrm{discarded}} s_i^2 \le \varepsilon$ (with the convention that the singular values are normalized), with $\varepsilon = 10^{-8}$, and the bond dimension is capped at $\chi_{\max}$.

\subsubsection{Benchmarking and results \label{sec:heisenberg_results}}

\begin{figure}[t]
    \centering
    \includegraphics[width=0.75\linewidth]{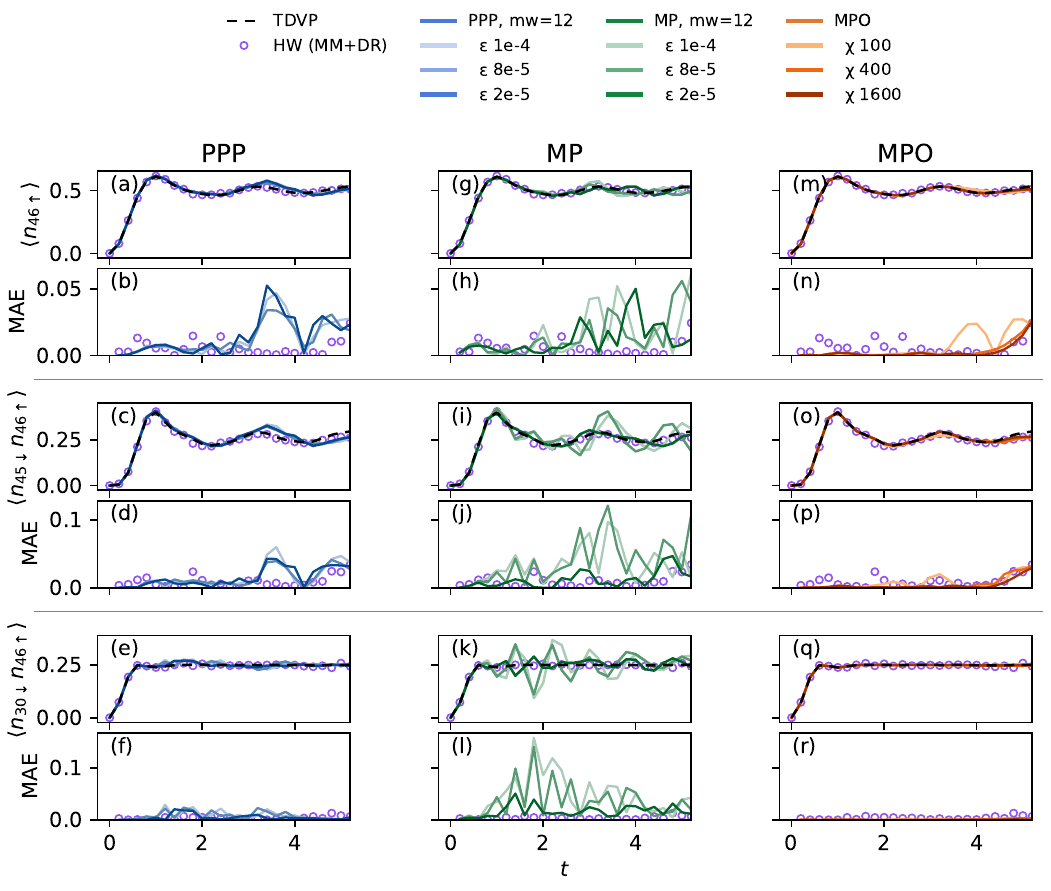}
    \caption{
    Time-series and mean absolute error (MAE) for Heisenberg, TDVP, and hardware simulation methods for select one- and two-point density correlations. Data correspond to an $L=60$ system with $U/t_h=-2$ and a N\'eel initial state. Shown is a representative subset of runs: for PPP and MP, those with the largest maximum weight; for MPO, those with bond dimension $\chi=100$, $400$, and $1600$. The time-series plots include the quantum hardware result with measurement mitigation and decay recovery applied, as well as the TDVP result with bond dimension $\chi = 4096$ as a reference (the same data shown in Fig.~\ref{fig:TDVP_summary_plot} of the main text). The MAE is computed relative to the TDVP simulation. Within each panel, lines are colored by simulation strength, controlled by the cutoff $\epsilon$ for PPP and MP and by the bond dimension $\chi$ for MPO. Data is shown for $t \le 5.2\,t_h^{-1}$, the maximum time for which the TDVP simulation is accurate.
    }    
    \label{fig:heisenberg_timeseries_mae}
\end{figure}

\begin{figure}[t]
    \centering
    \includegraphics[width=0.75\linewidth]{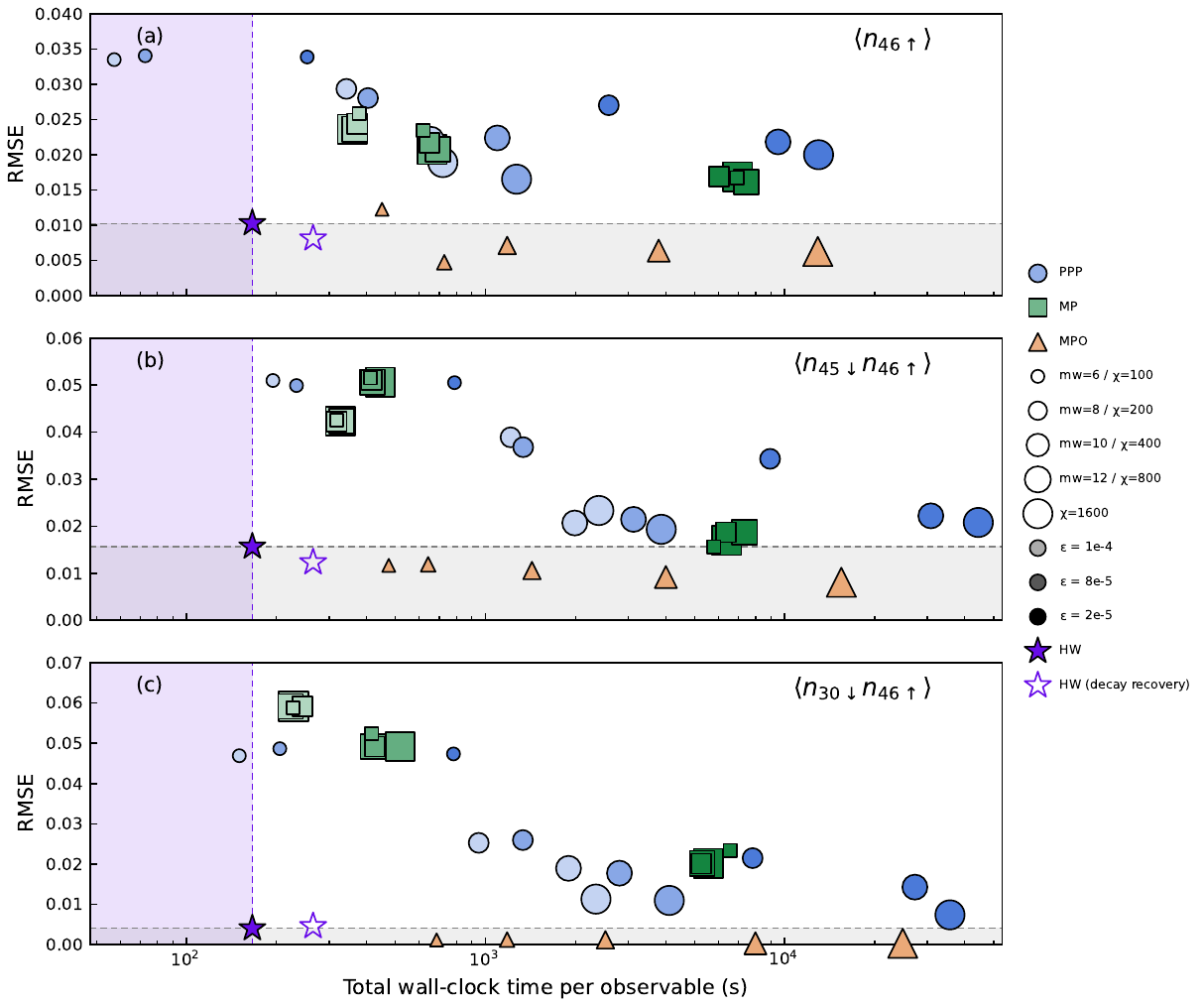}
    \caption{
    Accuracy--cost tradeoff for Heisenberg-picture methods.
    (a)~$\langle n_{46,\uparrow}\rangle$;
    (b)~two-point correlator $\langle n_{45,\downarrow}\, n_{46,\uparrow}\rangle$;
    (c)~two-point correlator $\langle n_{30,\downarrow}\, n_{46,\uparrow}\rangle$.
    Each panel shows the root-mean-square error (RMSE), averaged over all time steps, versus total wall-clock time for PPP, MP, MPO, and the quantum hardware execution, for a system with $L = 60$ sites, $U/t_h = -2$, and a N\'eel initial state. RMSE is computed relative to the reference TDVP simulation at $\chi = 4096$ (Fig.~\ref{fig:TDVP_summary_plot}). For PPP and MP, markers are colored by cutoff $\epsilon$ and sized by the corresponding algorithm parameter---the maximum string weight $mw$, or the bond dimension $\chi$ for MPO. For PPP, $mw$ refers to the full Pauli string, with a threshold of $mw/2$ applied to the XY-weight as elsewhere in this work. Wall-clock time is the cost of computing the stated observable only. Two quantum hardware results are shown, with and without decay recovery and measurement mitigation. Dashed lines and shading correspond to the raw hardware result for reference.
    }
    \label{fig:heisenberg_runtime_vs_error_rmse}
\end{figure}

We now present benchmarking results demonstrating performance and runtime for the three Heisenberg methods considered above. We target a single reference simulation---$L=60$ sites, N\'eel initial state, $U/t_h=-2$---identical to that shown in Fig.~\ref{fig:TDVP_summary_plot} of the main text, and three observables: one one-point, $\langle n_{46,\uparrow}\rangle$, and two two-point, $\langle n_{45,\downarrow}\, n_{46,\uparrow}\rangle$ and $\langle n_{30,\downarrow}\, n_{46,\uparrow}\rangle$. For each method, we scan the parameters controlling simulation strength: the maximum string weight $\mathrm{mw}$ and cutoff $\epsilon$ for the string-based methods (PPP, MP), and the bond dimension $\chi$ for the MPO method. All simulations are run on a 64-core/128-thread workstation (AMD Ryzen Threadripper PRO 3995WX, 1 TB RAM), each inside a Docker container configured with a 4-CPU time quota and 4 Julia threads, for reproducibility and dependency isolation.

Figure~\ref{fig:heisenberg_timeseries_mae} shows time-series data for a representative subset of runs, each compared against the baseline (validated) TDVP with maximum bond dimension $\chi = 4096$ (the same data shown in Fig.~\ref{fig:TDVP_summary_plot}). All Heisenberg-picture methods broadly agree with both the TDVP and hardware results. The MPO method is the most accurate, with the discrepancy only becoming apparent towards the end of the simulation. PPP and MP are less accurate, with MP in particular exhibiting appreciable discrepancies at intermediate times. Moreover, PPP and MP fail to exhibit a monotonic convergence as the cutoff $\epsilon$ is decreased; the pointwise most accurate simulation at a given time step is not necessarily the one with the tightest cutoff. This is most pronounced for the MP method -- for instance, the best MP method oscillates between the three cutoff values $\epsilon = 2\times10^{-5}$, $8\times10^{-5}$, and $10^{-4}$ in Fig.~\ref{fig:heisenberg_timeseries_mae}(h). This lack of monotonic convergence is consistent with previous observations for string-based Heisenberg-picture methods in Refs.~\cite{beguvsic2024fast, beguvsic2025real, rudolph2025pauli} and stands in contrast to the monotonic convergence typical of Schr\"odinger-picture methods such as the TDVP simulations used here.

Next, Fig.~\ref{fig:heisenberg_runtime_vs_error_rmse} gives a broad overview of our benchmarking results. The figures of merit for simulation quality are the RMSE, computed for each observable across all time steps against our reference TDVP simulation, and the computational wall-clock runtime. Note that this RMSE measure differs from the measure in Fig.~\ref{fig:TDVP_summary_plot}, for which the RMSE is calculated across all one-point expectation values for each time-step separately, whereas here the average is separately performed across all time-steps for each observable. For all methods, the wall-clock runtime grows with $\chi$, $1/\epsilon$, or $\mathrm{mw}$, as expected. For MP, points cluster by $\mathrm{mw}$, indicating that $\epsilon$, rather than $\mathrm{mw}$, is the dominant accuracy-controlling parameter in that case. Consistent with the earlier plot, the MPO method achieves substantially lower error than the string-based PPP and MP methods across all observables --in fact, the least accurate MPO simulation is still more accurate than the best PPP or MP simulation in all cases considered. 

We also compare against our quantum hardware results. Runtime comparisons across method classes---Schr\"odinger-picture, Heisenberg-picture, and quantum-circuit---require care, because the classes produce fundamentally different outputs. Each Heisenberg simulation yields a single expectation value, whereas a TDVP run produces an MPS from which any operator expectation value can be computed, and a quantum execution yields a bitstring distribution from which any of the $2^{N_Q}$ $Z$-basis expectation values can be estimated. Further complicating matters is the fact that not all such values are equally accurate in either case. Still, we note that restricting to Fock-basis density operators, the quantum execution gives access to all $2L = 120$ one-point operators $n_{i,\sigma}$ and all $\binom{2L}{2} = 7140$ two-point operators $n_{i,\sigma} n_{j,\sigma'}$ at once. A final point to note here is that the Heisenberg methods are all embarrassingly parallel, so the time to compute a given set of expectation values scales inversely with available compute.

With few exceptions, every individual-observable simulation takes longer than the corresponding quantum-circuit execution (2\,min 46\,s for the raw output, 4\,min 25\,s when the time for the readout error mitigation and decay recovery circuit executions are included, see Table~\ref{tab:gate_complexity}) -- and those exceptions are far less accurate than the quantum hardware results. MPO is the only method that meets or exceeds the accuracy of our quantum hardware results across all observables, though at the cost of significantly increased wall-clock time. The runtime of each Heisenberg method is sensitive to both the target observable and the simulation strength, making the quality of single-instance one-point correlator comparisons across method classes particularly susceptible to misinterpretation. In the case of MPO, the runtime ranges from 8 minutes ($\chi = 100$, one-point observable) to nearly 7 hours ($\chi = 1600$, two-point observable). A rough estimate for the total time required to compute all one- and two-point operators is $120\times$ the time required to compute ${\langle n_{46, \uparrow}}\rangle$ plus $7140 \times$ the average time to compute the two-point expectation values ${\langle n_{45,\downarrow} n_{46,\uparrow}\rangle}$ and ${\langle n_{30,\downarrow} n_{46,\uparrow}\rangle}$. This works out to be $\sim 51$, $170$, and $1700$ \textit{days} for $\chi = 100$, $400$, $1600$, respectively. This estimate is quite simple in that it assumes all one-point operators to take the same amount of time to simulate, which is certainly not the case, and it uses just two representative two-point operators to estimate the time required to compute all 7140 such terms. This estimate should therefore be understood as setting a rough order-of-magnitude estimate. And lastly, we reiterate the key point above that Heisenberg methods are embarrassingly parallel across observables: this parallelism reduces the total elapsed (wall-clock) time but leaves the total compute cost (CPU-hours) unchanged.

\section{Error analysis and validation \label{sec:validation}}
This section provides a characterization of the primary error sources inherent in our digital quantum simulations, specifically algorithmic Trotter error and temporal device drift. Before detailing these contributions, we briefly establish the treatment of statistical shot noise.

Consider a $Z$-basis Pauli operator $P_Z$. After $N_{\mathrm{shot}}$ shots, the sample proportions of the eigenvalue outcomes ($\pm 1$) yield the estimated probabilities ${\hat{p}_+ = N_+ / N_{\mathrm{shot}}}$ and ${\hat{p}_- = 1 - \hat{p}_+}$, where the hat denotes a statistical estimate. The binomial variance of the estimated expectation value $\widehat{\langle P_Z \rangle}$ is thus given by
\begin{equation}
    \mathrm{Var}(\widehat{\langle P_Z \rangle})=\frac{4\hat{p}_+(1-\hat{p}_+)}{N_{\mathrm{shot}}}=\frac{1-\widehat{\langle P_Z \rangle}^2}{N_{\mathrm{shot}}} \,.
\end{equation}
These errors are then propagated to other reported quantities, such as the orbital occupations and the spin-spin correlation function.

\subsection{Trotter error and step-size selection}
\label{sec:trotter_step_selection}
The two-qubit depth of the Trotterized time evolution circuit scales with the number of Trotter steps $n_{\mathrm{step}}$ as $D_{2\mathrm{Q}}=7+5(n_{\mathrm{step}}-1)$. Consequently, the two-qubit gate fidelities of contemporary hardware limit our simulations to a maximum of roughly 30 steps, or a two-qubit depth of 152, to ensure the physical signal remains distinct from hardware noise. With this limitation on the number of steps, the next consideration is the size of the time increment, $\Delta t$. This choice involves a fundamental trade-off: larger values of $\Delta t$ allow us to simulate longer total evolution times $T = n_{\mathrm{step}} \, \Delta t$, but at the cost of increased Trotter error. We choose the time step size $\Delta t$ to achieve a favorable balance: the circuit gives a total evolution time that is long enough to capture the relevant dynamical features, while keeping the Trotter error, which scales as $\mathcal{O}(\Delta t^{2})$ for odd steps and $\mathcal{O}(\Delta t^3)$ for even steps, reasonably small at each time step. The appropriate $\Delta t$ value depends on the strength of the onsite coupling $U$---stronger couplings correspond to larger commutator error terms in the Trotter approximation, which require smaller $\Delta t$ to compensate. 

\begin{figure}[t]
    \centering
    \includegraphics[width=0.75\linewidth]{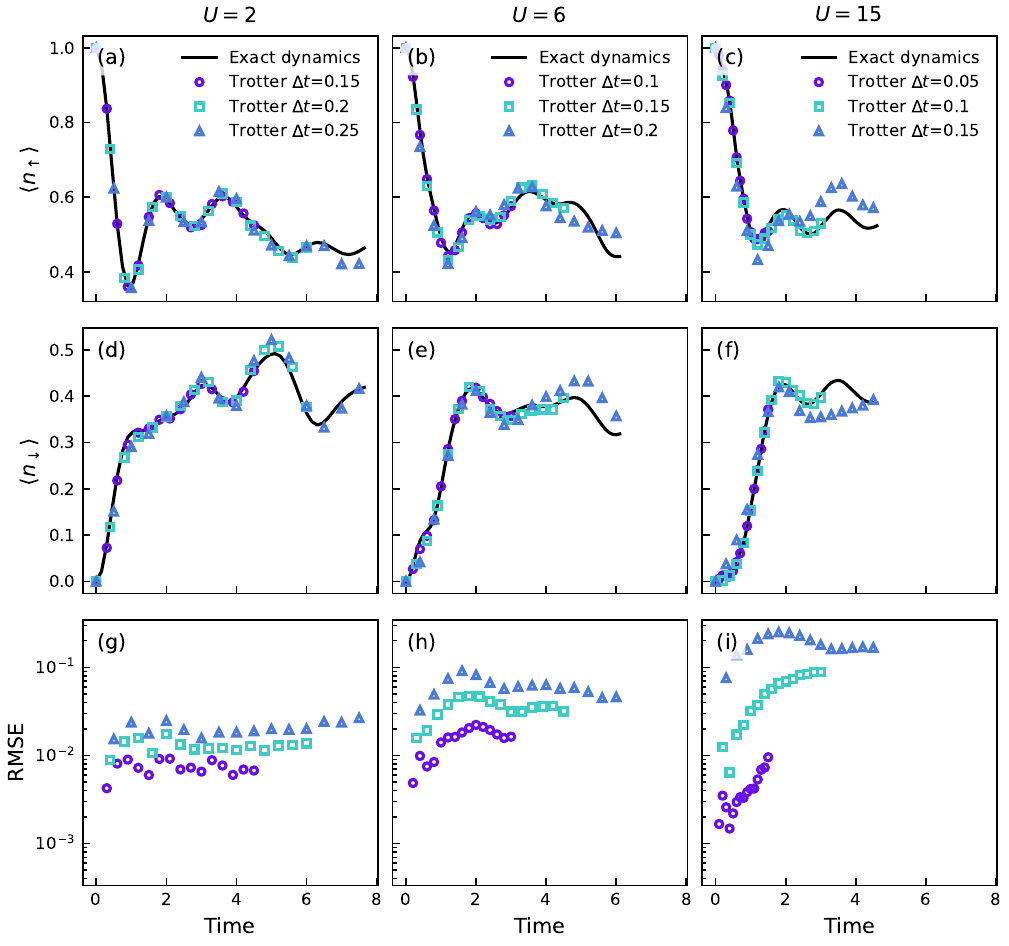}
    \caption{First-order Trotter error for the 1D Fermi--Hubbard model on $L=10$ sites with open boundaries, hopping $t=1$ and chemical potential $\mu=0$. Each column corresponds to a different on-site interaction strength: $U=2$ (left), $U=6$ (middle), and $U=15$ (right). The initial state is N\'{e}el-like $\ket{\uparrow\downarrow\uparrow\downarrow\cdot\downarrow\uparrow\downarrow\uparrow\downarrow}$ with a vacancy in the middle site ($\cdot$). All Trotter simulations use $30$ steps at varying step sizes $\Delta t$, giving different total evolution times $T = 30\,\Delta t$. The top and center rows show the spin-up ($\langle n_\uparrow \rangle$) and spin-down ($\langle n_\downarrow \rangle$) occupation at the middle site (site~5), comparing the exact dynamics (solid black line) with Trotter results (open markers) for each $\Delta t$. The bottom row shows the root-mean-square error (RMSE) $\sqrt{\sum_{L}\mathrm{SE}(t)/L}$, where $\mathrm{SE}(t) = \bigl(\langle n_\uparrow\rangle_{\mathrm{exact}}
    - \langle n_\uparrow\rangle_{\mathrm{Trotter}}\bigr)^{2}
    + \bigl(\langle n_\downarrow\rangle_{\mathrm{exact}}
    - \langle n_\downarrow\rangle_{\mathrm{Trotter}}\bigr)^{2}$. }
    \label{fig:fh1d_trotter_error}
\end{figure}

To determine the appropriate $\Delta t$ for each interaction strength $U$, we evaluate the Trotter error on small systems. Specifically, we classically simulate the Trotterized circuit at three step sizes—our selected $\Delta t$ alongside $\Delta t \pm 0.05$—and compare the resulting site occupations against the continuous-time dynamics obtained via exact diagonalization of the Hamiltonian (Fig.~\ref{fig:fh1d_trotter_error}).
We find that using $\Delta t + 0.05$ increases the Trotter error by about one order of magnitude, while using $\Delta t - 0.05$ significantly shortens the total evolution time. Although this analysis is performed for a small system size, we expect the same choice of $\Delta t$ to provide a similarly favorable balance at larger system sizes.

\begin{figure}[t]
    \centering
    \includegraphics[width=0.75\linewidth]{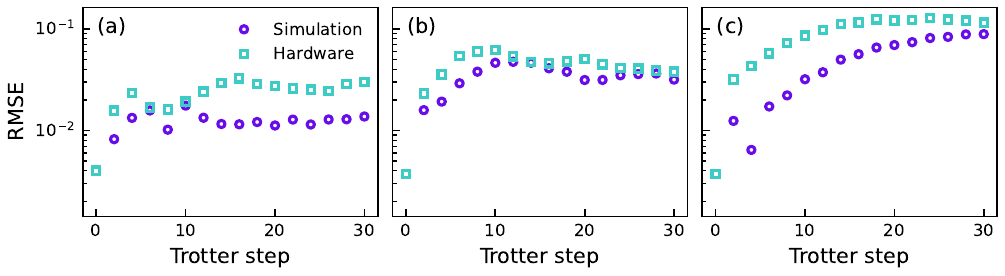}
    \caption{Same parameter setting and initial state as Fig.~\ref{fig:fh1d_trotter_error}. Each panel corresponds to a different on-site interaction strength and its corresponding Trotter step size: (a) $U=2,\, \Delta t=0.2$, (b) $U=6,\, \Delta t=0.15$ , and (c) $U=15,\, \Delta t=0.1$. We show the root-mean-square error (RMSE) $\sqrt{\sum_{L}\mathrm{SE}(t)/L}$, where $\mathrm{SE}(t) = \bigl(\langle n_\uparrow\rangle_{\mathrm{exact}}
    - \langle n_\uparrow\rangle_{\mathrm{target}}\bigr)^{2}
    + \bigl(\langle n_\downarrow\rangle_{\mathrm{exact}}
    - \langle n_\downarrow\rangle_{\mathrm{target}}\bigr)^{2}$, of two different targets: noiseless simulation and hardware execution. Error bars are smaller than the markers and have been omitted for clarity.}
    \label{fig:trotter_hardware_comparison}
\end{figure}

Furthermore, in Fig.~\ref{fig:trotter_hardware_comparison}, we compare the overall hardware error with the Trotter error for a $10$-site example. The hardware error, measured against the exact dynamics, contains contributions from Trotterization, sampling error, and hardware noise. Here, the sampling error---which is much smaller with $20{,}000$ shots---is already included in the Trotter simulation, and therefore forms only a small part of the Trotter error that can be neglected. From Fig.~\ref{fig:trotter_hardware_comparison}, by inspecting the ratio between the RMSE of noiseless simulation and that of hardware, we can see that the Trotter error can account for up to about $60\%$ of the overall hardware error in this example. At this small system size where exact dynamics is possible, we observe that the hardware error is comparable to the Trotter error; however, we expect hardware errors to be dominant at much larger system sizes.

\subsection{Assessment of hardware stability and temporal drift}
\begin{figure}
    \centering
\includegraphics[width=0.75\linewidth]{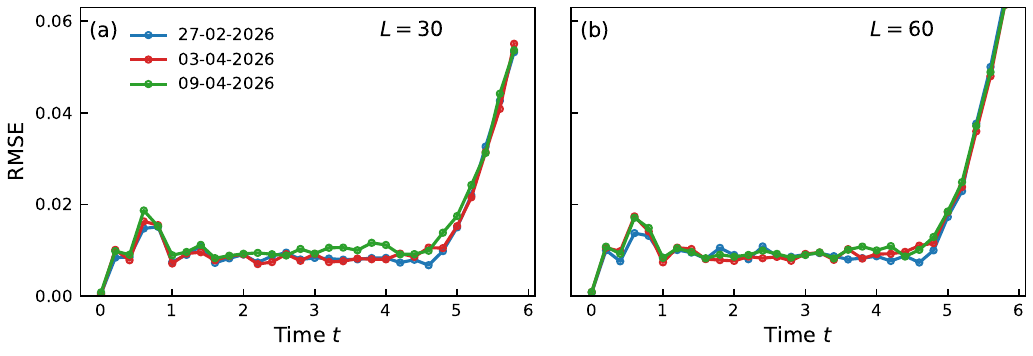}
    \caption{The root-mean-square error (RMSE) of the fermionic occupations $\langle n_{i\sigma} \rangle$ as a function of the evolution time is shown for the Néel initial state at $U=-2$ for (a) $L=30$ and (b) $L=60$ sites. For each case, the RMSE is computed across all sites and spins relative to classical TDVP benchmarks (max bond dimension $\chi=2048$). Each panel displays three independent datasets (colored lines) collected over a six-week period, illustrating the consistency of the results over time. All data was post-processed using the measurement mitigation and decay recovery protocols described in the text. Error bars are smaller than the markers and have been omitted for clarity.}
    \label{fig:device_drift_rmse}
\end{figure}

Superconducting quantum processors are susceptible to temporal drift and fluctuations, necessitating frequent recalibration by dedicated hardware teams~\cite{Klimov2018, Burnett2019, Carroll2022}. 
Our experiments were conducted via the standard IBM Quantum API \cite{qiskit} on a publicly available device, without performing custom calibrations immediately prior to execution.
Accessing the hardware at various intervals relative to the manufacturer's calibration cycle introduces temporal variability that might be absent in a dedicated hardware demonstration that includes full device calibration.
To quantify this variability and its impact on our results, we performed a temporal drift study by collecting multiple datasets for the same model parameters using the \texttt{ibm\char`_boston} device over several weeks. Crucially, our data acquisition was independent of the hardware’s calibration schedule. For each execution, we optimized the qubit layout based on the quality of the device at the time of data collection, following the protocol detailed in Sec.~\ref{sec:compilation_layout_selection}.

Figure~\ref{fig:device_drift_rmse} shows the root-mean-square error (RMSE) of the fermionic occupations $\langle n_{i\sigma}\rangle$ as a function of time for three distinct parameter sets, with each dataset collected at three separate time points over a six-week period. The RMSE is calculated across all sites and spins relative to classical TDVP benchmarks (max bond dimension $\chi=2048$). To maintain consistency, this data has undergone standard post-processing techniques---the measurement mitigation and decay recovery---we have used throughout our study. 
The stability of the RMSE values across these disparate time points, characterized by fluctuations of $\sim 2.7\%$ for $L=30$ [Fig.~\ref{fig:device_drift_rmse}(a)] and $\sim 0.8\%$ for $L=60$ [Fig.~\ref{fig:device_drift_rmse}(b)], demonstrates that hardware fluctuations do not qualitatively alter the observed dynamics. This consistency confirms the robustness of our experimental findings against temporal device drift.

\section{Post-processing \label{app:postprocessing}}

Quantum devices suffer from systematic control imperfections, readout infidelities, and unavoidable environment-induced decoherence, and deep quantum circuits naturally dampen toward a mixed state at the end of hardware execution. 
In addition to the hardware- and circuit-level error suppression and noise tailoring techniques (see Section~\ref{sec:error_suppression}), some other classes of error reduction techniques have been devised to mitigate this computational bottleneck---these include error mitigation techniques~\cite{Huggins_2021, van_den_Berg_2023, Kim2023, Liao2023} that mitigate errors on the expectation values by statistically averaging out some of the noise effect (e.g., readout error mitigation~\cite{Maciejewski2020, Bravyi2021, nation2021scalable, Mundada2023}, zero-noise extrapolation~\cite{Temme2017}, probabilistic error cancellation~\cite{van_den_Berg_2023}, and damping reversal~\cite{Arute2020SpinCharge, alam2026onsetergodicityscalesdigital, google2025observation}), and error detection techniques that post-select measurement shots based on violated symmetries~\cite{Liao2025, hartnett2026simulatingdynamicssu2matrix}; we employed some of the above-mentioned techniques or their variants through post-processing in our experiments, which we detail below. Altogether, they help extend the effective depth and coherence of the hardware execution. 

\subsection{Readout error mitigation}

\begin{figure}
    \centering
    \includegraphics[width=0.75\linewidth]{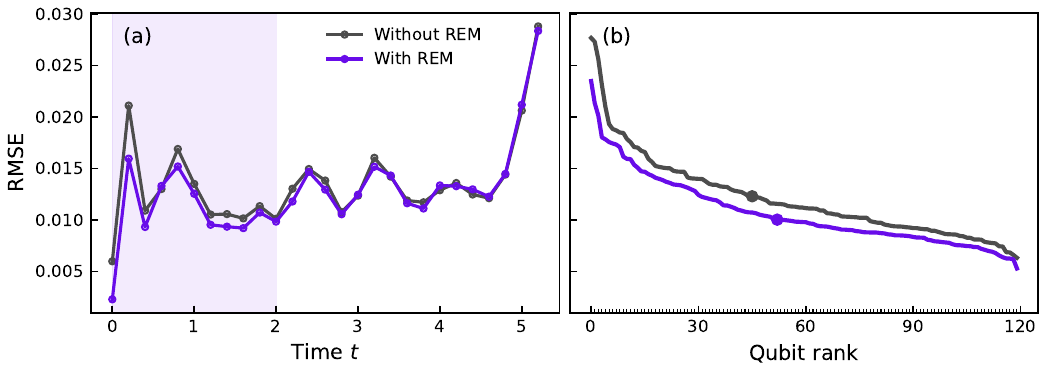}
    \caption{
    Impact of readout error mitigation (REM). Data is shown for the $U=-2$, $L=60$ system prepared in a Néel initial state with a central vacancy. The RMSE of the fermionic occupations $\langle n_{i\sigma} \rangle$ is calculated relative to classical TDVP benchmarks (maximum bond dimension $\chi=2048$) and presented as a function of evolution time (a) and qubit rank (b). All experiments were performed with deterministic error suppression and Pauli twirling. No additional post-processing or error mitigation techniques were applied.
    (a) The RMSE is computed across all qubits for each time step.
    (b) The RMSE is computed over the $t \in [0, 2]$ interval (corresponding to the purple shaded region in (a)) for each qubit, with both datasets (raw and REM-corrected) separately sorted by magnitude. To emphasize that the sorting is independent for each dataset, a single representative qubit (index 30) is marked with a distinct marker on both the raw and REM-corrected lines. Error bars are smaller than the markers and have been omitted for clarity.}
    \label{fig:meas_mit_rmse}
\end{figure}

To mitigate the impact of readout noise, we employ a simple readout error mitigation (REM) protocol \cite{Maciejewski2020, Bravyi2021, nation2021scalable, Mundada2023}. Immediately following the execution of the Trotter time evolution circuits, we execute characterization circuits to quantify the readout fidelity of each qubit. These circuits consist of preparing every qubit in either the $|0\rangle$ or $|1\rangle$ state, followed by measurement in the computational basis. For each experiment, we collect $32{,}768$ readout characterization shots ($16{,}384$ for each $|0\rangle$ and $|1\rangle$ initial state), which allow a set of independent single-qubit confusion (assignment) matrices to be constructed:
\begin{equation}
    C_i=
    \begin{bmatrix}
        p(0|0)_i & p(0|1)_i \\
        p(1|0)_i & p(1|1)_i
    \end{bmatrix}\,, \quad i \in \{0, \dots, 2L-1\}\, .
\end{equation}
Here, $p(\beta|\alpha)_i$ denotes the probability of measuring the outcome $\beta \in \{0,1\}$ given the preparation of state $|\alpha\rangle \in \{0,1\}$ on qubit $i$, such that $p(0|0)_i + p(0|1)_i = p(1|0)_i + p(1|1)_i = 1$.
We assume the full confusion matrix to be given by a tensor product of the single-qubit matrices, $C = \otimes_{i=0}^{2L-1} C_i$, thus omitting inter-qubit readout correlations, which we find to be negligible on the \texttt{ibm\char`_boston} processor.
In our experiments, typical error probabilities $p(1|0)_i$ and $p(0|1)_i$ range from approximately $0.06\%$ to $9.7\%$, with a median value of $0.4\%$.
By strategically selecting the circuit layout on the device (see Sec.~\ref{sec:layout}), we avoid qubits with exceptionally high readout errors, which in our characterization experiments reached values as high as $\sim 38\%$.

The raw data from the main circuit execution consists of a measured bitstring distribution.
Applying a global REM correction to this distribution is computationally prohibitive as the Hilbert space (and thus the number of produced bitstrings after correction) scales exponentially, i.e. the REM procedure explores the full Hilbert space.
However, since our goal is to estimate low-weight Pauli-$Z$ observables, $\langle P_Z \rangle$, we circumvent this exponential cost via marginalization.
For a given observable supported on a subset of qubits, we first marginalize the full output distribution down to that specific support.
We then apply the inverse confusion matrices to this reduced distribution to obtain a mitigated marginal distribution, from which the corrected expectation value $\langle \tilde{P}_Z \rangle$ is computed.

Figure~\ref{fig:meas_mit_rmse} quantifies the impact of the REM protocol on the observed fermionic occupations $\langle n_{i\sigma} \rangle$.
In Figure~\ref{fig:meas_mit_rmse}(a), we plot the RMSE relative to classical TDVP benchmarks as a function of the evolution time.
While REM consistently reduces the RMSE, the mitigation gain is most significant during the early stages of the evolution.
At later time steps, gate errors become the dominant error source, eventually overshadowing the improvements gained from correcting readout fidelity.

To further resolve the effect during the earlier time steps, Fig.~\ref{fig:meas_mit_rmse}(b) displays the RMSE computed for each qubit over the initial time interval $t \in [0, 2]$.
Both the raw and REM-corrected data are independently sorted by magnitude to illustrate the global improvement in fidelity across the entire register.
Quantitatively, the REM protocol yields a $4.6$\% reduction in the total RMSE across all time steps, and a more substantial $12.1$\% reduction within the $t \in [0, 2]$ window.
This relatively modest overall improvement reflects the high baseline readout fidelity of the \texttt{ibm\char`_boston} processor, as previously indicated by the reported values of the error probabilities within the single-qubit confusion matrices.

\subsection{Decay recovery}

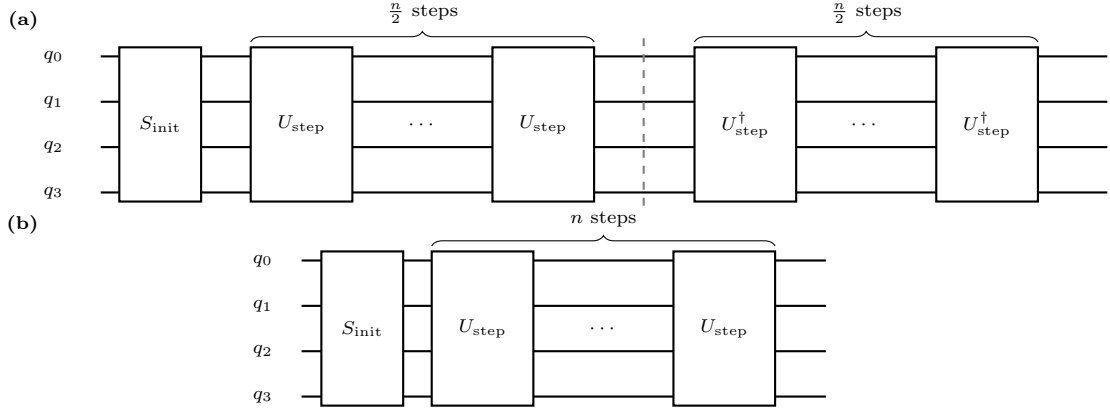
\begin{figure}[t]
\centering
\begin{tikzpicture}[x=1.64cm,y=0.6cm]

\tikzset{
  wire/.style={thick},
  qlabel/.style={anchor=east, font=\scriptsize},
  brace/.style={decorate, decoration={brace, amplitude=4pt, raise=2pt}},
}

\def\bh{0.2}
\def\bwi{0.33}
\def\bws{0.41}

\newcommand{\block}[5]{%
  \draw[thick, fill=white] (#1-#4,#2+\bh) rectangle (#1+#4,#3-\bh);
  \node at (#1,{0.5*(#2+#3)}) {\scriptsize #5};
}

\newcommand{\qlabels}[2]{%
  \foreach \i in {0,...,3} {
    \node[qlabel] at (#1,{#2-\i}) {$q_{\i}$};
  }
}


\def\ytop{0}
\def\ybot{-3}

\foreach \i in {0,...,3} { \draw[wire] (0.33,-\i) -- (8.45,-\i); }

\qlabels{0.1}{0}
\node[font=\scriptsize\bfseries, anchor=south east] at (-0.1,0.4) {(a)};

\block{0.81}{\ytop}{\ybot}{\bwi}{$S_{\mathrm{init}}$}
\block{1.95}{\ytop}{\ybot}{\bws}{$U_{\text{step}}$}
\node at (2.93,-1.5) {\scriptsize $\cdots$};
\block{3.90}{\ytop}{\ybot}{\bws}{$U_{\text{step}}$}

\draw[dashed, thick, gray] (4.71,0.4) -- (4.71,-3.4);

\block{5.53}{\ytop}{\ybot}{\bws}{$U_{\text{step}}^{\dagger}$}
\node at (6.50,-1.5) {\scriptsize $\cdots$};
\block{7.48}{\ytop}{\ybot}{\bws}{$U_{\text{step}}^{\dagger}$}

\draw[brace] (1.95-\bws,\ytop+\bh) -- (3.90+\bws,\ytop+\bh)
  node[midway, above=6pt, font=\scriptsize] {$\frac{n}{2}$ steps};
\draw[brace] (5.53-\bws,\ytop+\bh) -- (7.48+\bws,\ytop+\bh)
  node[midway, above=6pt, font=\scriptsize] {$\frac{n}{2}$ steps};


\def\off{-4.5}
\pgfmathsetmacro{\bytop}{0+\off}
\pgfmathsetmacro{\bybot}{-3+\off}

\foreach \i in {0,...,3} { \draw[wire] (1.95,{-\i+\off}) -- (6.18,{-\i+\off}); }

\qlabels{1.79}{\off}
\node[font=\scriptsize\bfseries, anchor=south east] at (-0.1,{0.4+\off}) {(b)};

\block{2.44}{\bytop}{\bybot}{\bwi}{$S_{\mathrm{init}}$}
\block{3.41}{\bytop}{\bybot}{\bws}{$U_{\text{step}}$}
\node at (4.39,{-1.5+\off}) {\scriptsize $\cdots$};
\block{5.36}{\bytop}{\bybot}{\bws}{$U_{\text{step}}$}

\draw[brace] (3.41-\bws,\bytop+\bh) -- (5.36+\bws,\bytop+\bh)
  node[midway, above=6pt, font=\scriptsize] {$n$ steps};

\end{tikzpicture}
\caption{
    Decay recovery circuits.
    (a) The $n$-th echo circuit used to measure the empirical decay rates. $S_{\mathrm{init}}$ prepares an initial Fock state, then $n/2$ forward Trotter steps are applied, followed by $n/2$ inverse Trotter steps. Here, $n$ is assumed to be even.
    (b) The corresponding time evolution Trotter circuit.
}
\label{fig:echo_circuit}
\end{figure}

To further mitigate noisy expectation values obtained from hardware, we apply a damping-reversal, post-processing pass which we dub \textit{decay recovery} and detail below.
Our primary goal with this technique is to empirically correct for the general damping of expectation values caused by hardware noise, as opposed to performing a full inversion of the underlying noise channel.
Much like standard mitigation techniques for a depolarizing noise channel, decay recovery relies on extracting damping factors to compute a rescaling factor for the expectation values~\cite{alam2026onsetergodicityscalesdigital, Seif_2024}. However, because actual hardware noise is rarely a simple depolarizing channel, the efficacy of these standard techniques is limited. To address this, our decay recovery protocol makes no assumptions about the specific hardware noise channel. We note the idea of extracting noise-damping factors to rescale observables has been explored previously for the one-dimensional Fermi-Hubbard model~\cite{Arute2020SpinCharge}. In that work, the authors observed that the damping is largely independent of the interaction strength; by comparing hardware outputs to classical simulations in the tractable, weakly interacting regime, they find the time-dependent noise damping factors and apply them to the strongly interacting regime. While we also compare hardware results against tractable references for calibration, we specifically utilize unitary inversion to isolate the effect of noise from the underlying physics, and our rescaling factors will generally depend on the interaction strength $U$.

We construct a family of \textit{echo circuits} $\{\mathcal{C}_n\}$ with an even number of Trotter steps $n=2k,\,k\in\mathbb{Z}$, and where $n=0$ represents the bare initial state preparation $\ket{\psi_0}$. The $n$-th echo circuit then applies $n/2$ forward Trotter steps followed $n/2$ inverse steps:
\begin{equation}
    \mathcal{C}_n = \bigl(U_\text{step}^{\dagger}\bigr)^\frac{n}{2} \bigl(U_\text{step}\bigr)^\frac{n}{2} S_{\mathrm{init}} \,.
    \label{eq:char_circuit}
\end{equation}
This is depicted in Fig.~\ref{fig:echo_circuit}. In the absence of noise, $\mathcal{C}_n$ implements the identity on $\ket{\psi_0}$ for all $n$. On noisy hardware, the deviation between measured and ideal expectation values grows with circuit depth and varies across physical qubits. By construction, the echo circuit $\mathcal{C}_n$ has the same total depth as a Trotter circuit of $n$ steps.

\begin{figure}[t]
\centering
\includegraphics[width=0.75\columnwidth]{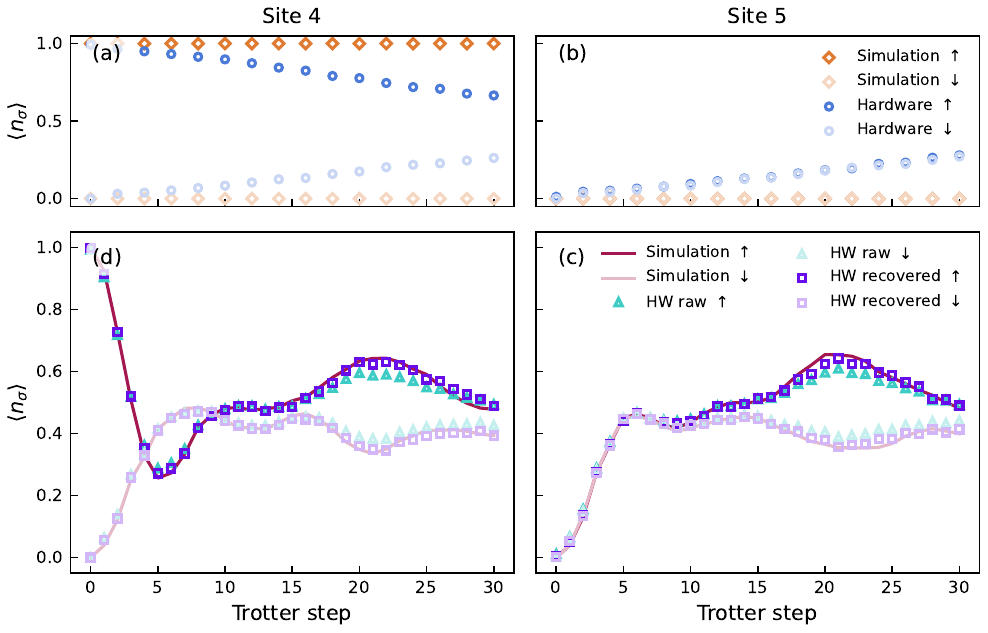}
\caption{
    Decay recovery for a small system with $L=10$ sites, coupling $U/t_h = 2$, and Trotter step $\Delta t = 0.2$, with the initial state taken to be the N\'{e}el state with a central vacancy. Arrows indicate the spin orbitals $\sigma$.
    (a, b) The echo circuit  expectation values for sites $4$ and $5$ (initially spin-up and vacancy, respectively).
    (c, d) The site occupation number expectation value for each spin. The raw and mitigated hardware (HW) results are shown, along with the noiseless simulation values. Error bars are smaller than the markers and have been omitted for clarity.
}
\label{fig:echo_example}
\end{figure}

Next, we detail the decay recovery procedure for a general non-identity Pauli observable $P$. We assume $\langle \psi_0 | P |\psi_0\rangle_n^{\mathrm{echo, ideal}}$ is known and non-vanishing. We note that any observable can be expanded in the Pauli basis, and thus, the decay recovery is applied to each non-identity Pauli term with non-vanishing expectation value individually. Let $\langle P \rangle_{n}^{\mathrm{hw}}$ denote the measured expectation value for the normal Trotter evolution at step $n$ on hardware, and let $\langle P \rangle_{n}^{\mathrm{echo, hw}}$ denote the corresponding noisy expectation value measured from the echo circuit on hardware. We note that $\langle P \rangle_{n}^{\mathrm{echo, hw}}$ can be obtained from hardware only at even Trotter steps $n=2k$. Therefore, we interpolate between the adjacent even-step echo expectation values to estimate the odd-step echo expectation values: $\langle P \rangle_{2k+1}^\text{echo, hw} := \big(\langle P \rangle_{2k+2}^\text{echo, hw} + \langle P \rangle_{2k}^\text{echo, hw}\big)/2$.

Subsequently, we extract the damping factor from the echo circuit at every Trotter step $n$ as ${d_n=\langle P \rangle_{n}^{\mathrm{echo, hw}}/\langle P \rangle_{n}^{\mathrm{echo, ideal}}}$, and normally $d_n$ is positive. This suggests the simple rescaling  $\langle P \rangle_{n}^{\mathrm{mit}} := \langle P \rangle_{n}^{\mathrm{hw}} / d_n$, where $\langle P \rangle_{n}^{\mathrm{mit}}$ is the mitigated Trotter evolution expectation values on hardware. In practice, this bare rescaling often over-compensates for the decay, producing mitigated values that exceed their ideal values. To regularize this behavior, we modify the rescaling procedure as follows:
\begin{equation}
    \label{eq:decay_recovery}
    \langle P \rangle_{n}^{\mathrm{mit}} := \frac{\langle P \rangle_{n}^{\mathrm{hw}}}{ c\, d_n + (1-c)} \,,\quad  d_n=\frac{\langle P \rangle_{n}^{\mathrm{echo, hw}}}{\langle P \rangle_{n}^{\mathrm{echo, ideal}}} \,,
\end{equation}
where $c \in [0,1]$ is a confidence parameter. By computing a convex combination of the absolute echo factor and $1$, we prevent the echo circuits from producing an overly aggressive decay recovery, interpolating between full mitigation ($c=1$) and raw hardware results ($c=0$). The confidence parameter $c$ in Eq.~\eqref{eq:decay_recovery} acts as a single, global scalar applied equally across all sites. To prevent overfitting our error mitigation to the target data, we calibrate this single parameter $c$ strictly out-of-sample. Specifically, we use an independent, small-scale experiment ($L=10$) where exact noiseless simulation remains tractable. This calibration yielded an optimal value of $c \approx 0.5$, which we then fixed for all subsequent decay recovery at larger system sizes. 

Figure~\ref{fig:echo_example} illustrates this method for the site occupation $\langle n_{i, \sigma} \rangle = (1 - \langle Z_{i,\sigma}\rangle)/2$, where $i$ and $\sigma$ index the site and spin. Applying the correction reduces the root-mean-square error (RMSE) (averaged over time) between the hardware and ideal simulation from $2.2\times10^{-2}$ to $1.3\times10^{-2}$, a $41\%$ improvement. Unless otherwise specified, we apply decay recovery with $c = 0.5$ across all larger-scale experiments to mitigate the noisy expectation values. We note that, in practice, generalizing this procedure requires executing dedicated echo circuits to extract the specific decay factors for every unique initial state and measurement basis.

\subsection{Symmetry post-selection}
The Fermi-Hubbard dynamics separately preserve the total number of spin-up and spin-down particles, $[H, N_{\sigma}] = 0$, where $N_{\sigma} = \sum_{i=0}^{L-1} n_{i, \sigma}$ is the total number of particles for each spin species $\sigma$. Equivalently, total particle number $N_{\uparrow} + N_{\downarrow}$ and the $z$-component of the spin $S^z = (N_{\uparrow} - N_{\downarrow}) / 2$ are conserved. This is true for both exact and Trotterized time evolution. As our hardware simulations always begin in a state with a definite number of particles, in particular, Fock states of the form 
\begin{equation}
    \ket{n_{0,\uparrow} \ldots n_{L-1, \downarrow}} = \prod_{i=0}^{L-1} \prod_{\sigma \in \{\uparrow, \downarrow\}} \left( c_{i, \sigma}^{\dagger} \right)^{n_{i, \sigma}} \ket{\mathrm{vac}} \,,  
\end{equation} 
the simulated time evolution should be restricted to $\mathcal{H}_{N_{\uparrow}, N_{\downarrow}}$, the subset of the full $4^L$-dimensional Hilbert space with exactly $N_{\sigma} = \sum_{i=0}^{L-1} n_{i, \sigma}$ particles for spin species $\sigma$. The number of such states is
\begin{equation}
    \dim \mathcal{H}_{N_{\uparrow}, N_{\downarrow}} = \binom{L}{N_{\uparrow}} \times \binom{L}{N_{\downarrow}} \,.
\end{equation}
Near half-filling, $N_{\uparrow} = N_{\downarrow} \approx L/2$, this subset is exponentially large but also exponentially reduced relative to the full Hilbert space. Of course, device noise and imperfections in quantum control cannot be expected to respect the symmetries of the model whose dynamics we wish to digitally simulate. Each gate or idling period will, with some probability, cause the state to exit the $\mathcal{H}_{N_{\uparrow}, N_{\downarrow}}$ subspace. We therefore expect that the degree to which the total particle count of each species is not preserved should grow with circuit depth, at least until a saturation point is reached.

\begin{figure*}[t]
    \centering
    \includegraphics[width=0.75\linewidth]{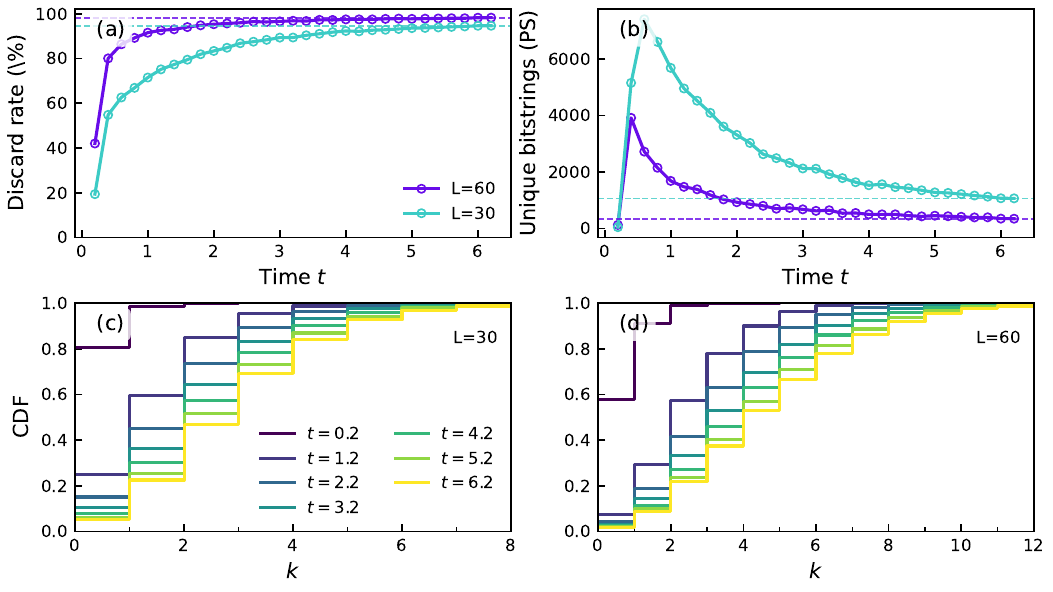}
    \caption{
    Hardware noise characterization via particle-number violation. (a) Post-selection discard rate---the fraction of measurements violating total particle-number conservation---as a function of simulation time, for $L=30$ and $L=60$ N\'{e}el-state simulations with repulsive onsite coupling $U=4$ and $20{,}000$ shots per step. (b) Number of unique post-selected bitstrings (threshold $K=0$); the initial count of 1 reflects the Fock-state initial condition. (c,d) Cumulative distribution function (CDF) of the particle-number violation $k$ [Eq.~\eqref{eq:k_definition}] at successive time snapshots for $L=30$ and $L=60$ respectively. Early in the simulation most measurements preserve particle number exactly ($k=0$), with violations confined to small fluctuations ($k \lesssim 4$). As the simulation progresses, the discard rate grows (cf.~panels (a,b)) and the $k$-distribution broadens, reflecting an accumulation of hardware errors that drive the system far from particle-number conservation.
    }
    \label{fig:discard_rate}
\end{figure*}

This broken symmetry may be put to use for a practical purpose. First, the measured violation rate, $r_{\mathrm{violation}}$, defined as the fraction of measured shots whose Fock states lie outside the particle-number sector of the initial state, provides an indirect diagnostic of device noise and its growth with Trotter step, or equivalently, circuit depth. Second, for Fock-basis observables, one can post-select on the correct particle-number sector by discarding violating shots. In this section, we examine both uses: as a diagnostic for hardware noise and as a possible error-mitigation strategy for Fock-basis expectation values. For the latter, we find that post-selection does not improve the results and often makes them worse; consequently, we do not use this method in the results reported elsewhere in this work. Lastly, we note that violations of a gauge symmetry were recently used to accomplish these tasks in recent work by Ref.~\cite{hartnett2026simulatingdynamicssu2matrix}.

Fig.~\ref{fig:discard_rate}(a) shows the rate of particle-preservation violation for two system sizes, $L=30$ and $L=60$. The discard rate quickly grows with simulation time, and eventually saturates at an $L$-dependent value of $94.7\%$ ($L=30$) and $98.3\%$ ($L=60$). The discard rates are higher for the larger system size, consistent with the larger discrepancy between the dimension of the half-filling Hilbert space and the full $2^{2L}$-dimensional space. Fig.~\ref{fig:discard_rate}(b) illustrates the number of unique bitstrings after post-selection. The simulation begins with a single Fock state, then initially grows as the Trotterized time evolution generates a superposition over Fock states. After the first few time steps, however, leakage from the symmetry-protected subspace $\mathcal{H}_{N_{\uparrow}, N_{\downarrow}}$ becomes significant, and the number of unique bitstrings decreases with the simulation time (or equivalently, with circuit depth).

There is additional structure in the symmetry-violating bitstrings beyond the overall violation rate. To quantify this structure, we define
\begin{equation}
\label{eq:k_definition}
k(s) := \left| N_\uparrow(s) - N_\uparrow^{(0)} \right|  + \left| N_\downarrow(s) - N_\downarrow^{(0)} \right|,
\end{equation}
where $N_\sigma(s)$ is the number of $\sigma$-spin particles in shot $s$, and $N_\sigma^{(0)}$ is the corresponding particle number in the initial state $\ket{\psi_0}$. When the shot $s$ is viewed as a measured bitstring, $k(s)$ measures the total particle-number excess or deficit across the two spin sectors. Figures~\ref{fig:discard_rate}(c,d) show the cumulative distribution of $k$ for the same data shown in panels (a,b). At early times, most symmetry-violating shots have small values of $k$, indicating relatively mild violations of the particle-number constraint. At later times, the distribution shifts toward larger values of $k$, and very few violating shots correspond to only a single particle-number error, $k=1$.

\begin{figure*}[t]
    \centering
    \includegraphics[width=0.75\linewidth]{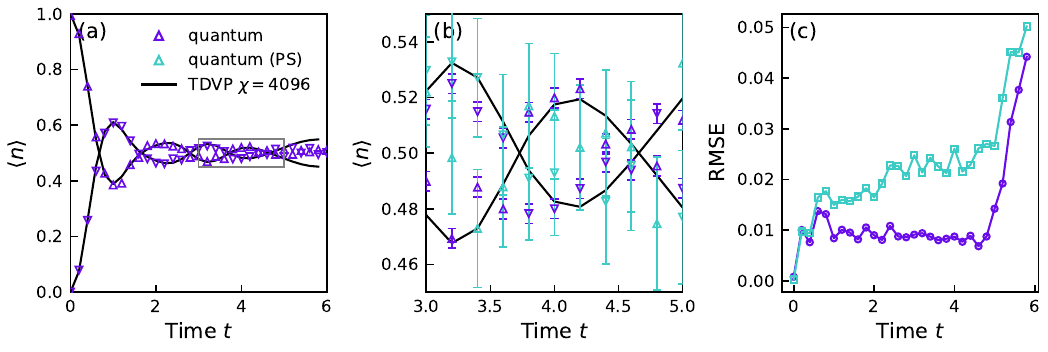}
    \caption{
    Impact of symmetry post-selection. All panels show data for a system with $L=60$ sites, Néel initial state, and coupling $U/t_h = -2$.
    (a) The spin-up and spin-down occupation numbers for the central site (site 29). The direction of the spin is indicated by the orientation of the triangular markers. Hardware results with measurement mitigation and decay recovery (quantum) are shown in purple circles, TDVP with $\chi=2048$ is shown in a solid line.
    (b) Zoomed-in version of (a), with post-selected data shown as well (quantum (PS)).
    (c) The RMSE between the hardware and TDVP results for both the full hardware data (purple) and symmetry post-selected data (teal).
    }
    \label{fig:occupation_rmse}
\end{figure*}

Symmetry post-selection amounts to discarding all shots with $k(s) \neq 0$. This is well motivated, since any such bitstring necessarily signals a detected error. However, when the discard rate is $\gtrsim 90\%$, strict post-selection also dramatically reduces the effective sample size. This tradeoff may be worthwhile only if the retained data improve enough to compensate for the increased statistical uncertainty. In our investigations, we found that symmetry post-selection does not improve the results, and in many cases can significantly reduce the accuracy. This is shown in Fig.~\ref{fig:occupation_rmse}, which uses the same dataset as Fig.~\ref{fig:TDVP_summary_plot} in the main text. Panel (a) shows the time evolution of the central-site occupation in an $L=60$ chain for both spin species, comparing the digital quantum simulation with TDVP. Panel (b) zooms in on the same data and includes the symmetry-post-selected results. Post-selection substantially increases the error bars, and the agreement with TDVP does not appear to improve. A more systematic comparison is shown in panel (c), which reports the RMSE between the quantum results and TDVP, with and without post-selection. Post-selection has little effect at early times, but at intermediate times it significantly worsens the agreement, increasing the RMSE by roughly a factor of two to three.

\section{Spin-charge separation}\label{SI:chargedenisty}
The one-dimensional Fermi-Hubbard model exhibits spin-charge separation: its Bethe-ansatz quasiparticle description separates spin and charge degrees of freedom into distinct excitation branches, conventionally associated with spinons and holons. These branches generally have different characteristic velocities. In dynamical settings, this separation can appear as distinct propagation speeds for spin- and charge-sensitive observable wavefronts. To experimentally measure this phenomenon, we consider dynamical quenches wherein the initial state is a highly excited, non-stationary state of the interacting Fermi-Hubbard system. We consider two such initial states -- the N\'{e}el state and the N\'{e}el state with a local vacancy defect inserted in the center of the 1d chain, corresponding to the absence of a single electron. The first case corresponds to a global quench, and the second to a global quench with an added localized defect. Under time evolution, the global quench evolves and the initial N\'{e}el order melts. Additionally, the localized defect, if present, also propagates outwards and mixes with the rest of the chain. In both cases, the spin and charge degrees of freedom travel at distinct velocities. To measure this, we utilize so-called tracer correlators, $C_i^c(t)$ and $C_i^s(t)$, which are correlation functions designed to be sensitive to just one of the two quantum numbers (charge and spin). These are:
\begin{subequations}
\label{eq:tracer_correlators}
\begin{equation}
    C_i^c(t) := \langle n_{i,\uparrow}(t) + n_{i,\downarrow}(t) \rangle - \langle n_{i,\uparrow}(0) + n_{i,\downarrow}(0) \rangle \,,
\end{equation}
\begin{equation}
    C_i^s(t) := 4 \left( \langle S_i^z(t) S_{i_*}^z(t) \rangle - \langle S_i^z(t) \rangle \langle S_{i_*}^z(t) \rangle \right) \,,
\end{equation}
\end{subequations}
where $S_i^z = (n_{i,\uparrow} - n_{i,\downarrow})/2$ is the $z$-component of the spin at size $i$, and $i_*$ is the site of the vacancy defect. The dynamics of these correlators are characterized by a spreading wavefront. We use a simple protocol to extract the wavefront velocity, which is detailed in Sec.~\ref{sec:wavefront}. 

Since no single closed-form expression remains valid across all coupling strengths $U$, we compare the extracted velocities against analytic predictions in the weak- ($U/t_h \ll 1$) and strong-coupling ($U/t_h \gg 1$) limits separately. The weak-coupling case corresponds to a perturbative treatment around the free fermion $(U=0)$ model and is treated in Sec.~\ref{sec:freefermion}. Exact expressions for $U=0$ are derived for both the charge and spin tracer correlators, from which wavefront velocities can be extracted and compared against our experimental measurements. The charge tracer correlator exhibits a propagating wavefront for the N\'{e}el-with-vacancy initial state only, and the wavefront velocity is found to be $2$ (henceforth, all velocities will be measured in hopping units). In contrast, the wavefront of the spin tracer correlator propagates at a velocity of $4$, independent of whether the initial state contained a vacancy. The calculation also reveals the presence of a second, slower wavefront caused by the localized vacancy that travels with velocity 2. 

These findings are consistent with prior work on spin-chain quenches. Ref.~\cite{bonnes2014light} showed that after a global quench in an integrable spin chain, the wavefront velocity depends on the initial state as well as the energy density injected by the quench. The relevant velocity is set by the maximum group velocity of quasiparticle excitations over the post-quench Generalized Gibbs Ensemble saddle-point state. Relatedly, Ref.~\cite{najafi2018light} showed that for a quench from a N\'{e}el initial state in the XY chain, the translational symmetry of the initial state modifies the effective quasiparticle dispersion, making the light-cone velocity of the connected spin-spin correlator state-dependent — the same mechanism responsible for the $v=4$ wavefront in the spin tracer here.  More broadly, the factor of 2 between the wavefront velocity driven by the global N\'{e}el quench and that driven by the local vacancy quench is consistent with the Calabrese-Cardy quasiparticle picture~\cite{calabrese2006time}: a global quench acts as a spatially extended source of entangled pairs propagating in both directions, doubling the effective light-cone velocity relative to a local quench.

In the strong-coupling regime ($U/t_h \gg 1$), the energetic penalty for double occupancy tightly constrains charge mobility, and the Fermi-Hubbard dynamics map onto an effective $t-J$ model. In the undoped background, this reduces to the antiferromagnetic Heisenberg model with an effective exchange coupling $J = 4t_h^2/U$. However, we do not pursue this angle here; rather, we consider instead the analytic predictions obtainable through the Bethe ansatz. The Bethe ansatz yields exact eigenspectra for the 1D Hubbard model at arbitrary $U$, filling, and system size. For the case of infinite $L$, half-filling, and low-energy excitations above the ground state, analytic expressions may be derived for the velocities of both quasiparticle species, see in particular Essler \emph{et al.}, Ch. 7 \cite{Essler2005}. Importantly, these assumptions are not strictly satisfied in our experiments. We work at finite system size, the vacancy causes a deviation from half-filling (the filling fraction becomes $1-1/L$), and the N\'{e}el background is a highly excited state for finite $U$. However, we expect these expressions to become approximately valid in the strong coupling limit $(U/t_h \gg 1)$ for which the N\'{e}el background becomes an energy eigenstate (in fact, in this limit it becomes one of many degenerate ground states). Therefore, in Sec.~\ref{sec:bethe}, we calculate the quasiparticle velocities using the Bethe ansatz to serve as a strong-coupling baseline. 

\begin{figure*}[t]
  \centering
  \includegraphics[width=0.75\linewidth]{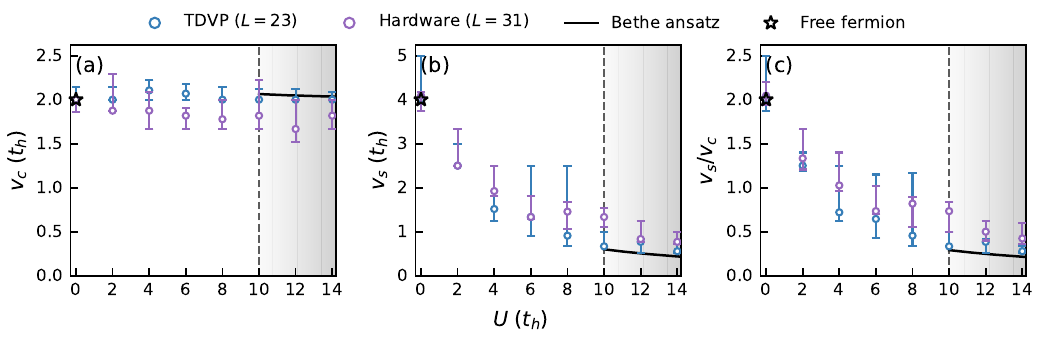}
  \caption{
  Charge and spin velocities extracted from TDVP and quantum hardware.
  (a) Charge velocity $v_c$.
  (b) Spin velocity $v_s$.
  (c) Ratio $v_s / v_c$. 
  Data are the same as in Fig.~\ref{fig:charge_spin_separation_hardware_and_tdvp}. The free-fermion result is shown at $U = 0$, and the Bethe-ansatz predictions are shown only for $U \geq 10$, where the strong-coupling expressions apply. Error bars reflect the sensitivity analysis described in Sec.~\ref{sec:wavefront}.}
  \label{fig:spin_charge_velocities_combined}
\end{figure*}

In Fig.~\ref{fig:spin_charge_velocities_combined}, we reproduce the extracted-velocity plot from Fig.~\ref{fig:charge_spin_separation_hardware_and_tdvp} of the main text and compare the weak- and strong-coupling predictions against our experimental results, each in its regime of validity. The Bethe-ansatz expressions hold in the large-$U$ limit; to indicate their regime of validity we plot them only over $U = 10$ to $14$ and shade this region. Since these values of $U$ are finite, the Bethe-ansatz curves should be understood as approximate even there, becoming exact only as $U \to \infty$. We also show the free-fermion values at $U = 0$. Error bars reflect the sensitivity analysis over the wavefront-detection and velocity-fitting parameters, described in Sec.~\ref{sec:wavefront}.

The remainder of this section is organized as follows. First, Sec.~\ref{sec:freefermion} considers the weak-coupling limit by calculating the tracer correlators in the free fermion theory. Then, Sec.~\ref{sec:bethe} reviews the Bethe ansatz analytic predictions. These are valid for any $U$ but assume low-energy excitations above a ground state. Sec.~\ref{sec:wavefront} provides further details about the wavefront detection algorithm and the velocity estimation method, and finally Sec.~\ref{sec:additional_results} provides additional experimental results.

\subsection{Free fermions and perturbative corrections \label{sec:freefermion}}
When the onsite coupling is set to zero, the model simplifies to a 1D chain of free fermions, and expressions for any correlation function of interest may be straightforwardly derived. Moreover, corrections to these free fermion expressions may be calculated in a perturbative treatment. Here, we calculate the free fermion expressions for the charge and spin tracer correlators and argue that the first order perturbative corrections vanish. The calculation presented here parallels that of Ref.~\cite{white2019correlations}, who computed density and spin correlators in the Fermi-Hubbard model evolving under the noninteracting Hamiltonian; the primary distinction is that we work with a pure Néel initial state rather than a mixed thermal state.

For convenience, we will work with periodic boundary conditions, as opposed to the open boundary conditions considered elsewhere in this work. The distinction will vanish in the thermodynamic limit $L \rightarrow \infty$, which we will eventually take. The Hamiltonian is naturally decomposed into a quadratic free term $H_0$ and the onsite interaction term:
\begin{equation}
    H = H_0 + U \, V \,,
\end{equation}
where
\begin{equation}
  H_0 = \sum_{\sigma} \bm{c}_{\sigma}^{\dagger} h \bm{c}_{\sigma} 
  \,, \qquad V = \sum_{i} n_{i \uparrow} n_{i \downarrow} \,.
\end{equation}
Here, $h$ is the $L \times L$ matrix encoding the nearest neighbor hopping terms, $h_{jl} = -(\delta_{j,l+1} + \delta_{j,l-1})$, and $\bm{c}_{\sigma}$ is the column vector of annihilation operators of spin $\sigma$. The quadratic term is diagonal in the momentum basis: the allowed momenta are $k_n = 2\pi n / L$ for $n = 0, 1, \ldots, L-1$ and the momentum space operators are related to the position space operators by the discrete Fourier transform
\begin{equation}
    \tilde{c}_{n,\sigma} = \frac{1}{\sqrt{L}}\sum_{j} e^{-ik_n j}\,c_{j,\sigma} \,.
\end{equation}
The change-of-basis unitary $W$ has matrix elements $W_{jn} = e^{ik_n j} / \sqrt{L}$ so that $\tilde{\bm{c}}_{\sigma} = W^{\dagger} \bm{c}_{\sigma}$. The free Hamiltonian in this basis is
\begin{equation}
    H_0 = \sum_{n,\sigma} \varepsilon_n\,\tilde{c}_{n,\sigma}^\dagger\,\tilde{c}_{n,\sigma}\,,
\end{equation}
where the $\varepsilon_n := \varepsilon_{k_n} = -2\cos k_n$ are the energies. The group velocity of mode $k$ is given in the continuum limit by $d \epsilon(k)/dk = 2 \sin(k)$, with the maximal velocity given by 2.

\subsubsection{Zeroth order}
For the free fermion system, all observables are determined by the single-particle density matrix
\begin{equation}
  \rho_{jl}^{\sigma}(t) := \langle c_{j,\sigma}^{\dagger}(t)\,c_{l,\sigma}(t)\rangle_0 \,,
\end{equation}
where $\langle \cdot \rangle_0$ denotes an expectation value taken in a free fermion state. We compute this for arbitrary times and initial states by solving the Heisenberg equation of motion for the corresponding operator, $O_{jl}^{\sigma} := c_{j,\sigma}^{\dagger}c_{l,\sigma}$. In the free theory, the equation closes,
\begin{equation}
  \dot{\rho}^{\sigma} = -i\,[\rho^\sigma,\, h] \,,
\end{equation}
where both $\rho^{\sigma}$ and $h$ are $L \times L$ matrices. The formal solution is
\begin{equation}
    \rho^\sigma(t) = e^{iht}\,\rho^\sigma(0)\,e^{-iht}\,.
\end{equation}
This also becomes diagonal in the momentum basis. Defining $\tilde{\rho}^{\sigma} = W^{\dagger} \rho^{\sigma} W$,
the solution is
\begin{equation}
    \tilde{\rho}^{\sigma}_{jl}(t) = e^{i (\epsilon_{k_j} - \epsilon_{k_l})t} \tilde{\rho}^{\sigma}_{jl}(0) \,.
\end{equation}

In this work, we have considered two initial states: the N\'{e}el state (${\rm N}$) and the N\'{e}el state with a vacancy ($\rm Nv$). We will therefore solve for $\rho^{\sigma}(t)$ for these two states. For concreteness, here we use the convention that the N\'{e}el state is $\uparrow \downarrow \uparrow \downarrow \ldots$, and that the vacancy removes a $\uparrow$ spin from the central site. We'll also assume $L$ is even so that the net magnetization is zero before the vacancy defect is added. The 1-pt density matrix of the N\'{e}el background is:
\begin{equation}
    \rho_{jl}^{\sigma, N} = \frac{1}{2} \left(1 + \eta_{\sigma} (-1)^j\right) \delta_{jl} \,,
\end{equation}
where here and throughout we use the convention that $\eta_{\uparrow} = 1$ and $\eta_{\downarrow} = -1$, and the sites are 0-indexed, ${j=0, 1, \ldots, L-1}$. The momentum-space matrix is
\begin{equation}
    \label{eq:singleparticledensitymatrix-Neel}
    \tilde{\rho}^{\sigma,N}_{jl} = \frac{1}{2} \delta_{jl} + \frac{1}{2} \eta_{\sigma} \delta_{k_j,k_l+\pi} = \frac{1}{2} \delta_{jl} + \frac{1}{2} \eta_{\sigma} \delta_{j,l + L/2} \,.
\end{equation}
The second term represents a $\pi$-coherence, as it contributes only when the momenta of the two sites are separated by $\pi$, or equivalently, when the site indices are related by a half-chain shift (with $l+L/2$ understood to be mod $L$).

Similarly, the N\'{e}el-with-vacancy (Nv) expressions are
\begin{equation}
    \rho^{\sigma, Nv}_{jl} = \frac{1}{2} \left(1 + \eta_{\sigma} (-1)^j\right) \delta_{jl} - \delta_{jl} \delta_{\sigma, \uparrow} \delta_{j,L/2} \,, \qquad \tilde{\rho}^{\sigma, Nv}_{jl} = \frac{1}{2} \delta_{jl} + \frac{1}{2} \eta_{\sigma} \delta_{k_j,k_l+\pi} - \frac{\delta_{\sigma, \uparrow}}{L} e^{-i(k_j - k_l) L/2} \,.
\end{equation}
Note that the vacancy term $e^{-i (k_j - k_l) L/2}$ couples all modes, whereas the N\'{e}el background terms only couple modes whose momenta are separated by $\pi$. 

The time-evolved position-space density matrices in the thermodynamic limit are obtained by taking $L\to\infty$, replacing $\frac{1}{L}\sum_n \to \frac{1}{2\pi}\int_0^{2\pi}dk$, and using the integral representation of the Bessel function. This motivates defining the free-fermion single-particle propagation kernel
\begin{equation}
    \mathcal{K}_{j}(t) := \frac{1}{2\pi}\int_0^{2\pi} dk\, e^{ikj}\,e^{-i\varepsilon_k t} = \frac{1}{2\pi}\int_0^{2\pi} dk\, e^{ikj}\,e^{2it\cos k} = i^j J_j(2t) \,,
\end{equation}
so that $(e^{-iht})_{jl} = \mathcal{K}_{j-l}(t)$. Using this, the time-evolved density matrices are:
\begin{align}
    \rho^{\sigma,N}_{jl}(t) &= \frac{1}{2}\delta_{jl} + \frac{\eta_\sigma}{2} (-1)^j \mathcal{K}_{j-l}(2t) \,, \\
    \rho^{\sigma,Nv}_{jl}(t) &= \frac{1}{2}\delta_{jl} + \frac{\eta_\sigma}{2}(-1)^j \mathcal{K}_{j-l}(2t) - \delta_{\sigma,\uparrow} (-1)^{j-l} \, \mathcal{K}_{j-L/2}(t) \, \mathcal{K}_{l-L/2}^*(t) \,.
\end{align}

Finally, the tracer correlators Eq.~\eqref{eq:tracer_correlators} may be computed. The charge tracer correlator is
\begin{equation}
    C_j^c(t) = 
    \begin{cases}
        0 \,, & \quad ({\rm N}) \\
        \delta_{j,L/2} - J_{j-L/2}(2t)^2 \,. & \quad ({\rm Nv})
    \end{cases}
\end{equation}

The spin tracer correlator is $C_i^s(t) = 4\langle S_i^z(t) S_{i_*}^z(t) \rangle_c$ -- four times connected spin-spin correlator with one of the indices taken to be the location of the vacancy defect. It is interesting to consider the case where the two site indices are left unfixed, i.e., the correlator $C^s(j,l;t) := 4 \, \langle S_j^z S_l^z \rangle_{0,c}$. Expanding:
\begin{equation}
    C^s(j,l;t) = \langle n_{j,\uparrow} n_{l,\uparrow} \rangle_0 + \langle n_{j,\downarrow} n_{l,\downarrow} \rangle_0 - \langle n_{j,\uparrow} \rangle_0 \langle n_{l,\uparrow} \rangle_0 + \langle n_{j,\downarrow} \rangle_0 \langle n_{l,\downarrow} \rangle_0 \,,
\end{equation}
where the $\uparrow-\downarrow$ cross terms have been dropped as they cancel because for $U=0$ the two spin species are uncorrelated. Assuming $j\neq l$, the 2-body terms can be simplified using Wick's theorem to yield $\langle n_{j,\uparrow} n_{l,\uparrow} \rangle_0 = \rho^{\uparrow}_{jj}\rho^{\uparrow}_{ll} - |\rho^{\uparrow}_{jl}|^2$ so that the connected correlator becomes simply $C^s(j,l;t) = - \sum_{\sigma} | \rho_{jl}^{\sigma} |^2$. For the autocorrelator case ($j = l$), the fermionic property $n_{j,\sigma}^2 = n_{j,\sigma}$ modifies the expansion to yield $C^s(j,j;t) = \sum_\sigma \left( \rho^\sigma_{jj} - (\rho^\sigma_{jj})^2 \right)$. The spin tracer correlator is therefore
\begin{equation}
    \label{eq:spin_tracer_freefermion}
    C^s(j,l;t) = 
    \begin{cases}
        - \frac{1}{2} J_{j-l}(4t)^2 \,, & \quad j \neq l, \quad ({\rm N}) \\[6pt]
        - \frac{1}{2} J_{j-l}(4t)^2 - J_{j-L/2}(2t)^2 J_{l-L/2}(2t)^2 + (-1)^l J_{j-l}(4t) J_{j-L/2}(2t) J_{l-L/2}(2t) \,, & \quad j \neq l, \quad ({\rm Nv}) \\[6pt]
        \frac{1}{2} \left( 1 - J_0(4t)^2 \right) \,, & \quad j = l, \quad ({\rm N}) \\
        \frac{1}{2} \left( 1 - J_0(4t)^2 \right) - J_{j-L/2}(2t)^4 + (-1)^j J_0(4t) J_{j-L/2}(2t)^2 \,. & \quad j = l, \quad ({\rm Nv})
    \end{cases}
\end{equation}

The $U=0$ calculation is complete; we now turn to analyzing the result. The charge tracer is trivially zero in the N\'{e}el background, and the dynamics are therefore entirely determined by the vacancy when present. The initially localized vacancy defect spreads outwards in a manner determined by the propagator term $J_{j-L/2}(2t)^2$. To understand this in more detail, consider the integral representation of the Bessel function:
\begin{equation}
    J_j(2t) = \frac{1}{\pi} \int_0^{\pi} d\theta\, \cos(j\, \theta - 2t \sin\theta) \,.
\end{equation}
At large $t$, the dominant contribution comes from the stationary phase point $\theta^*$ where $\phi'(\theta^*) = 0$, with ${\phi(\theta) = j \theta - 2t\sin\theta}$. This gives $\cos\theta^* = j/2t$, which has a solution only for $|j| \leq 2t$: the propagator is exponentially suppressed beyond the wavefront at $j = 2t$. Thus, the free propagator spreads with velocity $v=2$, which matches the maximum group velocity computed earlier. Furthermore, near the wavefront $j \approx 2t$, the stationary phase point approaches $\theta^* \to 0$, and the propagator can be approximated as:
\begin{equation}
    J_j(2t) \sim \frac{1}{t^{1/3}}\,\mathrm{Ai}\!\left(\frac{j - 2t}{t^{1/3}}\right), \qquad j \approx 2t \,,
\end{equation}
which indicates a broadening of the wavefront with time (this phenomenon was discussed in the context of the XY chain in Ref.~\cite{najafi2018light}).

The spin tracer correlator differs from the charge tracer in that both the N\'{e}el background and the localized vacancy contribute to the dynamics. The $J_{j-l}(4t)^2$ term in Eq.~\eqref{eq:spin_tracer_freefermion} corresponds to the melting of the initial N\'{e}el order. According to the above discussion, it travels at a velocity of $4$, twice that of the charge correlator. This can be attributed to the $\pi$-coherence term in Eq.~\eqref{eq:singleparticledensitymatrix-Neel}. When the vacancy defect is included, two additional terms contribute. The first, $J_{j-L/2}(2t)^2 J_{l-L/2}(2t)^2$, is entirely due to the defect and corresponds to a wavefront spreading at velocity $2$. The final term, $J_{j-l}(4t) J_{j-L/2}(2t) J_{l-L/2}(2t)$, is a cross term corresponding to the product of both the N\'{e}el background and localized vacancy, and it also spreads at velocity $2$ (the spreading of the product of two wavefronts will be limited by the slower of the two).

\begin{figure*}[t]
    \centering
    \includegraphics[width=0.75\linewidth]{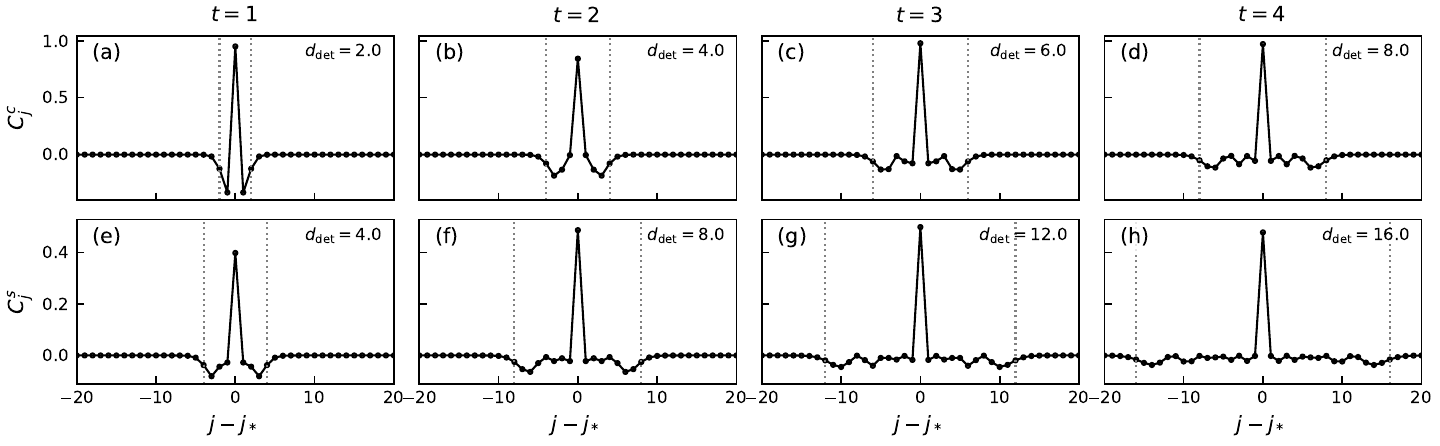}
    \caption{
    The charge and spin tracer correlators for the free fermion case. 
    Panels (a-d): the charge correlator spatial profile at different time snapshots. The wavefront detected by the Algorithm~\ref{alg:wavefront} and described in Sec.~\ref{sec:wavefront} is shown for reference. 
    Panels (e-h): the spin correlator spatial profile at different time snapshots, also with the detected wavefronts shown for reference.
    }
    \label{fig:spin_charge_separation_free_fermion}
\end{figure*}

Figure~\ref{fig:spin_charge_separation_free_fermion} depicts the spatial profile of the charge and spin tracer correlators at different time snapshots for the N\'{e}el-with-vacancy initial state. By definition, both correlators are initially uniformly zero across all sites. Under time evolution, both exhibit a growing central peak and a spreading wavefront. In the charge case, the central peak corresponds to the vacancy delocalizing outward, homogenizing the local charge density at the defect site, while the spreading wavefront at $v^{\rm wf} = 2$ reflects the charge deficit propagating outward. In the spin case, the growing central peak reflects increasing spin uncertainty at the vacancy site: initially empty, the site becomes progressively populated by spin-up and spin-down fermions hopping in from the surrounding N\'{e}el background with approximately equal weight, driving the local spin fluctuation $C^s(i_*, i_*; t)$ away from zero. The spreading wavefront at $v^{\rm wf} = 4$ reflects the melting of the initial N\'{e}el order: the $\pi$-coherence in the N\'{e}el density matrix causes modes $k$ and $k+\pi$ to constructively interfere at frequency $\varepsilon_k - \varepsilon_{k+\pi} = 4\cos k$, giving a maximum spin propagation speed twice that of the charge.

\subsubsection{First order}
Next, we will consider the onsite interaction to first order in perturbation theory around the free fermion point. We work in the interaction picture, where states evolve under the interaction and operators evolve under the free Hamiltonian $H_0$, i.e. $O_I(t) = e^{i H_0 t} O e^{-i H_0 t}$. To first order,
\begin{equation}
    \langle O(t) \rangle = \langle O_I(t) \rangle_0 - iU \int_0^t dt' \langle [O_I(t), V_I(t')] \rangle_0 + \mathcal{O}(U^2) \,,
\end{equation}
where as above $\langle \cdot \rangle_0$ denotes the expectation value in the free theory. 

The first-order corrections to both tracer correlators vanish. The calculation is explicit but tedious, so we omit it here and instead offer a brief justification. The first-order correction to the charge tracer reduces to an integral of $\langle [n_{j\sigma}(t), n_{m\sigma}(t')] \rangle_0$, which evaluates to $2i\,\mathrm{Im}[A^\sigma_{jm}(t,t') B_{jm}(t,t')]$, where $A^\sigma_{jm}(t,t') = \sum_a b_a^\sigma \mathcal{K}^*_{j-a}(t) \mathcal{K}_{m-a}(t')$ and $B_{jm}(t,t') = \mathcal{K}_{j-m}(t-t')$. Here $b_a^\sigma \in \{0,1\}$ is the occupation number of spin $\sigma$ at site $a$ in the initial state. As an aside, we note that $A^\sigma_{jl}(t,t') = -iG^{<,0}_{jl,\sigma}(t,t')$ and $B_{jl}(t,t') = i(G^{R,0}_{jl}(t,t') - G^{A,0}_{jl}(t,t'))$ in standard Keldysh notation, where $G^{<,0}, G^{R,0}, G^{A,0}$ are the lesser, retarded, and advanced free Green's functions, respectively.

The product $A^\sigma_{jm} B_{jm}$ is real, and hence the commutator vanishes identically. To see this, substitute ${\mathcal{K}_j(t) = i^j J_j(2t)}$:
\begin{align}
    A^\sigma_{jm}(t,t')\,B_{jm}(t,t') &= \sum_a b_a^\sigma\,(-i)^{j-a} J_{j-a}(2t)\cdot i^{m-a} J_{m-a}(2t')\cdot i^{j-m} J_{j-m}(2(t-t')) \\
    &= \sum_a b_a^\sigma\, J_{j-a}(2t)\,J_{m-a}(2t')\,J_{j-m}(2(t-t')) \,, \nonumber
\end{align}
where the three factors of $i$ cancel exactly, leaving a manifestly real expression since all Bessel functions are real-valued. Importantly, this result holds for \emph{any} initial state that is a product state in the position-space occupation basis — the N\'{e}el and N\'{e}el-with-vacancy states are two specific examples, but the vanishing is a general consequence of the diagonal structure of the initial density matrix and the propagator. This holds for each spin species separately. The spin tracer correction vanishes by an analogous argument. The first non-zero correction to both tracers therefore appears at $\mathcal{O}(U^2)$.

\subsection{Bethe ansatz \label{sec:bethe}}
Next, we review the calculation of quasiparticle velocities in the Bethe ansatz. As discussed above, although the Bethe ansatz may be applied to the Fermi-Hubbard model at any system size, filling fraction, or state, the analysis here assumes the infinite-size limit, half-filling, zero external magnetic field, and applies to excitations above the ground state. Our treatment follows Ch. 7 of Essler \emph{et al.} \cite{Essler2005} but differs in that we are specifically interested in the wavefront velocity, defined here as the maximal Bethe ansatz quasiparticle group velocity across rapidities.

First, we introduce the dimensionless coupling $u = U / 4 t_h$. Henceforth, all velocities below are in units of $t_h a_0 / \hbar$, with $\hbar =1$, and $J_{\alpha}(x)$ and $I_{\alpha}(x)$ denote the Bessel and modified Bessel functions of the first kind, respectively. The group velocity of a spinon at rapidity $\Lambda$ is
\begin{equation}
    v_{\rm spinon}^{\rm group}(\Lambda) = \frac{dE_s/d\Lambda}{dP_s/d\Lambda} \,,
\end{equation}
where $E_s, P_s$ are the dressed spin energy and momentum, respectively. The spinon dispersion spans $P_s \in [0, \pi]$, with gapless points at both endpoints and a maximum at $P_s = \pi/2$; in terms of rapidity, the endpoints correspond to $|\Lambda| \to \infty$ and the maximum to $\Lambda = 0$. The group velocity therefore vanishes at $\Lambda = 0$ and increases monotonically to its maximum as $|\Lambda| \to \infty$.
Hence $v^{\rm wf}_{\rm spinon} := v_{\rm spinon}^{\rm group}(\infty)$ is the wavefront velocity, and is given by Essler \emph{et al.} Eq. (7.21)
\begin{equation}
    \label{eq:vs_bethe}
    v^{\rm wf}_{\rm spinon} = 2 \frac{I_1\left( \frac{\pi}{2u} \right)}{I_0\left( \frac{\pi}{2u} \right)} \,.
\end{equation}

Similarly, the group velocity for the holon is given by 
\begin{equation}
    v_{\rm holon}^{\rm group}(k) = \frac{d E_h / d k}{d P_h / d k} = \frac{\kappa'(k)}{p'(k)} \,,
\end{equation}
where $\kappa = -E_h$ is the dressed energy and $p = - P_h$ is the dressed momentum. Differentiating Eqs.~(7.10) and (7.12) of Essler \emph{et al.} gives
\begin{equation}
    \kappa'(k) = 2\sin k + 2\cos(k)\int_0^\infty d\omega \, \frac{J_1(\omega)\sin(\omega \sin k)\,e^{-\omega u}}{\cosh(\omega u)} \,,
\end{equation}
and
\begin{equation}
    p'(k) = 1 + 2\cos k \int_0^\infty d\omega\, \frac{J_0(\omega)\cos(\omega \sin k)}{1 + e^{2\omega u}} \,.
\end{equation}
Maximizing over the rapidity gives
\begin{equation}
    \label{eq:vc_bethe}
    v^{\rm wf}_{\rm holon} = \max_k \frac{\kappa'(k)}{p'(k)} \,,
\end{equation}
which may be evaluated numerically. 

Next, to draw a distinction between the perturbative calculation of the previous subsection, we consider the small coupling expansion of the above wavefront velocities. We first state the result and then provide the derivation. To first order in $u$, the wavefront velocities are:
\begin{equation}
    \label{eq:vc_small_u_expansion}
    v^{\rm wf}_{\rm spinon}(u) = 2 - \frac{2}{\pi} \, u + O(u^2) \,, \qquad 
    v^{\rm wf}_{\rm holon}(u) = 2 + \frac{2}{\pi} \, u + O(u^2) \,.
\end{equation}
The spinon expression is obtained by making use of the asymptotic expansion of the Bessel functions. The maximization over rapidities in Eq.~\eqref{eq:vc_bethe} complicates the calculation for the holon; we will first perform the integrations in the expressions for $\kappa'(k)$ and $p'(k)$ and then perform the maximization.

First, define $a := \sin k$ for convenience and assume that $k \in (\pi/2, \pi)$, so that $\cos k < 0$. Next, introduce the integrals:
\begin{equation}
    K(a,u) = \int_0^{\infty} d\omega  J_1(\omega) \frac{2 \sin(a \omega)}{1 + e^{2 w u}} \,, \qquad P(a,u) = \int_0^{\infty} d\omega J_0(\omega) \frac{2\cos(a \omega)}{1 + e^{2w u}} \,,
\end{equation}
so that
\begin{equation}
    \kappa'(a,u) = 2a - 2 \sqrt{1-a^2}K(a,u)  \,, \qquad p'(a,u) = 1 - \sqrt{1-a^2} P(a,u) \,.
\end{equation}
We have that $K(a,0) = a/\sqrt{1-a^2}$ and $P(a,0) = 1/\sqrt{1-a^2}$, so that $\kappa'(a,0) = p'(a,0) = 0$. Define the differences
\begin{equation}
    \Delta K(a,u) \equiv K(a,0) - K(a,u) \,, \qquad \Delta P(a,u) \equiv P(a,0) - P(a,u) \,,
\end{equation}
so that the group velocity numerator and denominator can be written as
\begin{equation}
    \kappa'(a,u) = 2\sqrt{1-a^2}\,\Delta K(a,u) \,, \qquad p'(a,u) = \sqrt{1-a^2}\,\Delta P(a,u) \,.
\end{equation}
These are:
\begin{equation}
    \Delta K(a,u) = \int_0^{\infty} d\omega J_1(\omega) \sin(a \omega) \tanh(\omega u) \,, \qquad \Delta P(a,u) = \int_0^{\infty} d\omega J_0(\omega) \cos(a \omega) \tanh(\omega u) \,.
\end{equation}

\begin{figure}[t]
\centering
\resizebox{0.65\linewidth}{!}{\begin{tikzpicture}[>=Stealth, scale=1.5]
    \tikzmath{\w0 = 1; \w1 = 3; \w2 = 5; \R = 4; \arcR = 4;}

    \draw[thick, ->] (-5,0) -- (5,0) node[right] {$\mathrm{Re}(\omega)$};
    \draw[thick, ->] (0,-1.5) -- (0,6) node[above] {$\mathrm{Im}(\omega)$};
    \node at (0,0) [below left] {$0$};

    \draw[black, thick, postaction={decorate, decoration={
        markings,
        mark=at position 0.15 with {\arrow{>}},
        mark=at position 0.35 with {\arrow{>}},
        mark=at position 0.60 with {\arrow{>}},
        mark=at position 0.85 with {\arrow{>}}
    }}]
        (-\R, 0) -- (\R, 0)
        arc (0:180:\arcR) -- cycle;

    \node[black, thick] at (0.75*\arcR, 0.75*\arcR) {$C$};

    \draw[thick] plot[mark=x, mark size=4pt] coordinates {(0,\w0)};
    \node[black, anchor=west] at (0.2, \w0) {$\omega_0 = \frac{i\pi}{2u}$};

    \draw[thick] plot[mark=x, mark size=4pt] coordinates {(0,\w1)};
    \node[black, anchor=west] at (0.2, \w1) {$\omega_1 = \frac{3i\pi}{2u}$};

    \draw[dotted, thick, black] (0, \w2-0.5) -- (0, \w2+0.5);
    \node[black, anchor=west] at (0.2, \w2) {$\dots$};

    \draw[->, thick, gray] (-1.2, 2) -- (-1.2, 3.5);
    \node[gray, align=left, text width=3cm, font=\small] at (-1.2, 1.3) {
        Poles migrate to $+i\infty$ as $u \rightarrow 0$
    };

\end{tikzpicture}}
\caption{
\label{fig:integration_contour}
    Integration contour for the $\Delta K$ and $\Delta P$ integrals. There are an infinite tower of Matsubara frequencies which migrate to $\pm i \infty$ as $u \rightarrow 0$. The dominant contribution is due to $\omega_0$.
}
\end{figure}
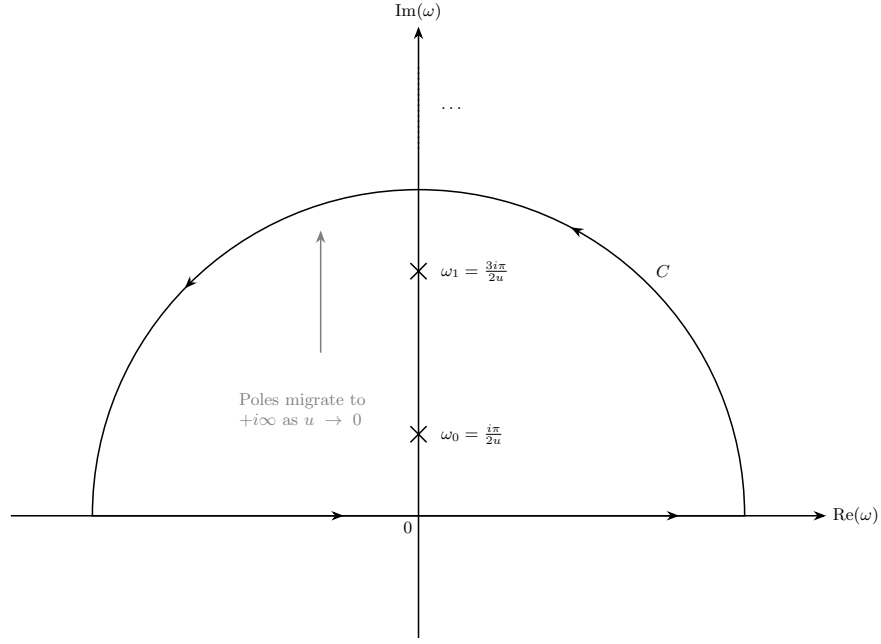

Each integral may be computed by deforming the integration contour to a closed semi-circle and applying the residue theorem, see Fig.~\ref{fig:integration_contour}. First, extend the limit from $\omega \in [0, \infty)$ to the entire real line $\omega \in \mathbb{R}$. To do so, write ${J_{\alpha}(\omega) = (H_{\alpha}^{(1)}(\omega) + H_{\alpha}^{(2)}(\omega))/2}$, where $H^{(1,2)}_{\alpha}$ are the Hankel functions of the first and second kind, respectively. Next, use the fact that $H_{\alpha}^{(2)}(\omega) = (-1)^{\alpha + 1} H_{\alpha}^{(1)}(-\omega)$ to obtain
\begin{equation}
    \Delta K(a,u) = \frac{1}{2} \int_{-\infty}^{\infty} d\omega H^{(1)}_1(\omega) \sin(a \omega) \tanh(\omega u) \,, \qquad \Delta P(a,u) = \frac{1}{2}  \int_{-\infty}^{\infty} d\omega H^{(1)}_0(\omega) \cos(a \omega) \tanh(\omega u) \,.
\end{equation}
The integration contour may then be closed in the upper half plane to form a semi-circle. This is possible because the Hankel functions dominate the behavior as $\mathrm{Im}(\omega) \rightarrow i \infty$ and ensure convergence, thus:
\begin{equation}
    \Delta K(a,u) = \frac{1}{2} \oint_{C} d\omega H^{(1)}_1(\omega) \sin(a \omega) \tanh(\omega u) \,, \qquad \Delta P(a,u) = \frac{1}{2} \oint_{C} d\omega H^{(1)}_0(\omega) \cos(a \omega) \tanh(\omega u) \,.
\end{equation}

The $\tanh(\omega u)$ factor leads to a series of poles at the frequencies:
\begin{equation}
    \tanh(\omega u) \sim \frac{1}{u} \frac{1}{\omega - \omega_n} \,, \quad \omega_n = \frac{i \pi}{2 u} (1 + 2n) \,, \quad n = 0, \pm1, \pm 2, \ldots
\end{equation}
The frequencies contained in the contour $C$ are those with $n=0,1,2,\ldots$. As $u\rightarrow 0$, these move to $+i \infty$. In this limit, the leading contribution is from the $n=0$ frequency. The integrals in this limit are therefore:
\begin{equation}
    \Delta K(a,u) \sim \frac{i \pi}{u} H_1^{(1)}(\omega_0) \sin(a \omega_0) \,, \quad \Delta P(a,u) \sim \frac{i \pi}{u} H_0^{(1)}(\omega_0) \cos(a \omega_0) \,.
\end{equation}
To evaluate this, we'll use the asymptotic form for the Hankel functions (assuming ${-\pi < \arg z < 2\pi})$:
\begin{equation}
    H_\alpha^{(1)}(z) \sim \sqrt{\frac{2}{\pi z}} e^{i\left(z - \frac{\alpha\pi}{2} - \frac{\pi}{4}\right)} \left[ 1 - \frac{4\alpha^2 - 1}{8iz} + O\left(\frac{1}{z^2}\right) \right] \,.
\end{equation}
This results in
\begin{equation}
    \Delta K \sim \frac{2}{\sqrt{u}} \sinh\left( \frac{a \pi}{2 u} \right) e^{-\frac{\pi}{2u}} \left(1 + \frac{3u}{4\pi} + \mathcal{O}(u^2) \right) \,, \qquad \Delta P \sim \frac{2}{\sqrt{u}} \cosh\left( \frac{a \pi}{2u} \right) e^{-\frac{\pi}{2u}} \left(1 - \frac{u}{4\pi} + \mathcal{O}(u^2) \right) \,.
\end{equation}
From these, the holon group velocity at rapidity $k$ is found to be:
\begin{equation}
    v_{\rm holon}^{\rm group}(k) = \frac{ \kappa'(k,u) }{ p'(k,u) } = \frac{2 \Delta K(k,u)}{\Delta P(k,u)} = 2 \tanh\left( \frac{\pi \sin(k)}{2 u} \right) \left(1 + \frac{u}{\pi} + \mathcal{O}(u^2) \right) \,.
\end{equation}
To find the wavefront velocity, we must maximize $v_{\rm holon}^{\rm group}(k)$ with respect to the rapidity $k \in (\pi/2, \pi)$. The $k$-dependence lies entirely within the $\tanh$ function, which monotonically approaches $1$ for large arguments. Therefore, the velocity is maximized when the argument is large, requiring $\sin k \gg u$. In the small coupling limit $u \to 0$, this condition is satisfied by a broad, continuous range of rapidities. Across this entire regime, the $\tanh$ factor saturates to $1$, making the velocity independent of $k$. Because of this flat plateau, there is no unique maximizing rapidity $k_*$; rather, an entire band of modes travels at the maximum speed. The final result is given above in Eq.~\eqref{eq:vc_small_u_expansion}. The leading coefficient of $2/\pi$ agrees with the slope at $u=0$ extracted from direct numerical evaluation of Eq.~\eqref{eq:vc_bethe}.

\subsection{Wavefront detection and velocity extraction \label{sec:wavefront}}
In this section, we detail the method used to identify the charge and spin wavefronts for the dynamical quench experiments where the initial state is taken to be the N\'{e}el state with a central vacancy defect, that is 
\begin{equation}
    \ket{\psi_0} = \ket{\uparrow, \downarrow, \ldots , \uparrow, \downarrow, \circ, \downarrow, \uparrow, \ldots, \downarrow, \uparrow} \,.    
\end{equation}
Under time evolution, the defect creates a spreading wavefront which travels outwards to the boundary, where it eventually reflects. Consistent with spin-charge separation, this disturbance is a composite structure with contributions from distinct wavefronts for both charge and spin degrees of freedom. 

The distinct wavefronts may be isolated by examining observables sensitive to spin and charge separately. For this purpose, we introduce tracer correlators. The tracer correlator for charge is
\begin{equation}
   C^{\rm c}_i(t) = \langle n_{i,\uparrow}(t) + n_{i,\downarrow}(t)\rangle  - \langle n_{i,\uparrow}(0) + n_{i,\downarrow}(0)\rangle \,,
\end{equation}
which is simply the difference between the per-site electron density at time $t$ from the value in the initial state. The tracer correlator for spin is the connected spin-spin correlation
\begin{equation}
    C^{\rm s}_i(t) = 4(\langle S_i^z(t) S_{i_*}^z(t) \rangle - \langle S_i^z(t) \rangle \langle S_{i_*}^z(t) \rangle) \,,
\end{equation}
where $i_*$ is the site index of the central defect. Because the initial state is a Fock state, all connected correlators vanish at $t=0$, so $C_i^s(0) = 0$ without any explicit subtraction — in contrast to the charge tracer, where the subtraction is needed to remove the nonzero background density. The values for these correlators across all sites and across all sampled time points are obtained either from classical TDVP simulations or from quantum hardware executions. 

\begin{figure*}[t]
  \centering
  \includegraphics[width=0.75\linewidth]{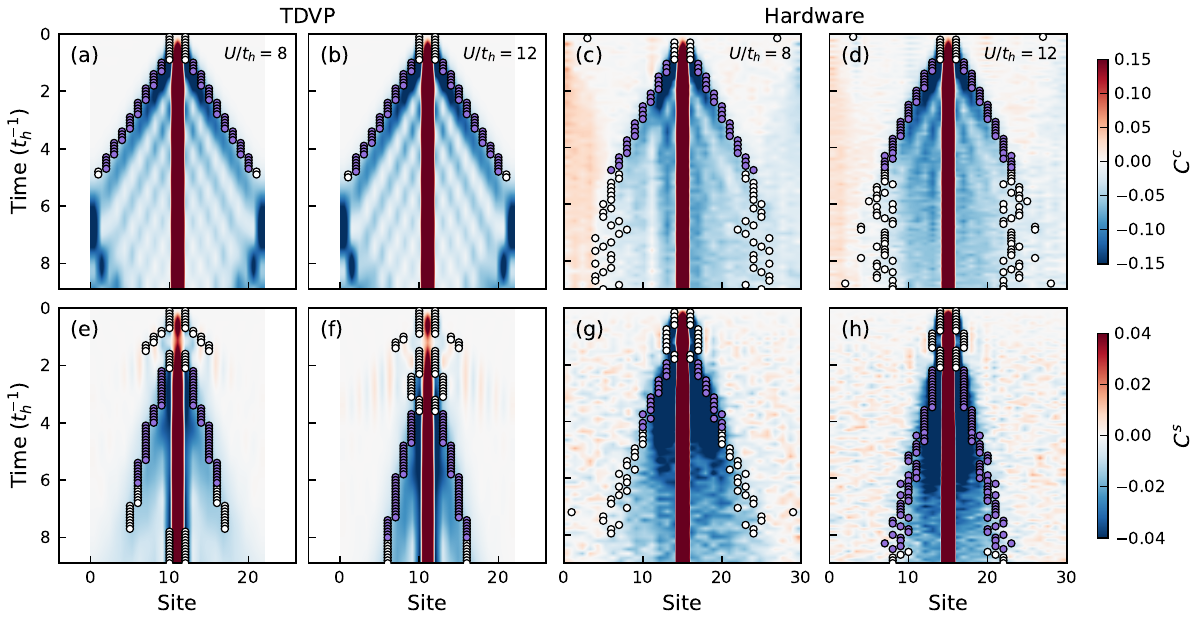}
  \caption{
  Wavefront detection algorithm applied to the dynamical quench of a N\'{e}el state initialized with a central vacancy defect. Markers indicate the identified left and right boundaries of the propagating wavefronts. The velocity fit uses only the purple markers; white markers are excluded. Heatmaps depict the charge tracer correlator $C_i^c(t)$ for (a, b) classical TDVP and (c, d) quantum hardware simulations. The corresponding spin tracer correlator $C_i^s(t)$ is shown for (e, f) TDVP and (g, h) hardware simulations. Data is presented for interaction strengths $U/t_h \in \{8, 12\}$ up to a total evolution time of $t = 9\, t_h^{-1}$. All TDVP simulations use a Trotter step size of $\Delta t = 0.1\, t_h^{-1}$. Hardware simulations use $\Delta t = 0.15\, t_h^{-1}$ for $U=8\, t_h$ and $\Delta t = 0.1\, t_h^{-1}$ for $U=12\, t_h$.
  } 
  \label{fig:wavefront_detection_points}
\end{figure*}

We use a wavefront-detection algorithm to identify the leading edge of the spreading charge or spin disturbance from the corresponding tracer correlator. The detection acts on the spatial profile $C_i(t)$ at fixed $t$. Since the underlying dynamics are reflection-symmetric about the vacancy site $i = i_*$ whereas hardware noise generically breaks this symmetry, we first symmetrize the profile about $i_*$, averaging out the antisymmetric noise component. We then take the absolute value and normalize by the maximum over all sites except $i_*$. On each side independently, the wavefront is identified as the outermost site whose normalized amplitude exceeds a fixed fraction $p$, and its distance from $i_*$ is recorded. This procedure is given in Algorithm~\ref{alg:wavefront}. We validated it against the analytic free-fermion ($U=0$) tracer correlators; see Fig.~\ref{fig:spin_charge_separation_free_fermion}. Next, a separate procedure extracts a velocity for each front (left and right) using the Theil--Sen estimator \cite{sen1968estimates, conover1999practical}, a robust non-parametric linear regression insensitive to outliers. Prior to fitting, detections too close to the vacancy or to the chain boundary are discarded, and the surviving detections are partitioned into contiguous blocks so that early-time transients and other anomalous detections are excluded from the largest block that is fit. Late-time points, where the wavefront signal deteriorates, are removed by trimming the tail of this block. The full procedure is detailed in Algorithm~\ref{alg:velocity}. Across both algorithms, we use the parameter values $p=0.3$ (wavefront detection threshold), $d_{\mathrm{min}} = b = 2$ (vacancy and boundary buffers), $n_{\mathrm{min}} = 5$ (minimum contiguous block size), and $q = 0.75$ (quantile). Figure~\ref{fig:wavefront_detection_points} illustrates the detected wavefronts and those retained by the velocity estimation algorithm (shown in purple).

\begin{algorithm}[H]
\caption{Wavefront Detection}
\label{alg:wavefront}
\begin{algorithmic}[1]
\State \textbf{Inputs:} tracer correlator $C_i(t)$; vacancy site $i_*$; threshold fraction $p$
\State \textbf{Output:} detected wavefront points $P_{\rm left}$, $P_{\rm right}$
\vspace{0.2cm}
\For{each time step $t$}
    \State Symmetrize the profile about $i_*$: $C_i \leftarrow \tfrac{1}{2}\bigl(C_i + C_{2i_* - i}\bigr)$
    \State Form the normalized profile $\hat{C}_i(t) = |C_i(t)| \,/\, \max_{i \neq i_*} |C_i(t)|$
    \State $x_{\rm left}(t) \leftarrow$ distance from $i_*$ to the outermost site \emph{left} of $i_*$ with $\hat{C}_i(t) > p$, if any
    \State $x_{\rm right}(t) \leftarrow$ distance from $i_*$ to the outermost site \emph{right} of $i_*$ with $\hat{C}_i(t) > p$, if any
\EndFor
\State $P_{\rm left} \leftarrow \{(t,\, x_{\rm left}(t))\}$, \quad $P_{\rm right} \leftarrow \{(t,\, x_{\rm right}(t))\}$
\State \Return $P_{\rm left}$, $P_{\rm right}$
\end{algorithmic}
\end{algorithm}
\begin{algorithm}[H]

\caption{Velocity Estimation}
\label{alg:velocity}
\begin{algorithmic}[1]
\State \textbf{Data:} detected points $P_{\rm left}$, $P_{\rm right}$
\State \textbf{Masking parameters:} minimum distance from vacancy $d_{\min}$; boundary buffer $b$; distance $x_{\rm edge}$ from $i_*$
\State \textbf{Fitting parameters:} minimum block size $n_{\min}$; velocity quantile $q$
\State \textbf{Output:} wavefront velocity $v$
\vspace{0.2cm}
\For{$s \in \{\mathrm{left},\,\mathrm{right}\}$}
    \State Discard points of $P_s$ with $x_s(t) < d_{\min}$ \Comment{too close to the vacancy}
    \State Discard points of $P_s$ with $x_{\rm edge} - x_s(t) < b$ \Comment{too close to the boundary}
    \State Partition the surviving points into maximal contiguous blocks of consecutive detections
    \State Let $B = (b_1, \dots, b_N)$ be the largest contiguous block
    \Statex \hspace{1em}$\triangleright$ Trim the block's final points, which can be anomalous as the wavefront dissipates:
    \For{$k = N, N-1, \dots, n_{\min}$}
        \State Fit the leading sub-block $B[1\!:\!k]$ by Theil--Sen regression to obtain velocity $v_k$
    \EndFor
    \State $\tau \leftarrow \mathrm{Quantile}_q\bigl(\{v_k\}_{k=n_{\min}}^{N}\bigr)$
    \State $k^\star \leftarrow \max\{\,k : v_k \ge \tau\,\}$ \Comment{largest sub-block reaching the $q$-th quantile velocity}
    \State $v_s \leftarrow v_{k^\star}$
\EndFor
\State \Return $v = \tfrac{1}{2}\bigl(v_{\rm left} + v_{\rm right}\bigr)$
\end{algorithmic}
\end{algorithm}

The extracted velocities are sensitive to the values of the parameters of both the detection and fitting stages. To quantify this sensitivity, we performed a scan over all combinations of the detection threshold fraction $p$ and the fitting velocity quantile $q$: $p \in \{0.15,\, 0.225,\, 0.30,\, 0.375,\, 0.45\}$ and $q \in \{0.60,\, 0.675,\, 0.75,\, 0.825,\, 0.90\}$, yielding 25 velocity estimates per dataset. The error bars in Fig.~\ref{fig:spin_charge_velocities_combined} and Fig.~\ref{fig:charge_spin_separation_hardware_and_tdvp} represent the envelope of this scan---the minimum and maximum velocity extracted across all parameter combinations---and thus reflect the systematic uncertainty due to the choice of detection and fitting parameters.

\subsection{Additional results \label{sec:additional_results}}
\begin{figure*}[t]
  \centering
  \includegraphics[width=\linewidth]{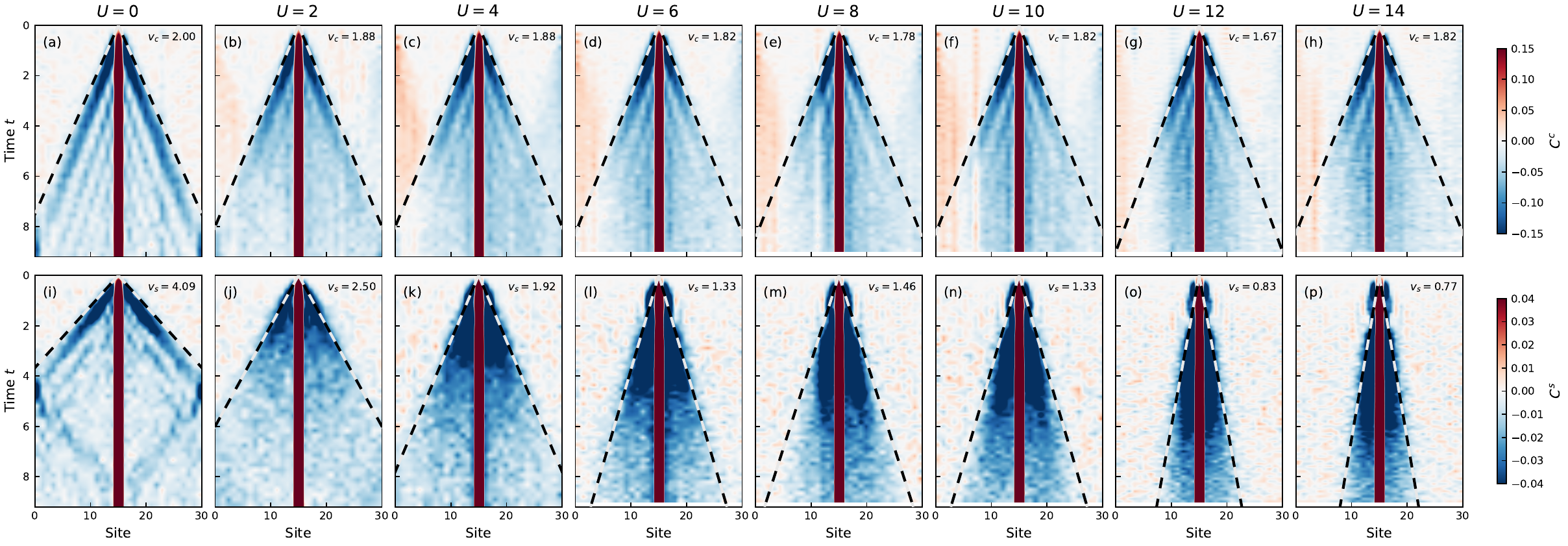}
  \caption{
  Quantum hardware simulation of spin-charge separation across varying interaction strengths. The time evolution of a $L=31$ N\'{e}el state initialized with a central vacancy over a range of repulsive couplings is simulated. The Trotter step size and number of steps depend on the coupling: for $U \in \{0, 2, 4\}\, t_h$, $\Delta t = 0.2\, t_h^{-1}$ over $46$ steps (total simulation time $t = 9.2\, t_h^{-1}$); for $U \in \{6, 8, 10\}\, t_h$, $\Delta t = 0.15\, t_h^{-1}$ over $60$ steps ($t = 9\, t_h^{-1}$); and for $U \in \{12, 14\}\, t_h$, $\Delta t = 0.1\, t_h^{-1}$ over $90$ steps ($t = 9\, t_h^{-1}$).
  (a--h) Space-time heatmaps of the charge tracer correlator $C^c_i(t)$. The wavefront boundary is indicated by the dashed line. The extracted charge wavefront velocity $v_c$ is reported in each panel.
  (i--p) Space-time heatmaps of the spin tracer observable $C^s_i(t)$. The same wavefront extraction method is applied to determine the spin wavefront velocity $v_s$. For both sets of panels, a mild Gaussian smoothing has been applied strictly for visualization purposes.
  }
  \label{fig:spin_charge_separation_hardware}
\end{figure*}

\begin{figure*}[t]
  \centering
  \includegraphics[width=\linewidth]{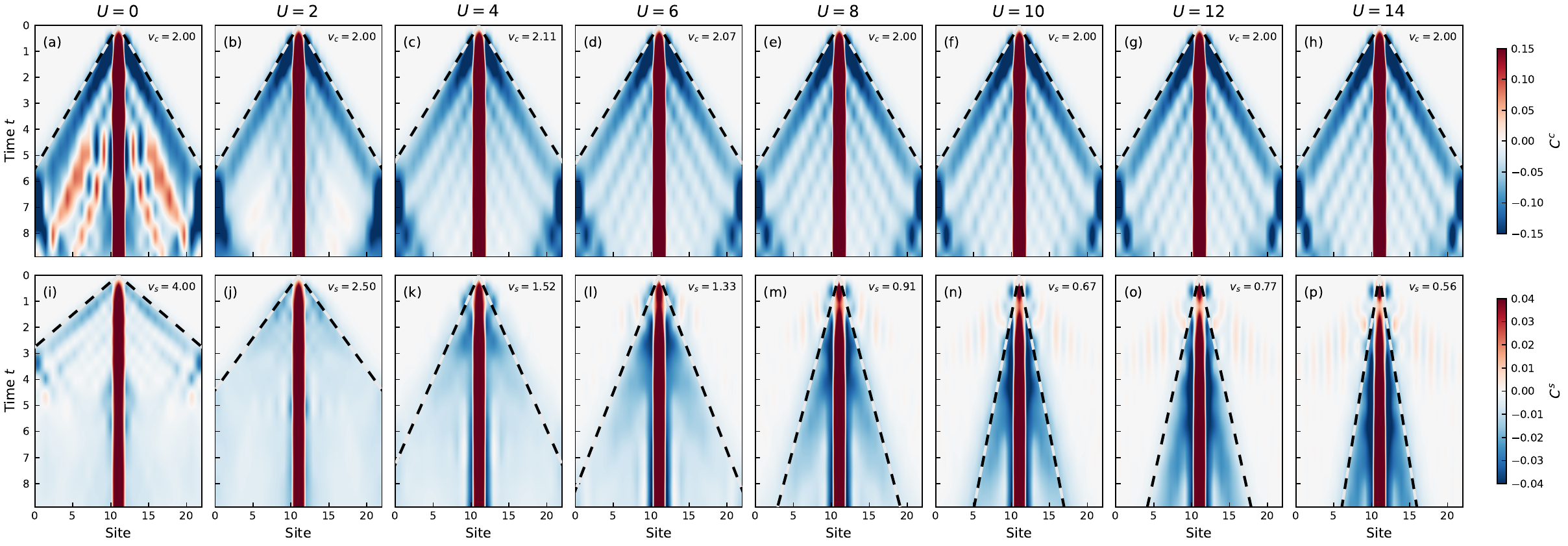}
  \caption{
  TDVP simulation of spin-charge separation across varying interaction strengths. All TDVP simulations used a maximum bond dimension of $\chi=1024$ and $L=23$. As in Fig.~\ref{fig:spin_charge_separation_hardware}, the time evolution of a N\'{e}el state initialized with a central vacancy over a range of repulsive couplings is simulated. All simulations are evolved to a total time of $T=9 \, t_h^{-1}$, with $\Delta t = 0.1 \, t_h^{-1}$.
  (a--h) Space-time heatmaps of the charge tracer correlator $C^c_i(t)$. The wavefront boundary is indicated by the dashed line. The extracted charge velocity $v_c$ is reported in each panel.
  (i--p) Space-time heatmaps of the spin tracer observable $C_i^s(t)$. The same wavefront extraction method is applied to determine the spin velocity $v_s$. For both sets of panels, a mild Gaussian smoothing has been applied strictly for visualization purposes.  
  }
  \label{fig:spin_charge_separation_tdvp}
\end{figure*}

In this section, we present extended data supporting our experimental observations of spin-charge separation. Figure~\ref{fig:spin_charge_separation_hardware} provides additional hardware results across a wider range of repulsive couplings than those featured in the main text (Fig.~\ref{fig:charge_spin_separation_hardware_and_tdvp}). To validate these findings, we performed classical TDVP simulations of identical quench dynamics across the corresponding interaction strengths, with the results detailed in Fig.~\ref{fig:spin_charge_separation_tdvp}. Velocities are extracted using the wavefront detection algorithm described above (Algorithms~\ref{alg:wavefront} and \ref{alg:velocity}). The extracted velocities are reported in Fig.~\ref{fig:charge_spin_separation_hardware_and_tdvp}(g-i) in the main text.

%

\end{bibunit}

\end{document}